\documentstyle[epsf,epsfig,12pt]{article}
\begin{document}

\def\gbb{$\Gamma_{b \bar{b}}\;$}
\newcommand{\ztobb}{\mathrm{Z^0 \rightarrow b\overline{b}}}
\newcommand{\micron}{\mbox{$\mu \mathrm{m}$}}
\newcommand{\Zz}{{\ifmmode Z \else $Z$\fi}\ }
\newcommand{\rphi}{\mathrm{$r\phi $ }}
\newcommand{\taub}{\mbox{$\tau_{\mathrm{B}}$}}
\newcommand{\rbb}{\mbox{$R_b\ $}}
\newcommand{\qq}{{\mathrm q\overline{\rm q}}}
\newcommand{\bb}{\mathrm{b\overline{b}}}
\newcommand{\cc}{{\rm c\overline{\rm c}}}
\newcommand{\Dstar}{{\mathrm D}^*}
\newcommand{\Dstarp}{{\rm D}^{*+}}
\newcommand{\Dstarm}{{\rm D}^{*-}}
\newcommand{\Dstarpm}{{\rm D}^{*\pm}}
\newcommand{\Dzero}{{\rm D}^0}
\newcommand{\Dplus}{{\rm D}^+}
\newcommand{\Dzerob}{{\overline{{\rm D}^0}}}
\newcommand{\Bzero}{{\mathrm B}^0}
\newcommand{\Bzerob}{{\overline{{\rm B}^0}}}
\newcommand{\Bplus}{{\mathrm B}^+}
\newcommand{\Bs}{{\rm{B_s}}}
\newcommand{\Bsb}{{\overline{{\rm B}_{\mathrm s}}}}
\newcommand{\Lb}{{\rm{\Lambda_b}}}
\newcommand{\clX}{\rm c \rightarrow \ell X}
\newcommand{\BrclX}{Br(\clX)}
\newcommand{\Brbl}{\mathrm{Br(b\rightarrow \ell)}}
\newcommand{\Brbcl}{\mathrm{Br(b\rightarrow c \rightarrow \ell)}}
\newcommand{\Brcl}{\mathrm{Br(c \rightarrow \ell)}}
\newcommand{\xE}{\left<x_E\right>}
\newcommand{\xEb}{\left<x_E({\rm b})\right>}
\newcommand{\xEc}{\left<x_E({\rm c})\right>}
\newcommand{\Ds}{{\rm{D_s}}}
\newcommand{\Lc}{{\rm{\Lambda_c}}}
\newcommand{\bclp}{\rm{b \rightarrow c \rightarrow \ell^+}}
\newcommand{\bclm}{\rm{b \rightarrow \overline{c} \rightarrow \ell^-}}
\newcommand{\cl}{\rm c \rightarrow \ell }
\newcommand{\bl}{\rm b \rightarrow \ell }
\newcommand{\bD}{\mathrm{ b \rightarrow D} }
\newcommand{\bu}{\mathrm{ b \rightarrow u} }
\newcommand{\ra}{\rightarrow }
\newcommand{\MeV}{\mathrm{MeV}}
\newcommand{\GeV}{\mathrm{GeV}}
\newcommand{\dRc}{\frac{R_c-0.172}{0.172}}

\pagenumbering{arabic}

\begin{titlepage}

\pagenumbering{arabic}
\vspace*{-1.5cm}
\begin{tabular*}{15.cm}{lc@{\extracolsep{\fill}}r}
& 
&
DELPHI 2000-069 PHYS 868 \\
& &
IEKP-KA/2001-3  
\\
\\
\end{tabular*}
\vspace*{1.cm}
\begin{center}
\Large 
{\bf \boldmath
 BSAURUS- A Package For Inclusive B-Reconstruction in DELPHI.
} \\
\vspace*{1.5cm}
\normalsize { 
   {\bf Z. Albrecht, T. Allmendinger, G. Barker}\\
   {\bf M. Feindt, C. Haag, M. Moch.}\\
   {\footnotesize University of Karlsruhe}\\
   
}
\vspace*{1.5cm}
\end{center}
\vspace{\fill}
\begin{abstract}
\noindent
BSAURUS is a software package for the inclusive reconstruction of B-hadrons
in ${\mathrm Z}^0 \rightarrow {\mathrm b} \bar{{\mathrm b}}$ events
taken by the DELPHI detector.
The BSAURUS goal is to reconstruct B-decays, 
by making use of
as many properties of b-jets as possible, with high efficiency and
good purity. This is achieved by exploiting the 
capabilities of the DELPHI detector to their extreme,
applying wherever possible physics knowledge about B production and
decays and combining different information sources with modern 
tools- mainly artificial neural networks.
This note provides a reference of how BSAURUS outputs are formed,
how to access them within the DELPHI framework, and the physics performance
one can expect.

\end{abstract}
\vspace{\fill}

\vspace{\fill}
\end{titlepage}
\setcounter{page}{1}
\section{Introduction}

BSAURUS is a package for inclusive B reconstruction of $Z^0$ decays
into b quark-antiquark pairs taken by the DELPHI detector.
Due to the large mass of B hadrons there are many thousand decay
channels, all with small branching ratio. This is a severe limitation
for studying B physics. BSAURUS tries to reconstruct {\em inclusively}
as many properties of b-jets as possible with high efficiency and
yet with good purity also. This is achieved by exploiting the 
capabilities of the DELPHI detector to their extreme,
applying wherever possible physics knowledge about B production and
decays and combining different information sources with modern 
tools- mainly artificial neural networks.

\subsection{Physics Motivation} 

High energy b-jets contain $\Bplus$,$\bar{\Bzero}$, $\Bs$ mesons
or B-baryons in an approximate mixture of 40:40:10:10.
To measure lifetimes one needs to be able to reconstruct the 
decay length, i.e. the distance between primary and 
secondary vertex, as well as the B-hadron energy.
The mean lifetime of B-hadrons have been measured with good accuracy
and together with the semileptonic branching ratio it can be used to
determine the Cabibbo-Kobayashi-Maskawa matrix element 
$V_{{\mathrm cb}}$.
The lifetimes of the individual B hadrons are predicted by the
Heavy Quark Effective Theory (HQET) to be only marginally different
from the mean lifetime. In order to test HQET, one needs to be
able to distinguish between the four B hadron species.    

The fraction of the available energy that B-hadrons attain after the 
fragmentation process is not directly calculable in perturbative QCD, 
and has to be phenomenologically modelled with fragmentation
functions. Measuring the b-quark fragmentation function, ideally
for the different B-types separately, is fundamental to the 
accurate simulation of B-physics.

The ${\mathrm Z}^0$ decay into b-quarks exhibits a forward-backward 
asymmetry between the b and $\bar{{\mathrm b}}$-jet.
A precise measurement of $A_{FB}^{\mathrm b}$ 
constitutes an important test of the Standard Model.
For an experimental study one needs to tag the production flavour
(i.e. decide whether a given jet is from a b-quark or $\bar{{\mathrm b}}$ 
antiquark). 
Neutral B
mesons can oscillate into their antiparticle (and back)
before they decay. The time dependence of ${\mathrm B_d}$ oscillations 
has been observed, whereas the oscillations of the ${\mathrm B_s}$ meson
are too fast to be experimentally resolved at present .
For experimental studies of
${\mathrm B^0}$-oscillations one needs to know the production flavour and the
decay flavour.
   
The weakly decaying B hadrons are the lowest mass, i.e. 
the ground states, of $b\bar{q}$ systems with quantum numbers 
$J^{PC}=0^{-+}$. For each of these systems a full spectroscopy
of excited states exists, e.g. $B^*$ vector mesons and 
orbital excitations which are predicted to appear as
two doublets for each unit of orbital angular momentum L, one
being broad decaying in S-wave, and the other narrow since
at least D-waves are necessary in the decay.
Signals of $L=1$ B-mesons are observed with a relatively 
large rate (about 20-30\%), but the composition of the
broad peak observed is still unclear.
Radial excitations are also predicted in the quark model.  
In B-baryon spectroscopy, 
in addition to the $\Lb$, the states   
$\Xi_{\mathrm b}^-$, $\Xi_{\mathrm b}^0$ and $\Omega_{\mathrm b}$ 
are expected to 
decay via weak interactions with an appreciable lifetime, whereas
the $\Sigma_{\mathrm b}$ and $\Sigma_{\mathrm b}^*$ 
should decay into $\Lambda_{\mathrm b}\pi$ via strong interactions.    
For testing B-spectroscopy one needs the reconstruction of B-hadron 
four-momenta and the identification of 
protons, pions and kaons originating from the fragmentation process. 
   
BSAURUS is designed primarily to allow analyses
in all these areas.  The algorithms used rely on a
satisfactory performance from all of the detector components in the
DELPHI barrel in conjunction with optimised reconstruction software.

\subsection{Detector-Specific Reconstruction: The Basis of BSAURUS} 

Charged particle reconstruction, especially the correct 
assignment of silicon vertex detector points to tracks 
reconstructed in the TPC, the Inner Detector and the 
Outer Detector, has been optimised in the years 1995/1996
by better alignment procedures and adding completely new 
algorithms including e.g. inside-out tracking, a complete event 
ambiguity resolution, and a Kalman filter track fit taking into account
multiple scattering and energy loss. All LEP I data taken from 
1992-1995 have been 
reprocessed with these new algorithms, and this has led to huge 
reconstruction improvements, especially in dense jets. As an example, the signal
of exclusively reconstructed $D^* $ mesons into $D^0\pi^+$ 
with $D^0\rightarrow K\pi\pi\pi $ was increased by a factor 2.4.    
In addition, optimised photon conversion algorithms \cite{ref:ELEPHANT},
reconstruction of secondary hadronic interactions in detector 
material and of  $\Xi^-$ decays, and a search for additional low-energy
tracks in the vertex detector \cite{ref:MAMMOTH} help to give
a clean reconstruction of complicated events. It should be noted that
since BSAURUS has been developed and tuned on these reprocessed data
the result of applying it to any other data set cannot be guaranteed. 
   
Particle identification is an important ingredient for many of
the algorithms applied. To achieve an optimal performance,
electron identification \cite{ref:ELEPHANT} is performed using
a neural network combining spatial and energy information from
the HPC calorimeter, tracking information, dE/dx measurements 
of the TPC, and searches for kinks in the track in known layers 
of material and for photons radiated tangentially in them.
In DELPHI, hadron identification information is gathered from the
Ring Imaging Cerenkov Counters(RICH) with both 
gas and liquid radiator, and from energy loss measurements in the TPC and the
silicon vertex detector. To optimise hadron classification at
various momenta, detailed information from all of these detectors
has been combined using neural networks in the program package
MACRIB \cite{ref:MACRIB}.

\subsection{Overview Of BSAURUS Outputs}
BSAURUS provides a host of information relating to the properties 
of B-hadron production, fragmentation and decay within an
event hemisphere. This information is available via output variables
stored in COMMON blocks, described in Appendix C.

Among the most important outputs for B-physics analysis topics are:
\begin{itemize}
\item estimates of the  B-hadron {\bf energy}(see Section~\ref{sec:ecor}) and 
{\bf decay length}(see Section~\ref{sec:proptime}),
\item the {\bf TrackNet} (see Section~\ref{sec:tracknet}) that discriminates
between tracks originating from the primary and a secondary vertex,
\item the {\bf BDnet} (see Section~\ref{sec:BDnet}) that discriminates
between tracks originating from the B and the cascade 
B$\rightarrow$D~vertex,
\item neural networks (see Section~\ref{sec:bspec})
that provide a probability that a particular
{\bf B-hadron type} (i.e. $\Bplus,\Bzero,\Bs$~or B-baryon) was present in the hemisphere,
\item neural networks (see Section~\ref{sec:flavtag}) 
that tag the charge or {\bf flavour of the b-quark}
in the B-hadron type at both the production and decay time.
\end{itemize}




\section{BSAURUS Definitions}
\subsection{Multihadronic Events}
There is no event selection applied inside BSAURUS i.e. it is left 
to the user to decide on which events to call BSAURUS. However, for all
development and tuning work and for all results presented in this note,  
events in real data and Monte Carlo had to pass a 
multihadronic event selection. The criteria used are from the DELPHI pilot record
hadron selection bits:
\begin{itemize}
\item team 4 selection, $>5$ charged particles (bit 1)
\item team 4 selection, charged energy $>0.12E_{LEP}$ (bit 2)
\item $E_{forward}$,$E_{backward}>0.03E_{LEP}$ (bit 12).
\end{itemize}

\subsection{Standard Particle Selection}
\label{sec:trsel}
Charged particles used by the BSAURUS algorithms must pass the following
track selection criteria: 
\begin{itemize}
\item Impact parameter, with respect to the origin, in the $r\phi$ plane
 $\left| \delta_{r-\phi} \right| < 4.0$~cm
\item Impact parameter, with respect to the origin, in the $z$ plane 
$\left| \delta_{z} \right| < 6.0$~cm
\item $\left| \cos \theta \right| < 0.94$
\item $\frac{\Delta E}{E} < 1.0$
\item At least 1 $r-\phi$ track hit from the silicon vertex detector(VD)
\item Tracks must not have been flagged as originating from material
interactions (via any of the PXPHOT codes: -124, -123, -75, -85, -84 or -74).
\end{itemize}

Neutral particles are used that are flagged by the DELPHI mass code as
being a photon, $\pi^0$, ${\mathrm K}_s^0$ or $\Lambda^0$ 
(codes 21, 47, 61 and 81 respectively).  

\subsection{Event Jets}
Event jets are reconstructed via the routine LUCLUS \cite{ref:jetset}.
All reconstructed particles, that pass the standard selection, are used  with a 
transverse momentum cutoff value of $d_{join}$=PARU(44)=5.0.

\subsection{Event Hemispheres}
All reconstructed particles, that pass the standard selection,  are
used to calculate 
the event thrust axis via routine LUTHRU \cite{ref:jetset}. 
Event hemispheres are then defined by the plane perpendicular to the 
thrust axis. 

 Each hemisphere has associated with it two axes: the {\bf thrust axis}
and the {\bf reference axis} used for the calculation of
rapidity (see Section \ref{sec:y} below) for tracks in that hemisphere.
The reference axis is a jet axis selected in the following way:
\begin{itemize}
\item If the event is a two jet event, the reference axis for the
 hemisphere is the jet axis in that hemisphere.
\item If a hemisphere contains 2 or more jets (i.e. an event with 3
or more jets) :
\begin{itemize}
\item if one of the jets is the highest energy jet in the event, that jet
axis forms the reference axis.
\item if the highest energy jet is in the opposite hemisphere, 
the combined probability ($P_{jet}$)
for the tracks in a jet to have originated
from the event primary vertex is formed
(via the AABTAG package algorithms\cite{ref:AABTAG}).

The jet that is most `B-like', i.e. with the smallest
probability, is then selected if $P_{jet}<0.05$.
\item if no jet in the hemisphere satisfies the above criteria, 
the jet with the highest energy is selected to form the reference axis. 
\end{itemize}
\end{itemize} 

 In addition, internally to BSAURUS, the hemispheres are numbered
 as 1  or 2 with the convention that the jet forming the 
reference axis for hemisphere 1 is of higher energy than that for
hemisphere 2.  

\subsection{Track Rapidity}
\label{sec:y}
The rapidity of a track is defined as follows,
\begin{displaymath}
y=\frac{1}{2}\ln \left( \frac{E+P_L}{E-P_L} \right)
\end{displaymath}
for $E$ the track energy and $P_L$ the longitudinal momentum component
along the reference axis for the hemisphere.

 The BSAURUS rapidity algorithm also returns an {\bf initial estimate} of the
B-hadron candidate 4-vector. This is defined simply to be the sum of
individual track momentum vectors in a hemisphere for tracks with
rapidity greater than 1.6. The value of the cut provides 
a good rejection of tracks originating from the primary vertex while
accepting the majority of tracks from the B-decay.
 
\subsection{Particle Identification}
\label{sec:partid}
Unless otherwise stated, particle identification in BSAURUS is based on the
following definitions:
\begin{itemize}
\item {\bf Charged kaons}: the kaon network output, or kaon probability, from the
MACRIB package \cite{ref:MACRIB} and stored in BSAURUS output 
\verb+BSPAR(IBP_MACK,IH)+. 
\item{\bf  Protons}: the proton network output, or proton probability, from the
MACRIB package \cite{ref:MACRIB}and stored in BSAURUS output 
\verb+BSPAR(IBP_MACP,IH)+.  
\item {\bf Electrons}: the electron network output, or electron probability, returned
by the ELEPHANT routine \verb+ELNETID+ \cite{ref:ELEPHANT}.
\item {\bf Muons}: the \verb+MUCAL2+ definitions of a very loose, loose, standard and
tight muon.
\end{itemize}

\subsection{Performance of Networks}
\label{sec:perf}
The main tool used to quantify and compare BSAURUS network  performance 
is the {\it purity} against {\it efficiency} plot. Given some definition
of `signal' and `background', purity is defined as,
\begin{equation}
Purity=\frac{correctly\,\,\,classified\,\,\,signal\,\,\,events}
{all\,\,\,classified\,\,\, events},
\end{equation}
and the efficiency defined as,
\begin{equation}
\epsilon=\frac{correctly\,\,\,classified\,\,\,signal\,\,\,events}
{all\,\,\,signal\,\,\, events}.
\end{equation}
In this note, the efficiency is calculated after 
all selection cuts, e.g. from Section~\ref{sec:trsel} and Section~\ref{sec:nn}, 
have been applied. 
When considering network outputs that involve symmetric binary decisions 
(i.e. the TrackNet and all flavour tagging networks), we use an
alternative measure of efficiency termed $f=$`fraction classified'
and defined as,
\begin{equation}
f=\frac{all\,\,\,classified\,\,\,tracks/hemispheres}
{all\,\,\,classifiable\,\,\,tracks/hemispheres}.
\end{equation} 
For this case, in order to make the purity-efficiency plot, $f$~is varied
by making successive cuts in a band around the equal probability
point for tagging `signal' or `background'.

\section{General BSAURUS Quantities}
\subsection{Secondary Vertex Finding}
\label{sec:sv}
Per-hemisphere, an attempt is made to fit a secondary vertex to tracks
with rapidity $>1.6$~that pass the standard track selection criteria of 
Section~\ref{sec:trsel}.

\begin{figure}[p]
\begin{center}
\epsfig{file=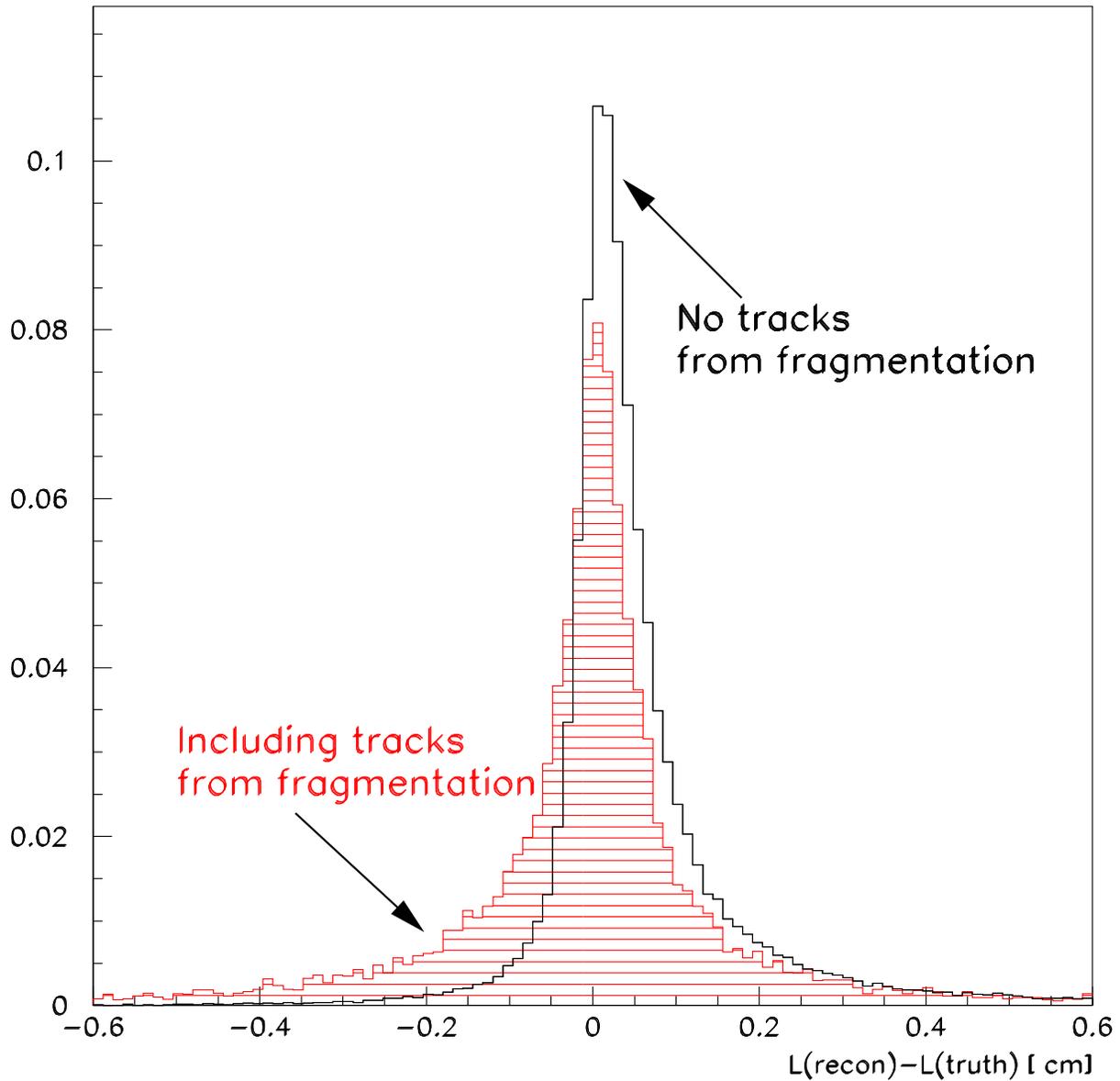,
width=16.0cm}
\caption[]
{\label{fig:pullfrag} The pull on the decay length
residual $L(recon)-L(truth)$ introduced if fragmentation tracks
are allowed in the secondary vertex fit definition. The open
histogram shows the case where no tracks from fragmentation are in the 
fit and the hatched histogram the case where fragmentation tracks
are included. }
\end{center}
\end{figure}

From this class of track, additional criteria are applied with the aim
of selecting tracks for the vertex fitting stage that are likely to 
have originated from the decay chain of a weakly decaying B-hadron
state. As Figure~\ref{fig:pullfrag} illustrates, it is important to 
reject as far as possible tracks from the 
fragmentation in order to avoid large pulls in the vertex position
toward the primary vertex.
There must be a minimum of two tracks for the vertex fit to be attempted
and the selection process is as follows:

\begin{itemize}

\item[{[1]}] The highest energy muon or electron candidate is selected if
$E_{lepton} > 3$~GeV.  

\item[{[2]}] The crossing point of each 
  track with the initial B-candidate direction vector (see Section~\ref{sec:y}) 
  is found and the distance, $L$, 
  between primary vertex \footnote{Unless otherwise stated, the primary vertex
  definition in BSAURUS is the event primary vertex from the AABTAG package.}
  and crossing point is calculated. Tracks are 
  selected if $L/\sigma_L > 2$ or, for cases where $L/\sigma_L < 2$,
  tracks must also satisfy $y>2.5$ and $L>0.1$cm.
\end{itemize}
If at this point the number of tracks selected for the fit is less than 2,
 further attempts are made to add tracks to the fitting
 procedure.
 When this happens in $\bb$ events,  it is likely that the true decay length of 
 the B-hadron was small.
  
 If only one track was added from stages 1 and 2:

\begin{itemize}
\item[{[3]}]  Add tracks with $y>3$.

\item[{[4]}] If still only one track selected, add the track of highest
         rapidity from the remaining track list of the hemisphere.   
\end{itemize}

If no track was selected from stages 1) and 2):

\begin{itemize}
\item[{[5]}] A search is made for the best kaon candidate in the hemisphere
based on the kaon probability defined in Section~\ref{sec:partid}. 
If this
kaon candidate has $y>2$ it is added to that track in the 
remaining track list of the hemisphere that has the highest rapidity. This
track pair alone will then form the starting track list
for the vertexing procedure. 

\item[{[6]}] Finally if no kaon candidate exists in the hemisphere,
the two tracks of highest rapidity are selected.

\end{itemize}

Using the track list supplied by this track selection process,
a secondary vertex fit is performed in 3-dimensions (using routine
DAPLCON from the ELEPHANT package~\cite{ref:ELEPHANT}) 
constrained to the direction of the B-candidate momentum vector.
The event primary vertex is used as a starting point and 
if the fit did not converge\footnote{Here, non-convergence 
means the fit took more than 20 iterations. A further iteration is 
deemed necessary if 
the $\chi^2$ is above 4 standard deviations during the first 10 iterations
or above 3 standard deviations during the next 10 iterations.},
the track making the largest $\chi^2$ contribution is stripped away
in an iterative procedure, and the fit repeated.

In addition to returning the secondary
vertex position, the procedure also fits the primary vertex position and 
updates the B-candidate direction according to the vector joining the
primary and secondary vertex points. 
This
information is used in forming some of the track net inputs as described
in the next section.

Once a convergent fit has been attained, the final stage of the
secondary vertex fitting procedure involves
an attempt to add into the fit tracks that failed the initial track selection
criteria but nevertheless are consistent with originating from the vertex.
These tracks are identified on the basis of an {\bf intermediate} version of the
TrackNet. This is a  neural network output that discriminates between
tracks originating from the primary vertex and those likely to have
come from a secondary vertex and is described in Section~\ref{sec:tracknet}.
A version of the TrackNet is constructed,
specifically for the  purpose of  use in this final stage of vertex fitting,
based on secondary vertexing information available before this 
final stage has run. In general, secondary vertex tracks have 
TrackNet $\sim 1$~whereas tracks from
the primary vertex have TrackNet values close to zero.
The track of largest TrackNet in the hemisphere is added to the existing
track list and retained if the resulting fit converges. This process continues
iteratively for all such tracks with TrackNet $>0.5$. 



\subsection{B-Candidate Four-Vectors Available from BSAURUS}

BSAURUS provides currently four different definitions of the B-candidate
four-vector in a hemisphere:
\begin{itemize}
\item The initial estimate from the rapidity algorithm outlined in Section \ref{sec:y}
and stored in the four BSAURUS output array elements starting at 
\verb+BSHEM(IBH_BY,IH)+ for hemisphere number \verb+IH+.
\item The three momentum vector is then transformed to lie along the direction
joining primary to secondary vertex as mentioned in Section \ref{sec:sv}.
The energy component is the
rapidity energy and this vector is stored in location starting at 
\verb+BSHEM(IBH_BF,IH)+.
\item In location starting at \verb+BSHEM(IBH_BT,IH)+, the energy 
component is then corrected to account for neutral energy losses   
(see Section \ref{sec:ecor}). The momentum vector
derives from a TrackNet weighted momentum sum
for the case of 2-jet events and is the rapidity vector for $>2$-jet events.
\item The four-vector starting
at location  \verb+BSHEM(IBH_BTT,IH)+ is, in principle, the 
same as the  \verb+BSHEM(IBH_BT,IH)+ vector but contains a more
detailed energy correction, described in detail in  Section~\ref{sec:ecor}. 
In addition, the momentum vector derives from the rapidity algorithm
for $>2$-jet events that also pass $E_{HEM}/E_{BEAM} > 0.6$ and from
the TrackNet weighted sum otherwise. 

\end{itemize}

\subsection{B-Candidate Energy Correction}
\label{sec:ecor}
We describe here the energy correction procedure that went into making vector 
\verb+BSHEM(IBH_BTT,IH)+. (Note that the \verb+BSHEM(IBH_BT,IH)+ vector
contains a very similar correction which, although still used internally
inside BSAURUS, is older and less optimal.)

 Separate correction functions are derived for 2-jet and $>2$-jet events.  
Within these two classes, 
because of the explicit energy ordering of the hemispheres, we treat
hemispheres 1 and 2 completely separately.
To be used in the correction procedure, hemispheres must pass the 
following cuts:
\begin{itemize}
\item The initial reconstructed B-candidate energy 
$E_{raw}$ is 20GeV or more.
\item The corresponding initial reconstructed B-candidate mass $m_{raw}$
lies within two standard deviations of the total data sample median value.  
\item The ratio, $x_h$, of the hemisphere energy $E_{hem}$ to beam energy
 $E_{beam}$ lies in the range $0.6<x_h<1.1$.
\end{itemize} 
In addition, for the case of 
hemispheres with larger missing energy, i.e. $x_h<0.6$, a 
further correction is derived separately for hemispheres 1 and 2.

 The starting point is the initial estimates of the B energy and mass,
 $E_{raw}$~and $m_{raw}$. Following studies, we choose these estimates 
to be from the rapidity algorithm for events with $>2$-jets 
and where $E_{HEM}/E_{BEAM} > 0.6$,
and to be derived from the sum of `B-weighted' four-vectors otherwise. 
This involves weighting
(via a sigmoid threshold function)
the momentum and energy components of charged tracks by the 
TrackNet output (see section \ref{sec:tracknet}) and those of neutrals
by their rapidity. In this way the effects of tracks from the B-decay
are enhanced and those of tracks from the primary vertex are suppressed.

 The correction procedure is motivated by the observation in Monte Carlo
of a correlation between the energy residuals $\Delta E=E_B^{raw}-E_B^{truth}$
and $m_{raw}$ (which is approximately linear in $m_{raw}$), 
and a further correlation between $\Delta E$ and $x_h$ resulting
from neutral energy losses and inefficiencies. 
The correction proceeds in the following way:  
the data are divided into several samples according to the measured 
ratio $x_h$ and for each of these classes the B-energy residual
$\Delta E$ is plotted as function of $m_{raw}$.
The median values of $\Delta E$ in each bin of $m_{raw}$ are calculated   
and their $m_{raw}$ dependence fitted by a third order polynomial
\begin{displaymath}
\Delta E(m_{raw};x_h)=a+b(m_{raw}-\left<m_{raw}\right>)+
c(m_{raw}-\left<m_{raw}\right>)^2+d(m_{raw}-\left<m_{raw}\right>)^3    
\end{displaymath}
The four parameters ${a,b,c,d}$ in each $x_h$ class are then plotted as a 
function of $x_h$ and their dependence fitted with third and second-order
polynomials. Thus one obtains a smooth correction function describing the mean
dependence on $m_{raw}$ and on the hemisphere energy as determined
from the Monte Carlo. Finally, a {\bf small} bias correction is applied
for the remaining mean energy residual as a function of the corrected 
energy. 

\begin{figure}[p]
\begin{center}
\mbox{\epsfig{figure=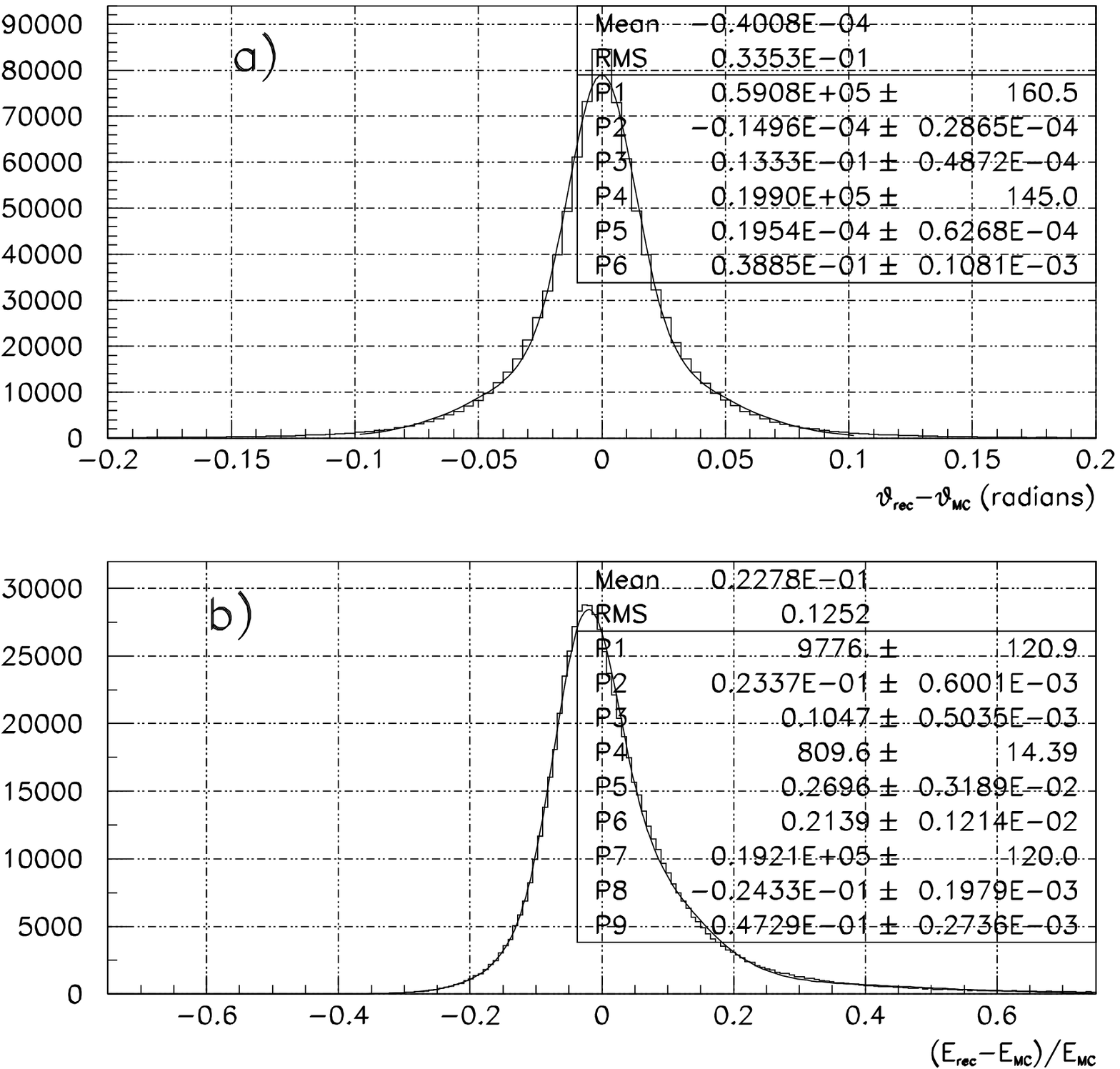,height=16cm,width=16cm}}
\caption[]
{\label{fig:eresol}For 2-jet events, (a) shows the residual between 
reconstructed and Monte Carlo truth for the B-direction and (b) the
resolution of the reconstructed corrected-energy for 
hemispheres inside the selection cuts described in Section \ref{sec:ecor}.}
\end{center}
\end{figure}

Figure \ref{fig:eresol} shows typical resolutions on the reconstructed
B-direction and energy attainable from the four-vector
output at location \verb+BSHEM(IBH_BTT,IHEM)+.

\subsection{Decay Length and Decay Time}
\label{sec:proptime}
The B-hadron decay length estimate in the $r-\phi$~plane, $L_{r\phi}$, is
defined as the (positive) distance between the primary and 
secondary vertex positions reconstructed from the secondary vertex search 
described in Section~\ref{sec:sv}. The three-dimensional decay length
is constructed as $L=L_{r\phi}/\sin(\theta)$, where the 
B-hadron candidate momentum vector, $p_B$ detailed in Section~\ref{sec:ecor},
defines the direction.  
The resulting proper decay time estimate is given by
$\tau=L m_B / p_B c$, where the B rest mass is taken to be $5.2789$ GeV$/c^2$.

\subsection{Track and Hemisphere Quality}
\label{sec:qual}
When presenting neural networks with track and hemisphere-based discriminating
variables, it is advantageous at the network learning stage to provide
the network with additional information relating to the potential
quality of the input information. For this reason BSAURUS defines both 
track and hemisphere quality words that are used extensively in 
many network definitions.

The track quality word stored in \verb+BSPAR(IBP_QUAL,IT)+ for track  \verb+IT+
is defined as follows:
\begin{itemize}
\item 0, means the track is good.
\item 1, means the track contains some ambiguous track hit information flagged
by the DELPHI track hit ambiguity word.
\item 10, means that the track has been flagged as originating from an
interaction, defined by PXPHOT code=-120 returned by routine  PXGECO.   
\item 20, means that the track did not pass the internal track quality
criteria of the AABTAG package.   
\end{itemize}

The hemisphere quality word, stored in, \verb+BSHEM(IBH_QUAL,IH)+ for hemisphere 
\verb+IH+, has the following definition:
\begin{itemize}
\item $N \times 1$, where $N$~is the number of tracks in the hemisphere rejected by the 
standard track cuts (described in Section~\ref{sec:trsel}) 
\item {\bf +} $N \times 100$, where $N$~is the number of tracks in the hemisphere flagged
as originating from an interaction in the same way as for the track quality word 
\item {\bf +} $N \times 1000$, where $N$~is the number of tracks in the hemisphere
flagged as being ambiguous in the same way as for the track quality word 
\item {\bf +} $N \times 100000$, where $N$~is the number of tracks in the hemisphere
flagged as failing the quality cuts of the AABTAG package.
\end{itemize}
For use in network training, the hemisphere quality is better used in 
a real, continuous form rather than as an integer. Stored in \verb+BSHEM(IBH_QUAL2,IH)+
therefore, is the transformation of the integer word into a 
continuous variable ranging from 0.0(good quality) to 10.0(bad quality).

\section{Use of Neural Networks in BSAURUS}
\label{sec:nn}
Events and event hemispheres used to prepare the training samples for 
BSAURUS networks must pass the following set of requirements:
\begin{itemize}
\item 2-jet events with the jets back-to-back to better than $10^\circ$,
\item $|cos \theta_{thrust}|<0.65$,
\item combined AABTAG event b-tag $>0.02$,
\item both gas and liquid parts of the RICH must be functioning,
\item defining $E_{HEM}$~as the total energy contained
      in charged and neutral particles in the hemisphere, then
      a hemisphere must satisfy $0.6<E_{HEM}/E_{BEAM}<1.1$,
\item a hemisphere must have an initial B-energy estimate 
       from the rapidity algorithm $>20$GeV,  
\item only hemispheres with a fully converged secondary vertex fit are 
      considered.
\end{itemize}

BSAURUS uses the JETNET\cite{ref:jetnet} neural network package run in a 
feed-forward mode nearly always with three layers and a sigmoid function 
as the  input-output(or transfer) function.
The hidden layer typically contains one more node than the number of inputs.

The network is trained using two statistically independent
samples; the {\it training} sample and the {\it performance} sample. Usually 
each sample consists of an equal number of `signal' and `background'
hemispheres/tracks.
The training sample is used to teach the network to separate signal from
background, but it is the 
network output error\footnote{Defined as $\sqrt{\sum_i (NO_i-TAR_i)^2}$, where
index $i$~labels the network output node, $NO_i$~is the network output value
at output node $i$~and  $TAR_i$~is the target value for output node $i$.}
based on the 
performance sample that determines when the end-point 
of the training procedure has been reached. 

Variables used as inputs to the networks are selected based on 
(a)the discriminating power between signal and background  
and (b)the agreement between simulation and data. Variables
are often also transformed and/or their range truncated so that
the network is presented with variables that show, ideally, a 
linear, continuous variation in the ratio of `signal' to
`background' across the full range.   


\section{The B-Track Probability Network(TrackNet)}
\label{sec:tracknet}
The aim of the B-track probability network is to discriminate between 
the class of tracks, in a $\bb$-event hemisphere, originating from the 
weak decay of a B-hadron from all other tracks. 

Only tracks passing the standard quality cuts of Section~\ref{sec:trsel}
are considered.
The discriminating variables that form the inputs to the network per 
track are:
\begin{itemize}
\item The track total momentum.
\item The track momentum in the B-candidate rest frame.
\item The helicity angle of the track defined as the angle between 
the track vector in the B-candidate rest frame and the B-candidate momentum vector
in the lab frame.
\item A flag to identify whether the track was in the secondary vertex
      fit or not.
\item The probability that the track originates from the fitted primary vertex
      (AABTAG algorithm). 
\item The probability that the track originates from the secondary vertex
      (AABTAG algorithm). 
\item The probability that the track originates from the fitted primary vertex
      (BSAURUS algorithm). 
\item The probability that the track originates from the secondary vertex
      (from the BSAURUS algorithm).
\item The track rapidity.
\end{itemize}
In addition, input variables that gave no inherent discriminating power 
were included to inform the network of the potential 
quality of the other input variables: 
\begin{itemize}
\item The decay length or distance between the primary and secondary
      vertex in the $r-\phi$ plane. This quantity is then scaled by the
      reconstructed error to form a decay length significance.
\item The track quality word from Section~\ref{sec:qual}.
\end{itemize}

\begin{figure}[p]
\begin{center}
\begin{tabular}{lll}

 \mbox{\epsfig{figure=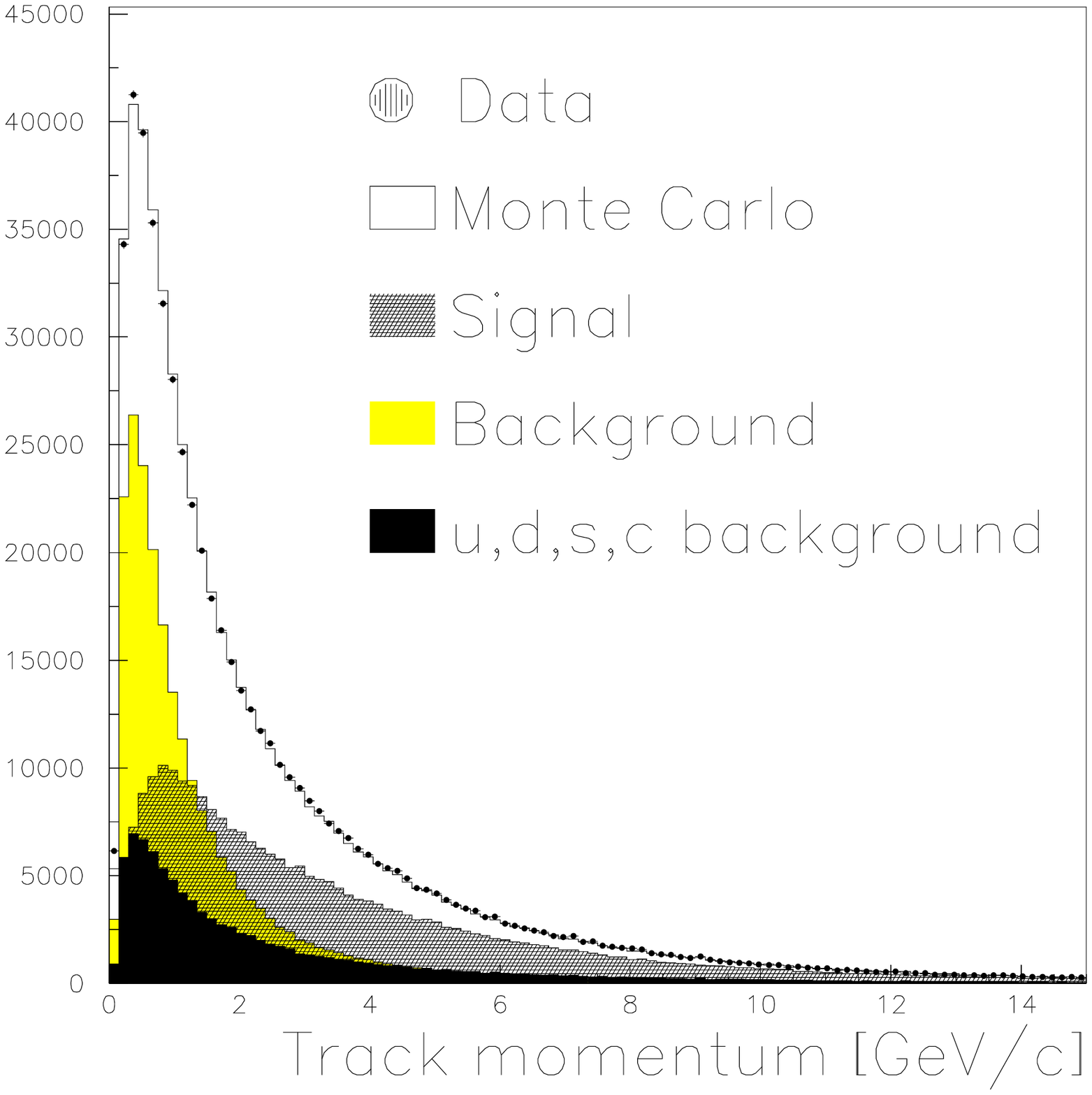,height=6cm,width=5.2cm}}
&
 \mbox{\epsfig{figure=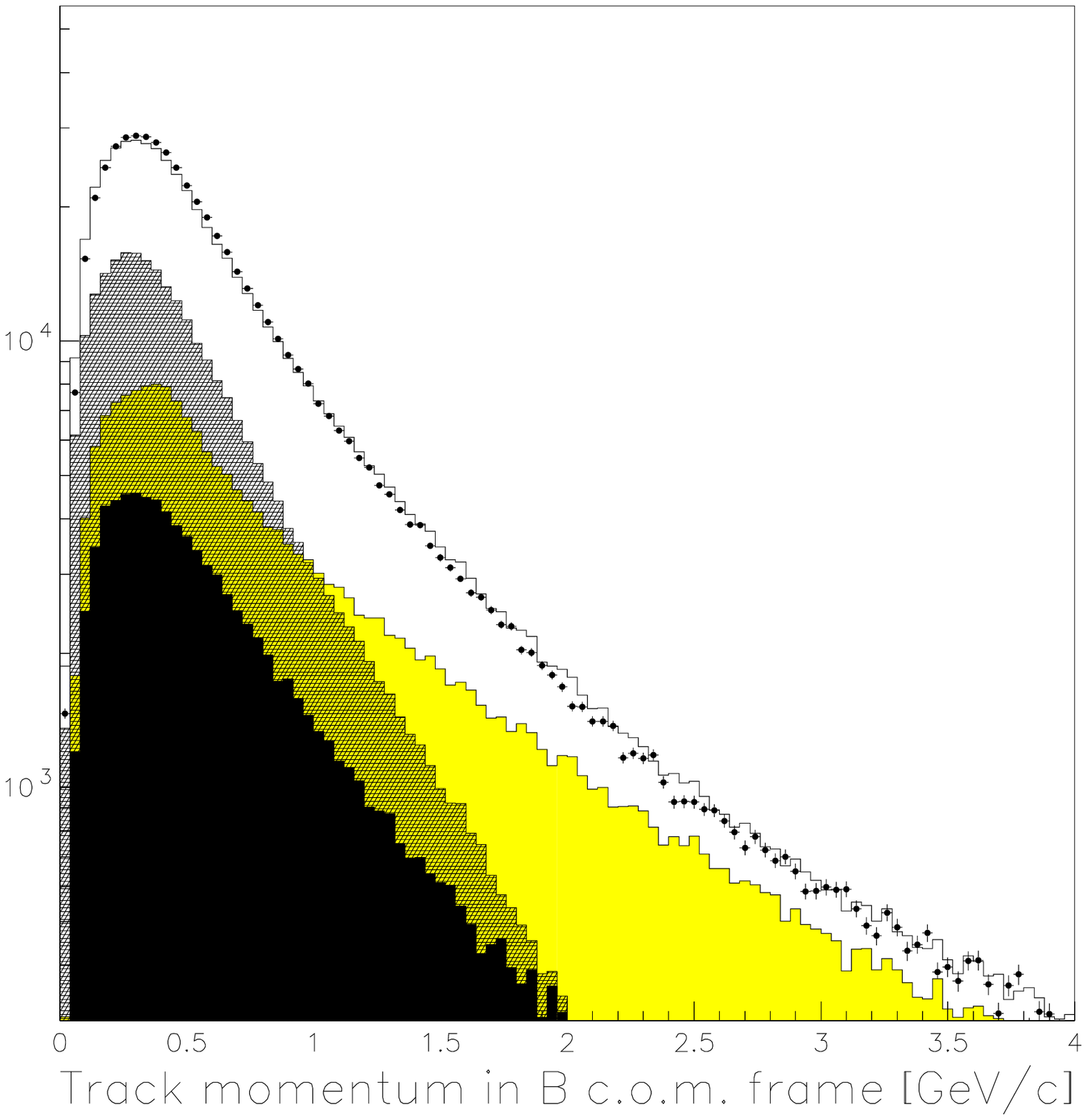,height=6cm,width=5.2cm}}
&
 \mbox{\epsfig{figure=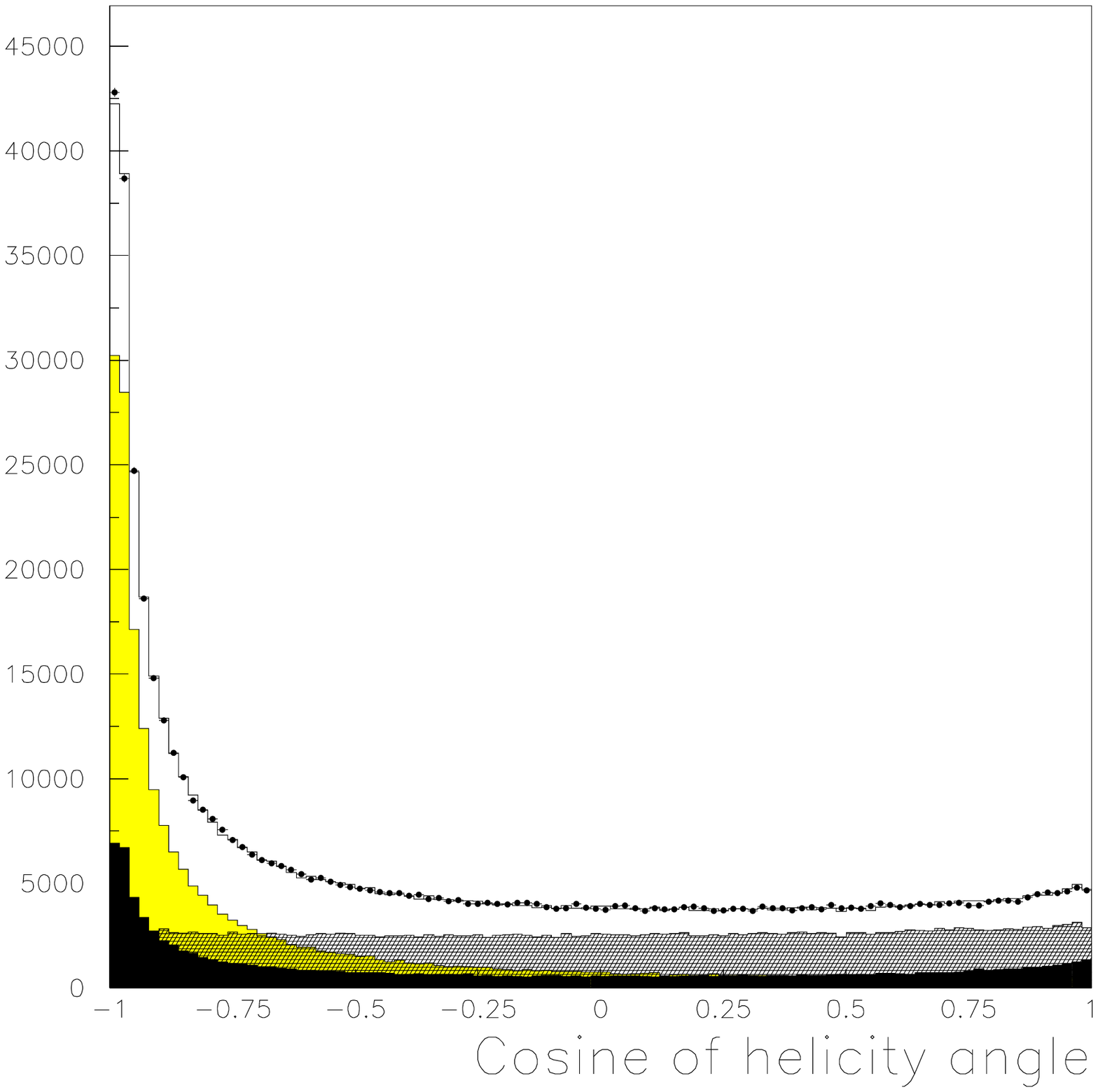,height=6cm,width=5.2cm}}
\\

  \mbox{\epsfig{figure=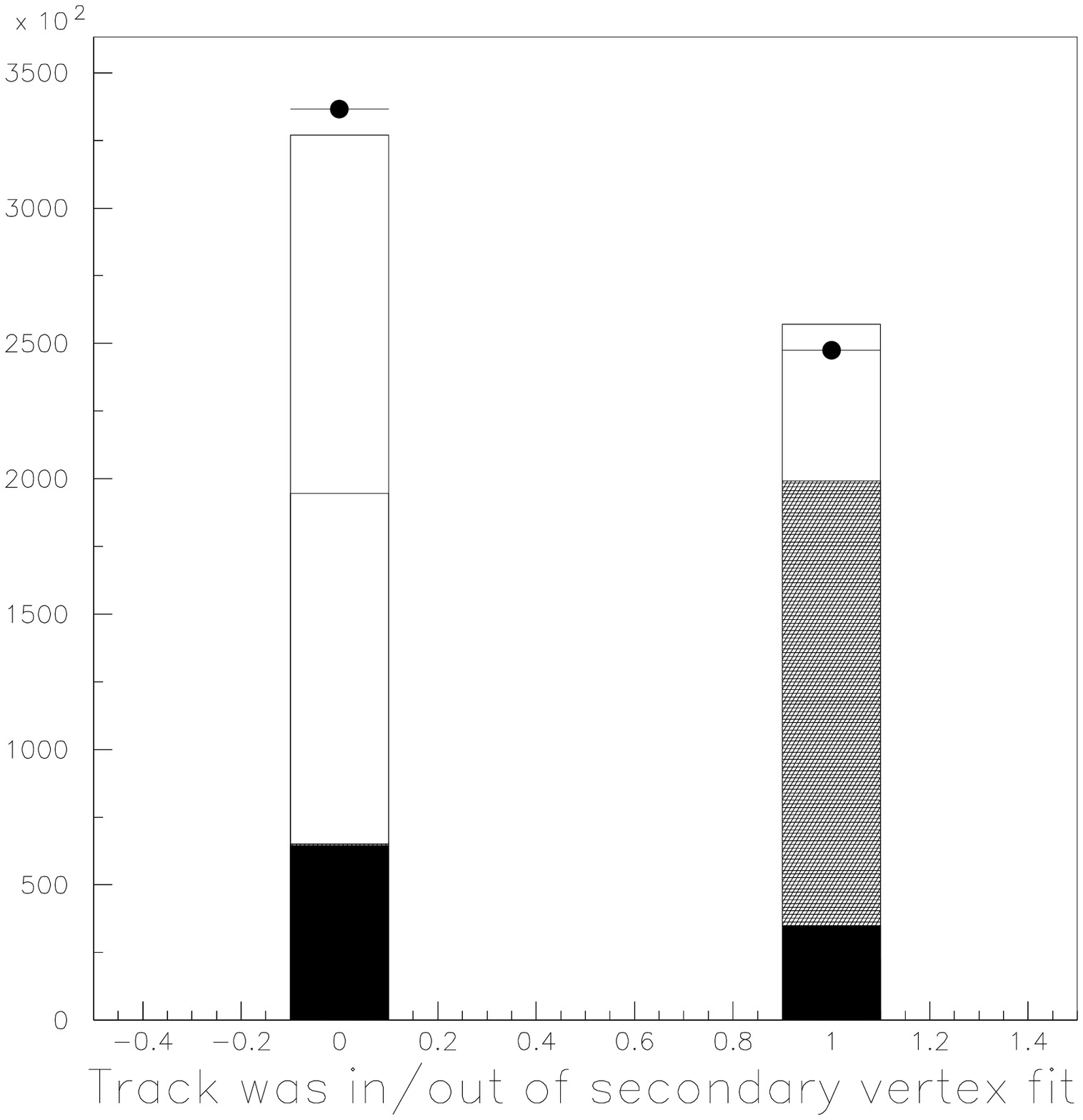,height=6cm,width=5.2cm}}
&
 \mbox{\epsfig{figure=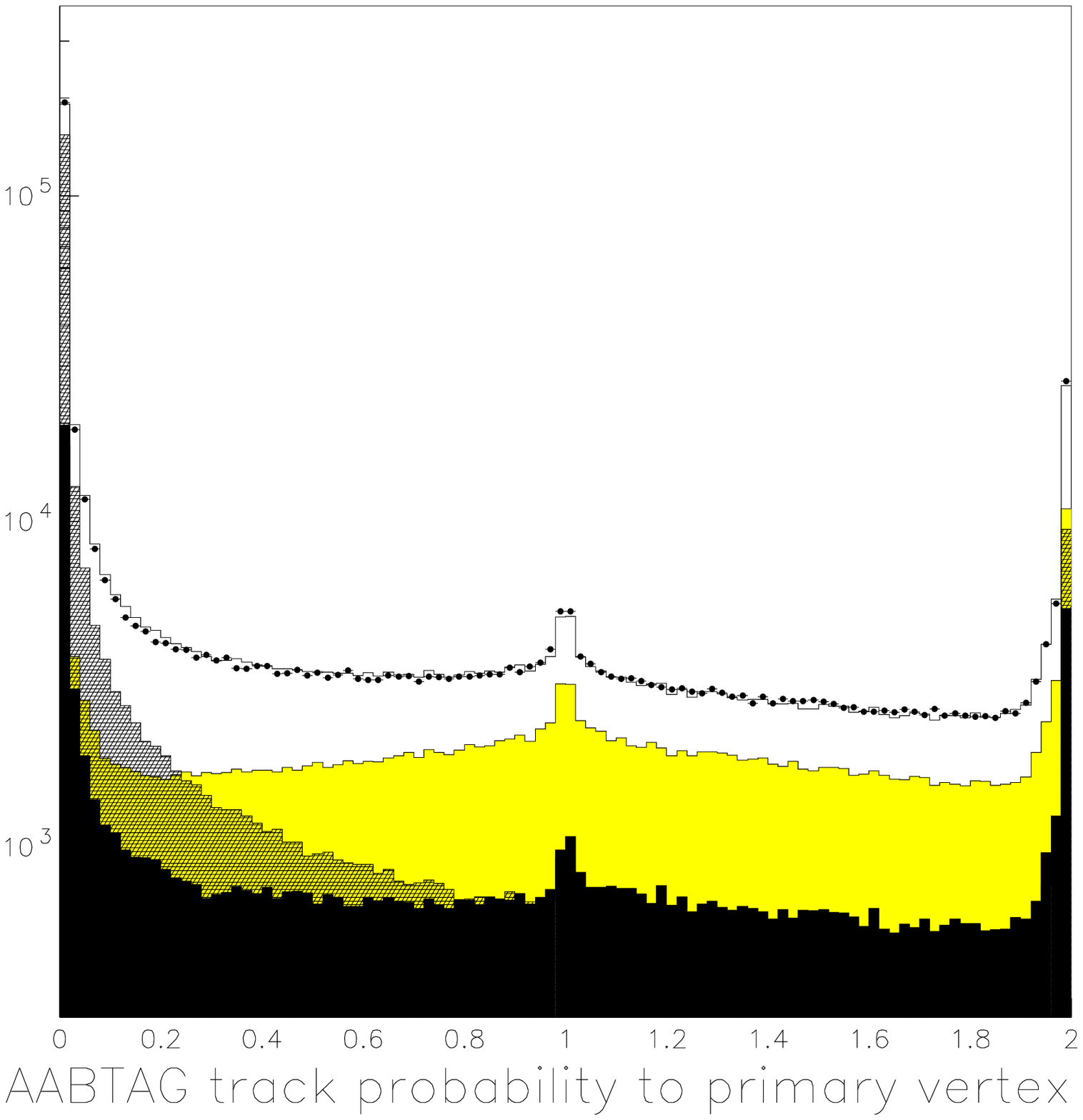,height=6cm,width=5.2cm}}
&
 \mbox{\epsfig{figure=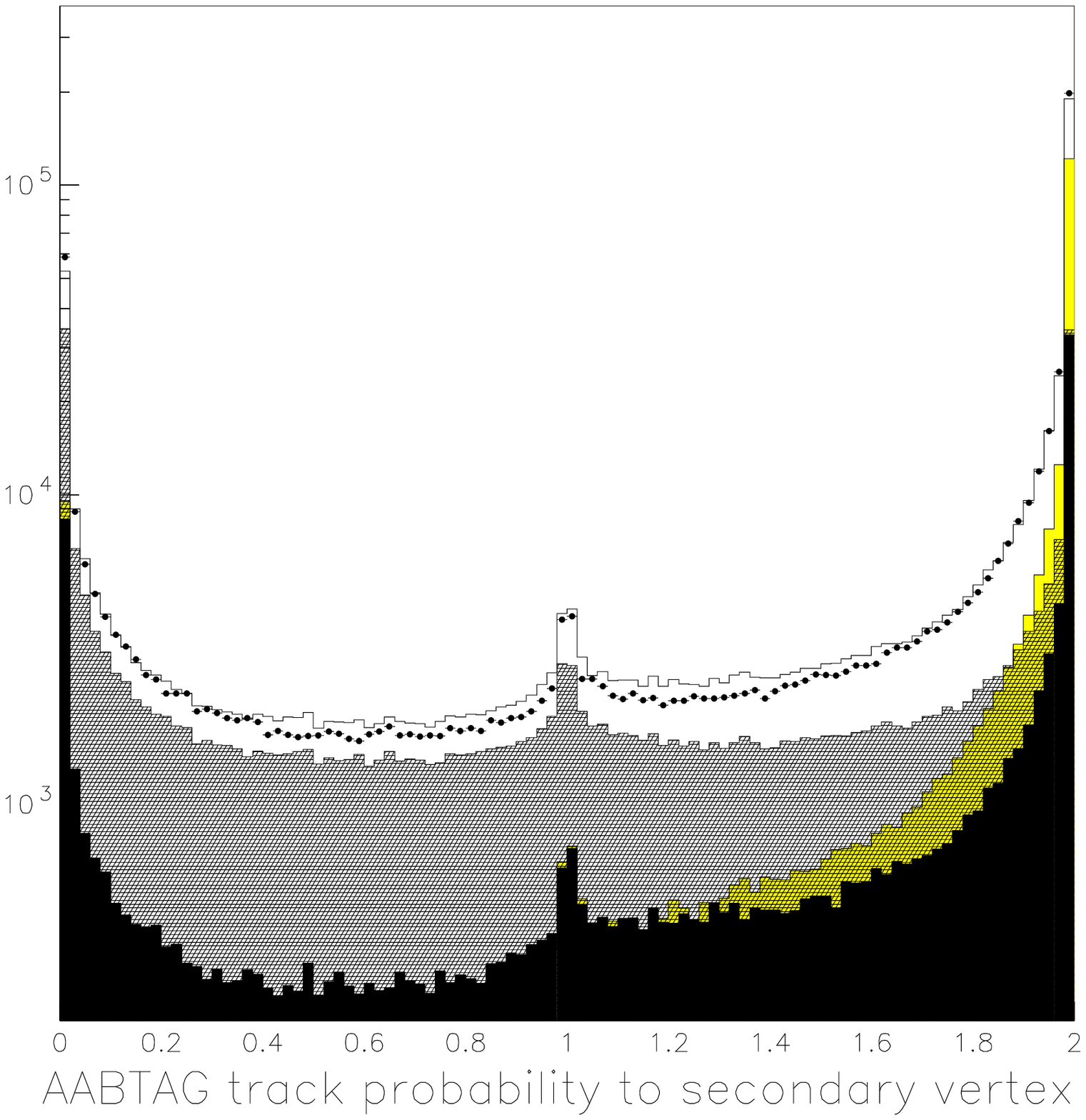,height=6cm,width=5.2cm}}

\\
 \mbox{\epsfig{figure=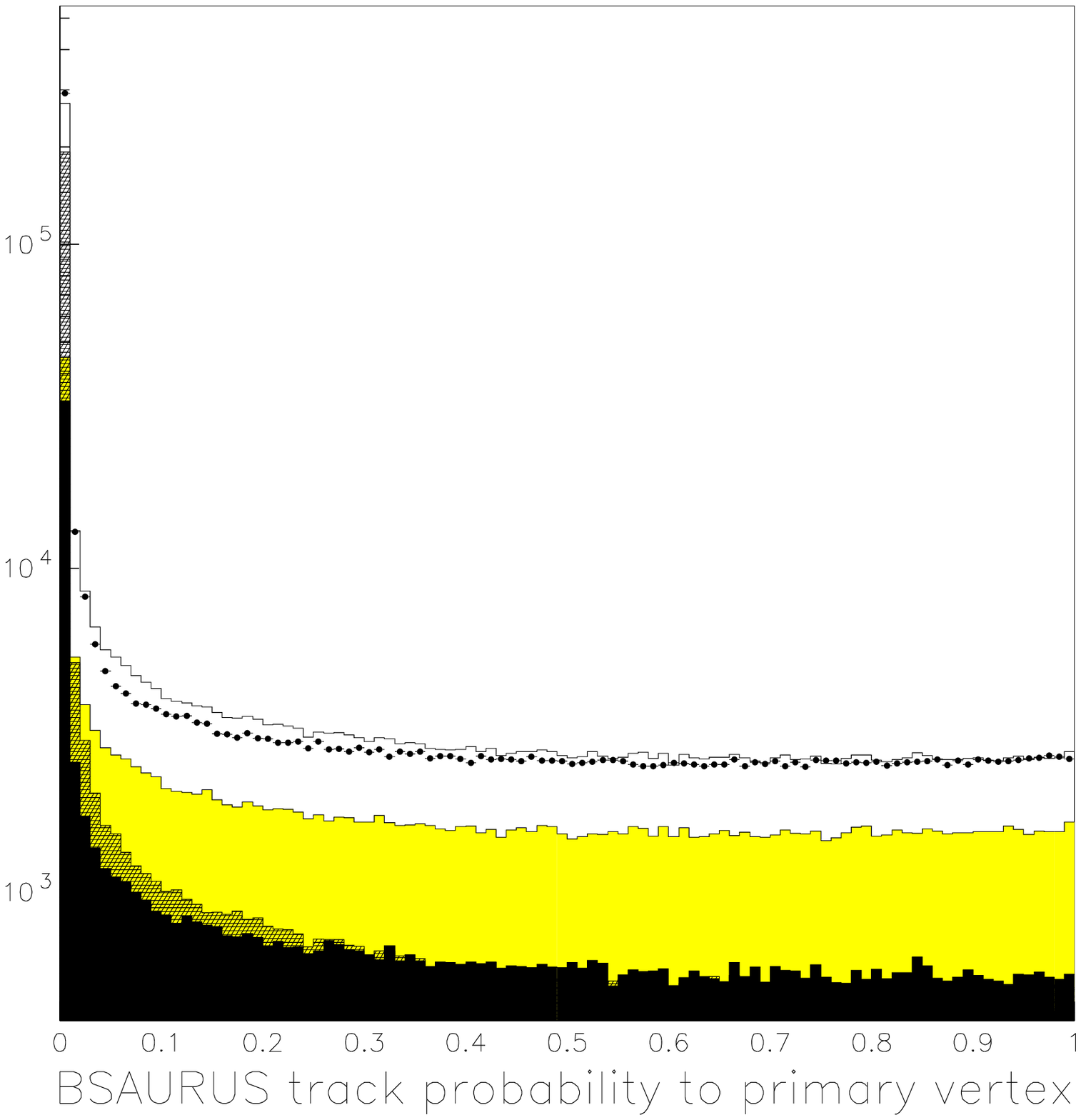,height=6cm,width=5.2cm}}
& 
 \mbox{\epsfig{figure=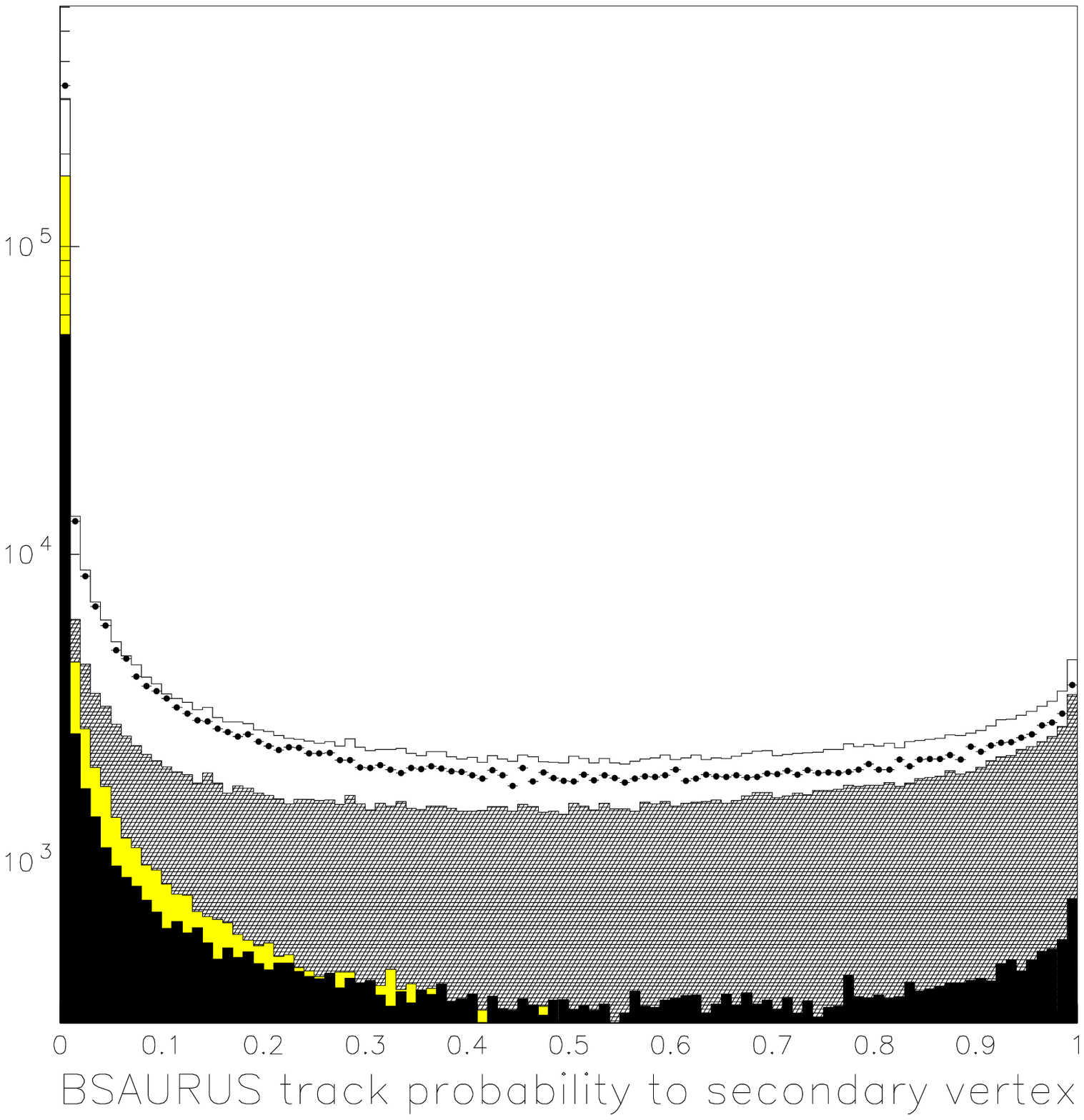,height=6cm,width=5.2cm}}
&
 \mbox{\epsfig{figure=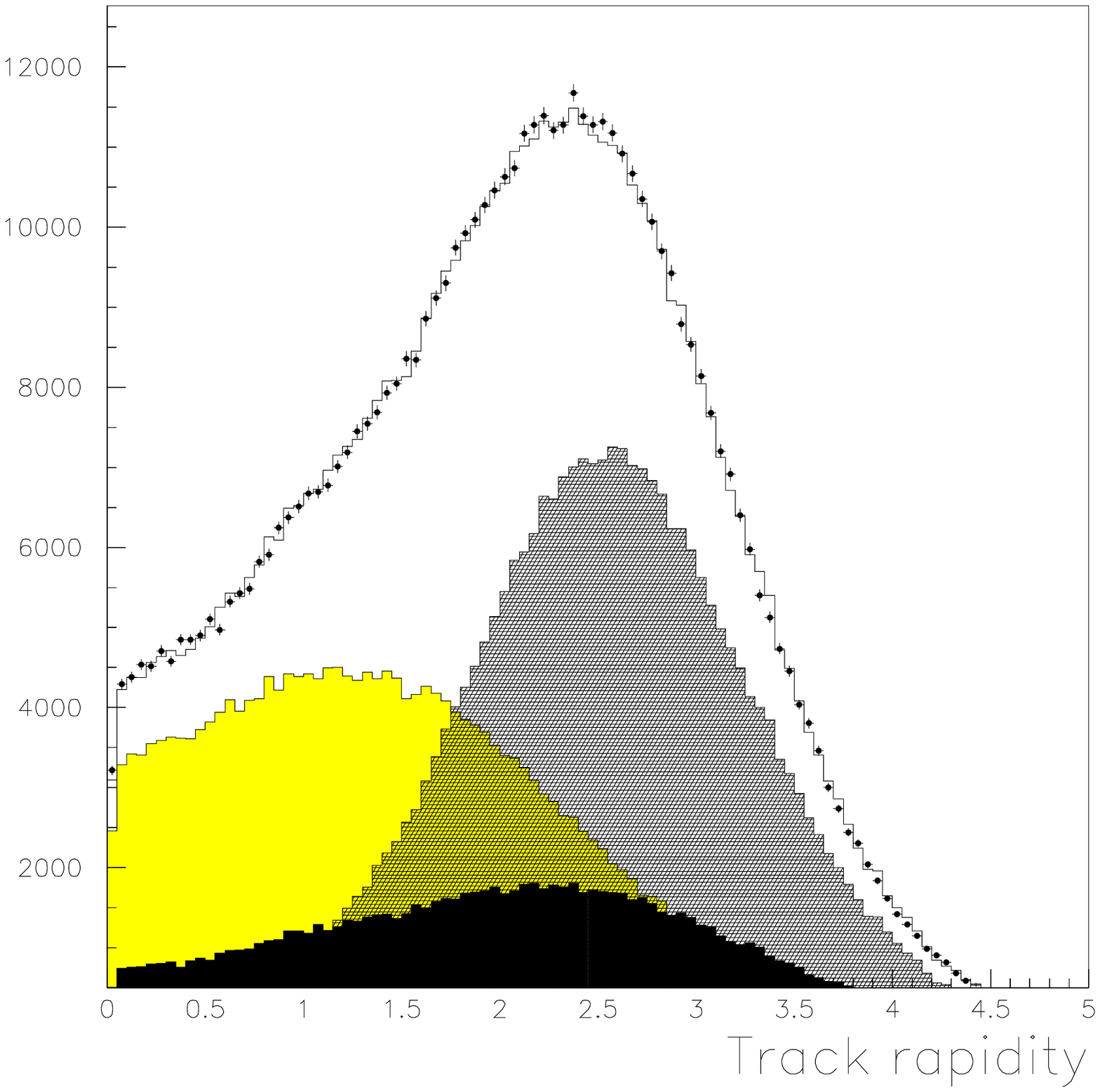,height=6cm,width=5.2cm}}

\end{tabular}
\caption[]
{\label{fig:trknet_inputs} Distributions of the input variables to the TrackNet,
compared to data. 
Tracks are selected from events that are contained in the barrel,
with 2-jets and are about 80\% pure in $\ztobb$~events. 
The key to the symbols is given in the top left plot. Here, `signal' refers to tracks
originating from the B-hadron decay chain, `background' are tracks from fragmentation
or excited B-hadron decay, and u,d,s,c background are any tracks in non-b decays of the 
${\mathrm Z^0}$.}
\end{center}
\end{figure}

Distributions of all discriminating variables are shown in Figure~\ref{fig:trknet_inputs}.
Two points should be noted.
Firstly, to first approximation the track probabilities
to originate from the primary or secondary vertex as calculated by the 
BSAURUS or the AABTAG packages, are calculations of the same quantity.
There are differences however in the way track errors are handled in the two
cases and, in addition, the AABTAG package returns a probability tuned to the
resolution seen in the data. For these reasons, the correlation in these
variables between the two methods is not 100\% and there is extra information
to be gained by including all definitions as input variables to the network.
Secondly, the decay length variable (which is the same for all tracks in a given 
hemisphere) and the track quality word do not {\it a priori} distinquish
between B-decay products and the background. These variables do however
provide the network with information about whether the hemisphere has a 
well separated secondary vertex and whether the information 
associated with a  particular track is reliable or not. This additional
information has been found to be 
useful during the network training procedure.

\begin{figure}[p]
\begin{center}
\epsfig{figure=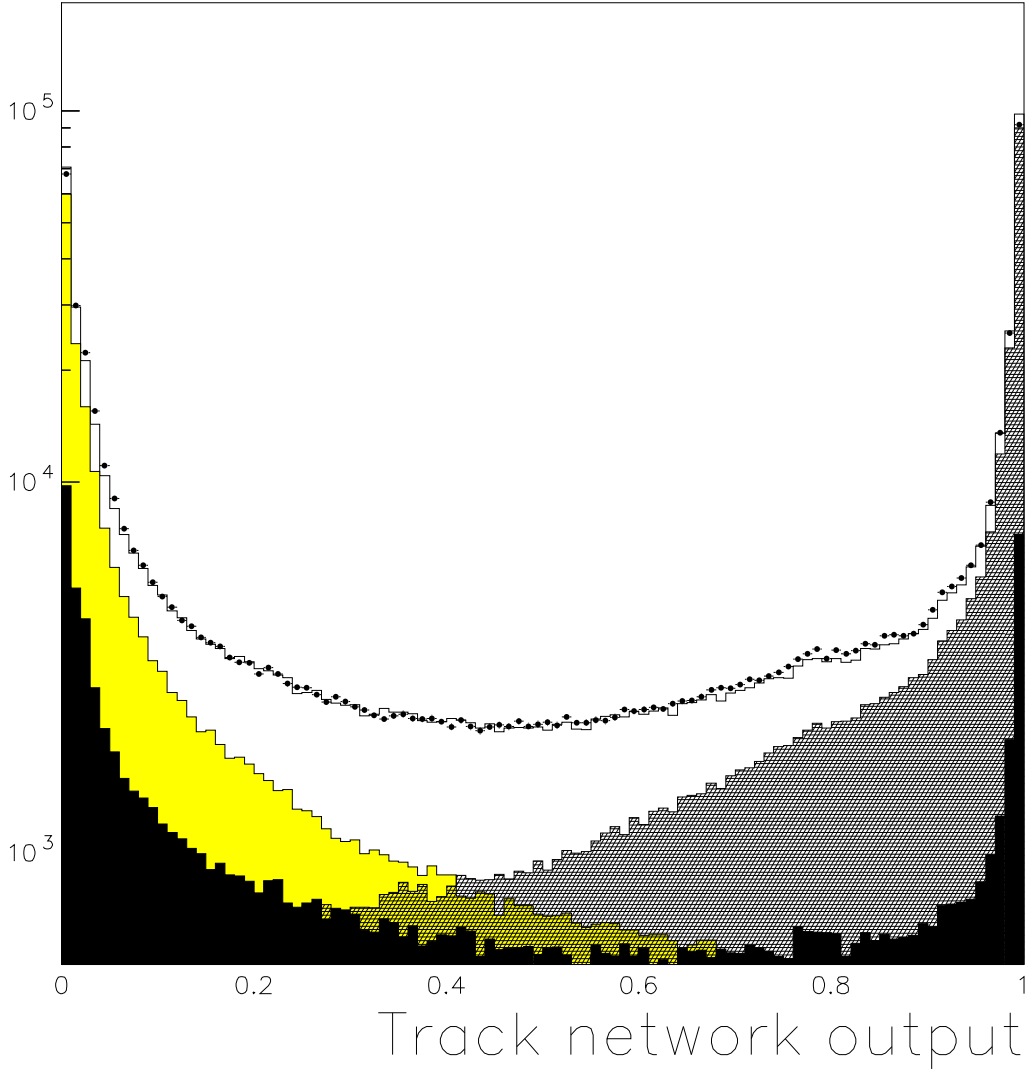,width=16cm}
\caption[]
{\label{fig:trknet_result} The TrackNet output distribution for Monte Carlo 
compared to the data. 
Tracks are selected from events that are contained in the barrel,
with 2-jets and are about 80\% pure in $\ztobb$~events. 
The shaded distributions are defined as in Figure~\ref{fig:trknet_inputs}.
Here, `Signal' refers to tracks
originating from the B-hadron decay chain, `background' are tracks from fragmentation
or excited B-hadron decay, and u,d,s,c background are any tracks in non-b decays of the 
${\mathrm Z^0}$.}
\end{center}
\end{figure}

\begin{figure}[p]
\begin{center}
\epsfig{figure=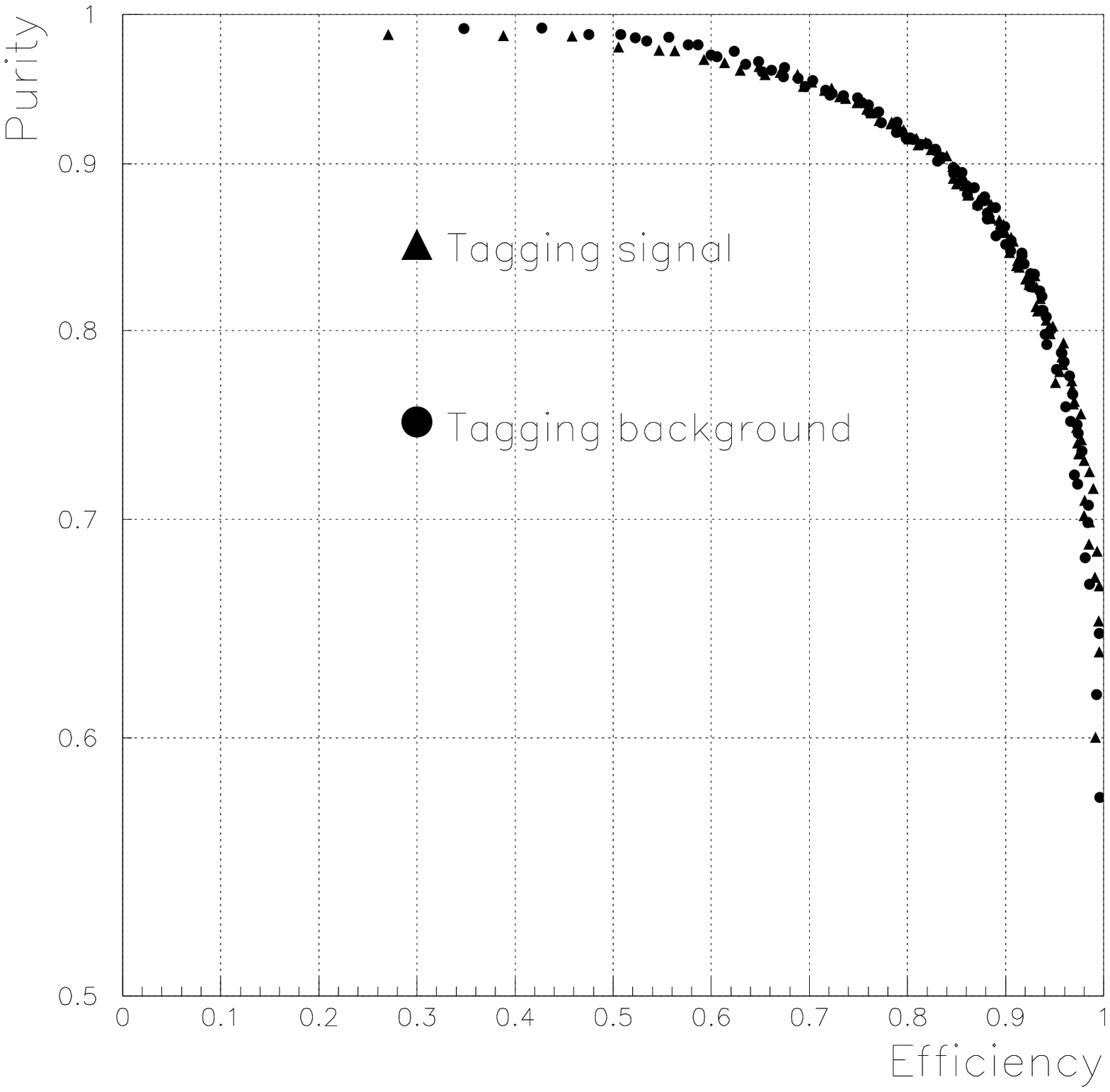,width=16cm}
\caption[]
{\label{fig:trknet_perf} The TrackNet performance.
Tracks are selected from events that are contained in the barrel,
with 2-jets and are about 80\% pure in $\ztobb$~events. 
The plot is based on equal numbers of `signal' tracks that 
originate from the B-hadron decay chain, and `background' tracks 
originating from fragmentation or excited B-hadron decay.}
\end{center}
\end{figure}

The network therefore consists of 11 input nodes, one for each of the
variables listed above, 12 nodes in the hidden layer, and a single output node
with target value 1 for signal tracks and 0 for background. The training
and performance samples are drawn only from $\ztobb$ Monte Carlo
where signal tracks are defined as coming from any point along the weak
B-decay chain, and background as coming from any source 
located at the primary vertex e.g. fragmentation and excited B-hadrons.
The network training was performed using 30K signal
and 30K background tracks in both the training and performance samples. 
The network output and performance for tagging tracks from B-decay are 
shown in Figures~\ref{fig:trknet_result} and~\ref{fig:trknet_perf}.



\section{B-Species Identification(SHBN)}
\label{sec:bspec}
BSAURUS returns in the output variables, \verb+BSHEM(IBH_PRBBP,IH)+, \verb+BSHEM(IBH_PRBB0,IH)+, 
\verb+BSHEM(IBH_PRBBS,IH)+ and \verb+BSHEM(IBH_PRBLB,IH)+ the probability for the 
hemisphere \verb+IH+ to contain a $\Bplus,\Bzero,\Bs$~or B-baryon respectively.
\footnote{The charge conjugate states are also implied.}
The approach
employed is to supply all discriminating variables to a neural network 
with four output nodes, one for each B-hadron type. 

The resulting {\bf S}ame {\bf H}emisphere {\bf B}-species enrichment {\bf N}etwork or SHBN, 
consists of 15 input nodes(described below) and  17 nodes in the single hidden layer.
Each output node delivers a probability for the hypotheses it is trained on i.e. 
the first supplies the probability for a $\Bs$~meson to be produced in the hemisphere, 
the second for a $\Bzero$~meson, the third for charged $\Bplus$~mesons and the fourth for all 
species of B-baryons. The following input variables are used:
\begin{itemize}
\item Using the TrackNet value as a  probability ($P_B$) for each track to originate from 
the B-hadron decay vertex rather than from the primary vertex, 
the weighted vertex charge is formed
\begin{equation}
\label{eqn-qv}
\left|Q_v\right|=\sum_{i}^{tracks} P_B(i)\cdot Q(i)
\end{equation} 
that distinguishes between charged and neutral B-hadrons
and is illustrated in Figure~\ref{fig:qver}.
\begin{figure}[p]
\begin{center}
\epsfig{file=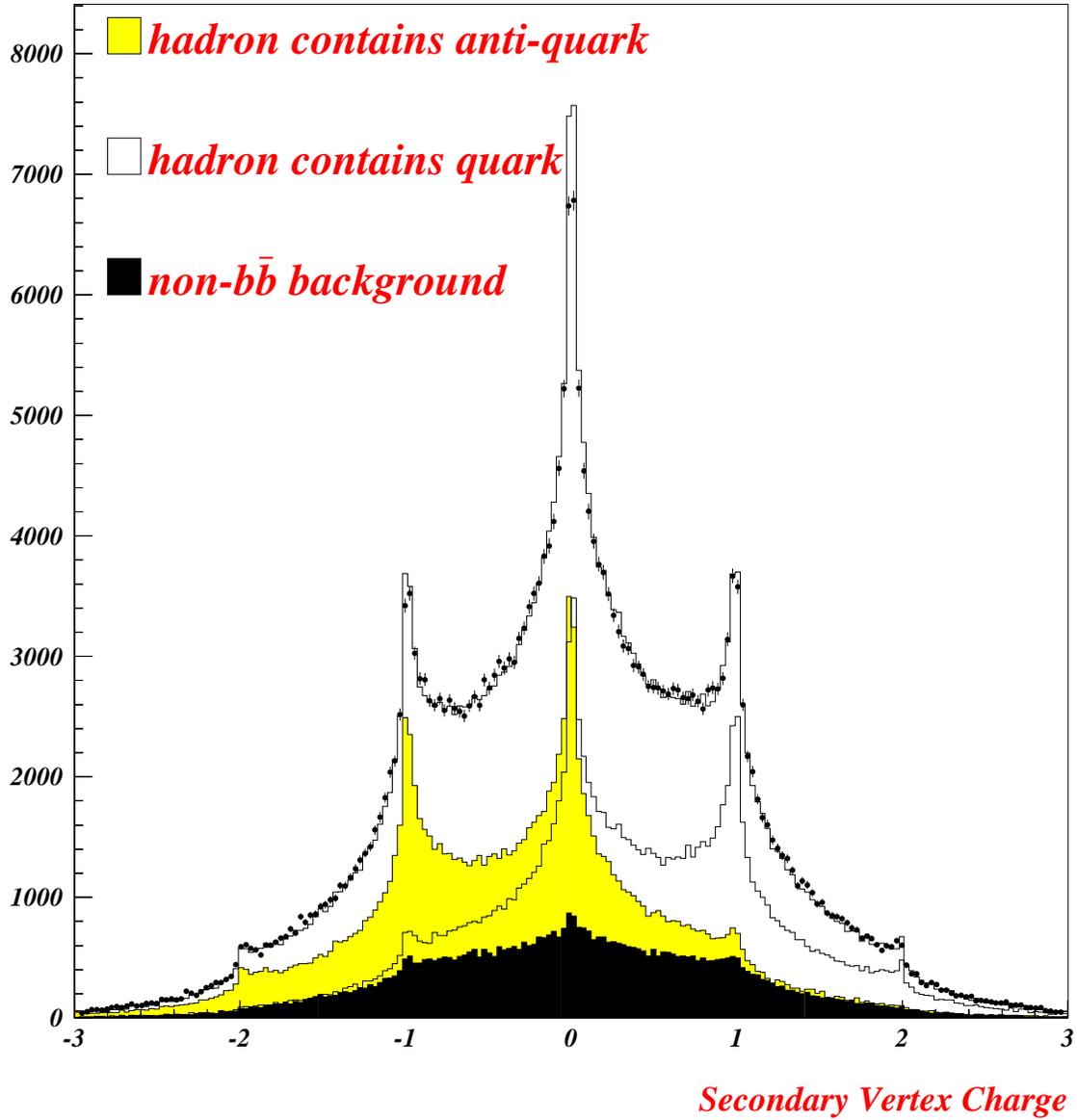,width=16.cm}
\caption[]
{\label{fig:qver} 
The vertex charge  variable comparing data(points) to simulation(histogram)
and showing
separately the contributions from neutral, positively and negatively
charged B-hadrons.
Selection cuts and weights have been applied as detailed in Appendix A.}
\end{center}
\end{figure}

\item The binomial error on the vertex charge, defined as 
\begin{equation}
\label{eqn-sqv}
 \sigma_{Q_v}=\sum_{i}^{tracks} \sqrt{P_B(i)(1 - P_B(i))}
\end{equation}
gives a measure of the reliability of the vertex charge information.

\item The number of charged pions in the hemisphere. 
This is most powerful for the case of
B-baryons and $\Bs$ mesons, which have a higher content of non-pion particles, 
i.e. neutrons, protons and kaons in comparison to other B-species.
\item The total energy deposited in the hemisphere by charged and neutral particles
scaled by the LEP beam energy.
This is sensitive to the presence of
B-baryons and $\Bs$~mesons, due to the fact that associated neutrons and $K_L^0$ are often not 
reconstructed in the detector, with the consequence that the total hemisphere energy 
tends to be smaller compared to $\Bzero$ or $\Bplus$ mesons.
\item $\Bs$~mesons are normally produced with a charged kaon as leading\footnote{The term {\it leading}
refers to the neighbouring hadron to the B-hadron that emerges from the fragmentation chain. 
This particle is identified in practice from the fact that it often has the largest rapidity with
respect to the B-hadron flight direction.}
fragmentation particle with a further kaon emerging from the weak decay 
(the same applies to the associated production of protons with B-baryons). 
Exploiting this fact, input variables are constructed giving the likelihood 
for the presence of a leading fragmentation kaon/proton and a
kaon/proton weak decay product, for each of the two hypotheses; $\Bs$~or B-baryon.

Utilising the Monte Carlo truth information, normalised track rapidity 
distributions were parameterised with Gaussians
separately for the case of leading fragmentation tracks and B-decay products. 
These were then used to form 
a weight per reconstructed track which was summed over to give a hemisphere level
variable. Cuts were applied for kaon or proton identification 
(see Section~\ref{sec:partid}) and in order 
to separate the fragmentation tracks from the B-decay tracks via
the  TrackNet output. 

\item The leading fragmentation track also can often be neutral
e.g. production of a
 ${\mathrm K}^0_s$~associated with  $\Bs$~meson production or
 ${\mathrm \Lambda}^0$~associated with  B-baryon production.
An input was therefore constructed based on the presence of a reconstructed  ${\mathrm K}^0_s$
or ${\mathrm \Lambda}^0$
in the hemisphere in the same way as for charged kaons and protons described above.
In this case,
the rapidity distribution was formed from Monte Carlo truth information 
for neutral 
particles originating from the primary vertex. The resulting weights were 
then summed over for reconstructed neutral particles with DELPHI mass codes 61 or 81 
identifying ${\mathrm K}^0_s$~and  ${\mathrm \Lambda}^0$ candidates.  

\item The probability for the leading fragmentation particle to be a charged kaon. Specifically,
the maximum kaon net output from the three tracks with highest rapidity originating from 
the primary vertex (via the condition TrackNet$<0.5$) was used. The three kaon 
probabilities can be found in BSAURUS outputs, 
\verb+ BSHEM(IBH_KRANK1,IH),BSHEM(IBH_KRANK2,IH),BSHEM(IBH_KRANK3,IH)+.

\item A charge correlation between the leading fragmentation particle charges 
provided a flag for the presence of $\Bplus$~mesons. Specifically, the 
rapidity-weighted track charge sum over all tracks in the hemisphere
is formed, and this is then scaled by the measured vertex charge.

\end{itemize}
In addition, input variables that gave no inherent separation power between
different B species  were included to inform the network of the potential 
quality of the other input variables: 
\begin{itemize}
\item the invariant mass of the reconstructed vertex,

\item the hemisphere quality factor described in Section \ref{sec:qual},

\item the energy of the B-hadron, to provide information on how hard the fragmentation was 
and therefore inform the network of how the available energy is expected to be shared 
between B hadron and fragmentation products.

\end{itemize}

\begin{figure}[p]
\begin{center}
\epsfig{file=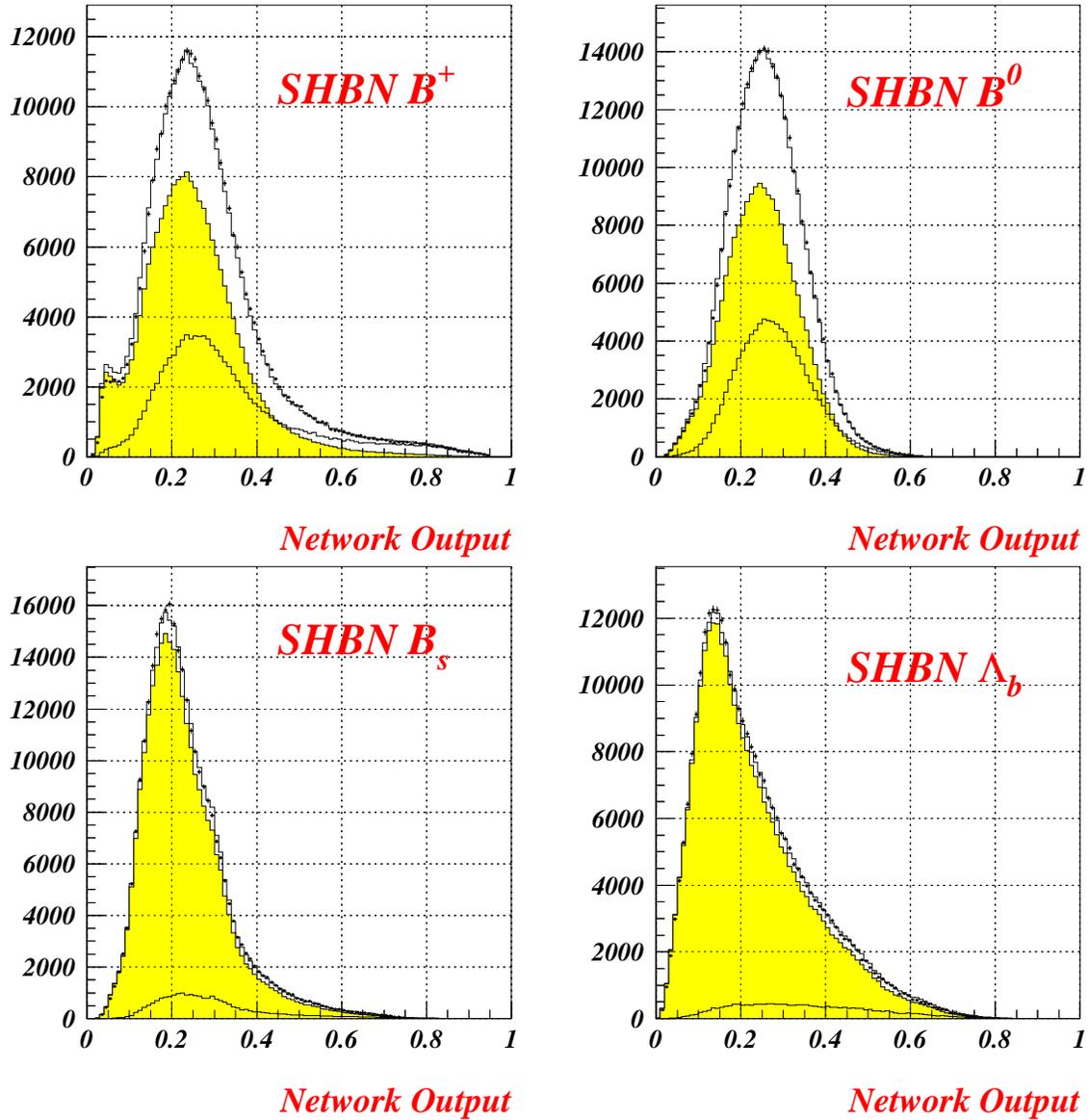,width=16.cm}
\caption[]
{\label{fig:SHBN_out} 
Output of the SHBN network for the $\Bplus,\Bzero,\Bs$~and B-baryon
hypotheses in the simulation(histogram) compared to the 
data(points). The component histograms
in each case show the distribution for the hypothesis being 
considered, i.e. the `signal'(open histogram), compared
to the distribution for everything else, 
i.e. the `background' (shaded histogram).
Selection cuts and weights have been applied as detailed in Appendix A.}
\end{center}
\end{figure}

\begin{figure}[p]
\begin{center}
\epsfig{file=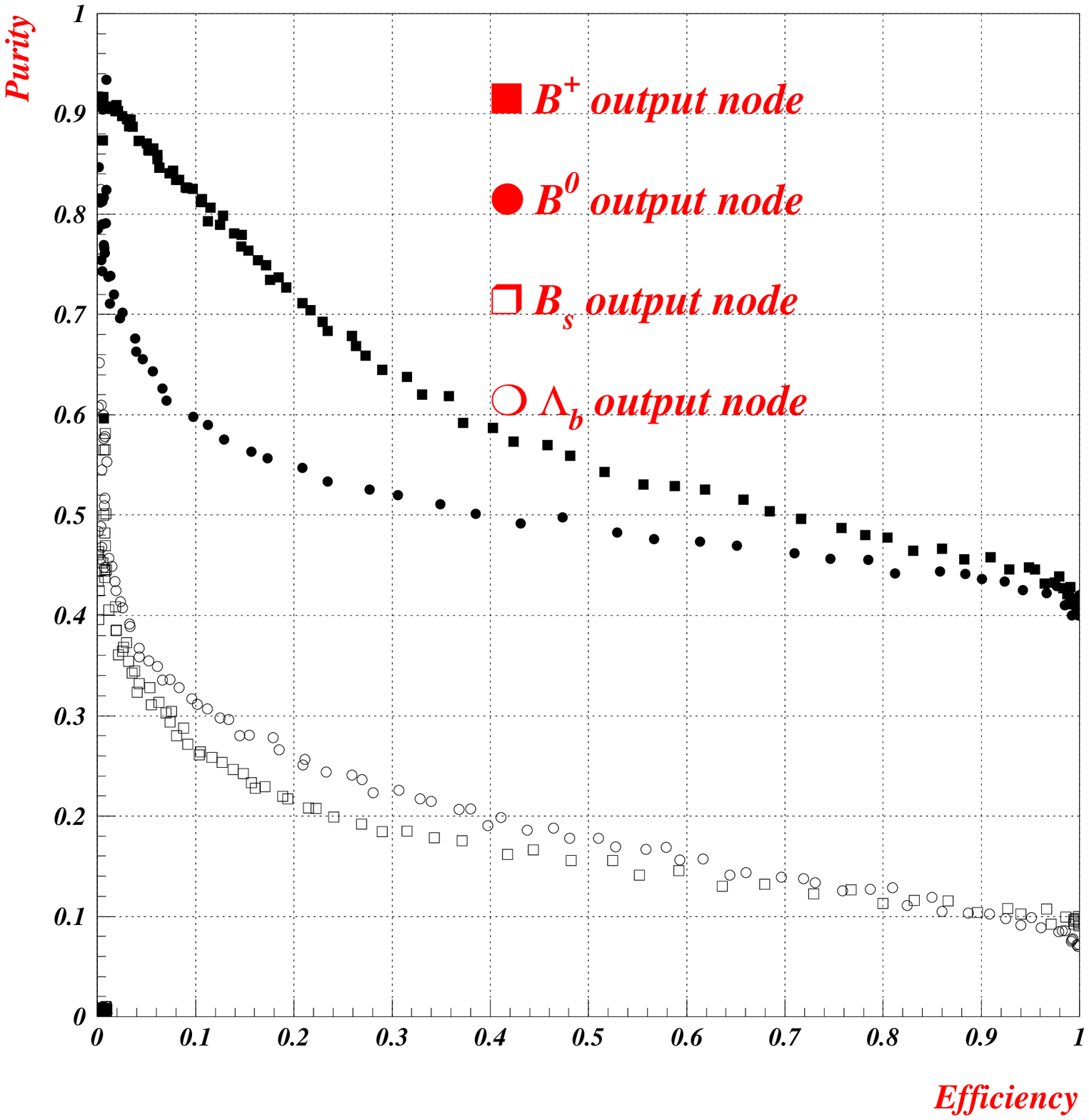,width=16.cm}
\caption[]
{\label{fig:SHBN_per} Performance of the SHBN for enriching samples in the
various B-hadron types. The plot is based on $\bb$~Monte Carlo
making successive cuts in the network output
and where `signal' and `background' are defined as in Figure~\ref{fig:SHBN_out}.
Selection cuts and weights have been applied as detailed in Appendix A.}
\end{center}
\end{figure}

The output of the SHBN in both simulation and data is shown in 
Figure~\ref{fig:SHBN_out}
and the resulting performance of the SHBN is shown in Figure~\ref{fig:SHBN_per}.
The performance can be further improved  by the combination
of charge correlation information from the opposite hemisphere 
to form the 
{\bf B}oth {\bf H}emispheres {\bf B}-species enrichment {\bf N}etwork or BHBN. This
network is described in Section~\ref{sec:BHBN}.

\section{B-D Separation(BDnet)}
\label{sec:BDnet}
\begin{figure}[p]
\begin{center}
\epsfig{file=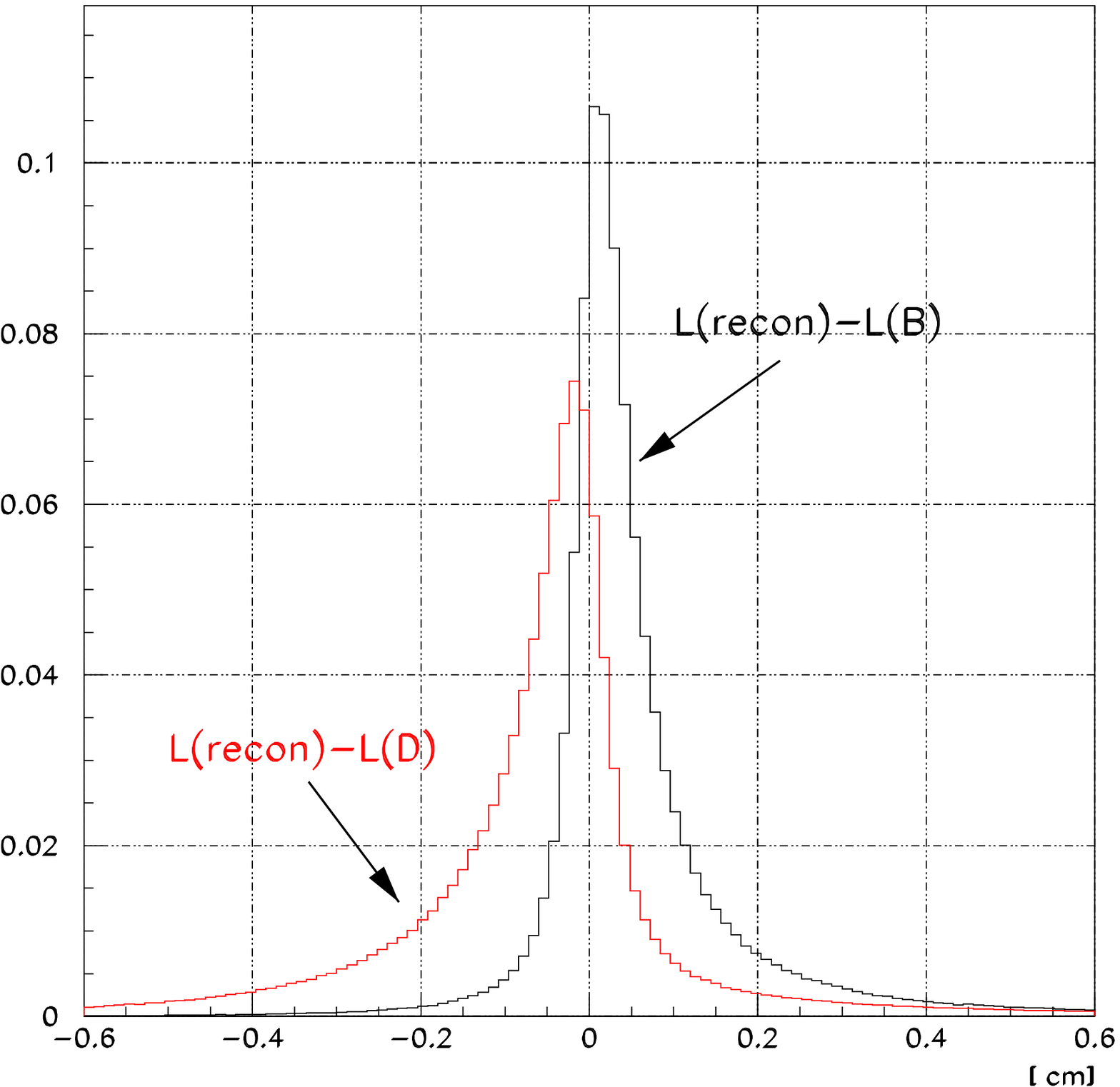,width=16.0cm}
\caption[]
{\label{fig:pullcasc} Comparing
$(L(recon)-L(B))$ to $(L(recon)-L(D))$ illustrating that the reconstructed
decay length is often larger than the true B-decay length but less than
the true D-cascade decay length. The origin of this pull to larger
decay lengths is the inclusion of one or more tracks from the 
D-vertex.} 
\end{center}
\end{figure}

In selecting tracks for inclusion in the B-secondary vertex fit
described in Section~\ref{sec:sv}, there is inevitably some 
background from tracks that originate not from the B-decay vertex
directly, but from the subsequent D-cascade decay point. The effect
of including such tracks is illustrated in Figure~\ref{fig:pullcasc} 
which shows that, on average, the B-decay length estimate
reconstructed by BSAURUS lies somewhere between the true B decay
point and that of the cascade D.

The BDnet is a neural network designed to 
discriminate between tracks originating from the weakly decaying B-hadron 
and those from the subsequent cascade D-meson decay. This information is
clearly crucial to any algorithm wishing to eliminate the positive
decay length bias just described, and is also used extensively throughout
BSAURUS, particularly in the construction of B-flavour tags 
(e.g. see Section~\ref{sec:trkflav}).

\begin{figure}[p]
\begin{center}
\epsfig{file=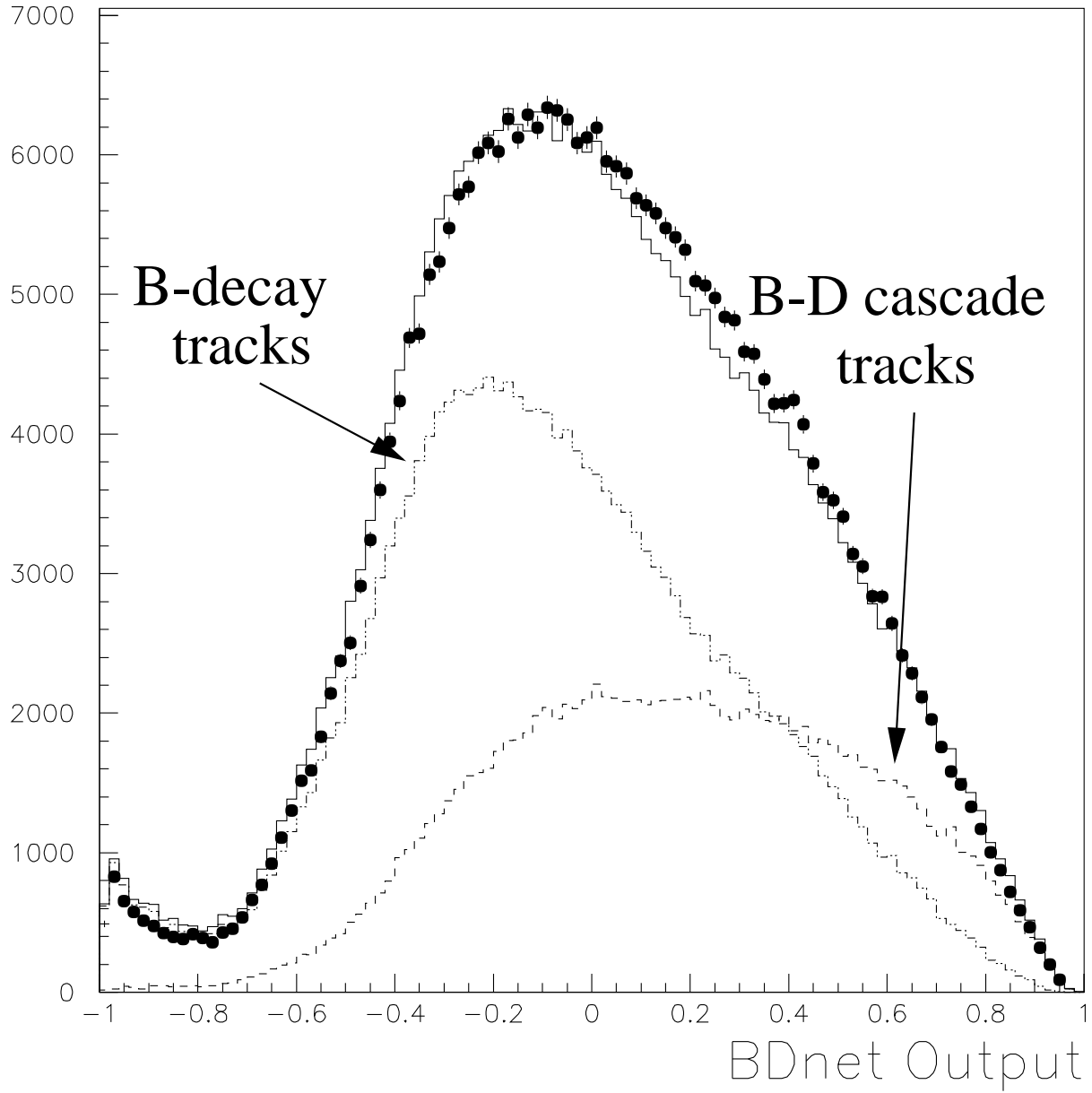,width=16.cm}        
\caption[]
{\label{fig:bdout} Output of the BDnet for tracks in simulation(histogram) 
and in data(points), for tracks with TrackNet$>0.5$. The
component histograms show the two classes of track that the network was
trained on, namely tracks originating from cascade D-decay(`signal')
and all other tracks which are mainly tracks from B-decay(`background').  
Tracks are selected from events that are contained in the barrel,
with 2-jets and are about 80\% pure in $\ztobb$~events.} 
\end{center}
\end{figure}

\begin{figure}[p]
\begin{center}
\epsfig{file=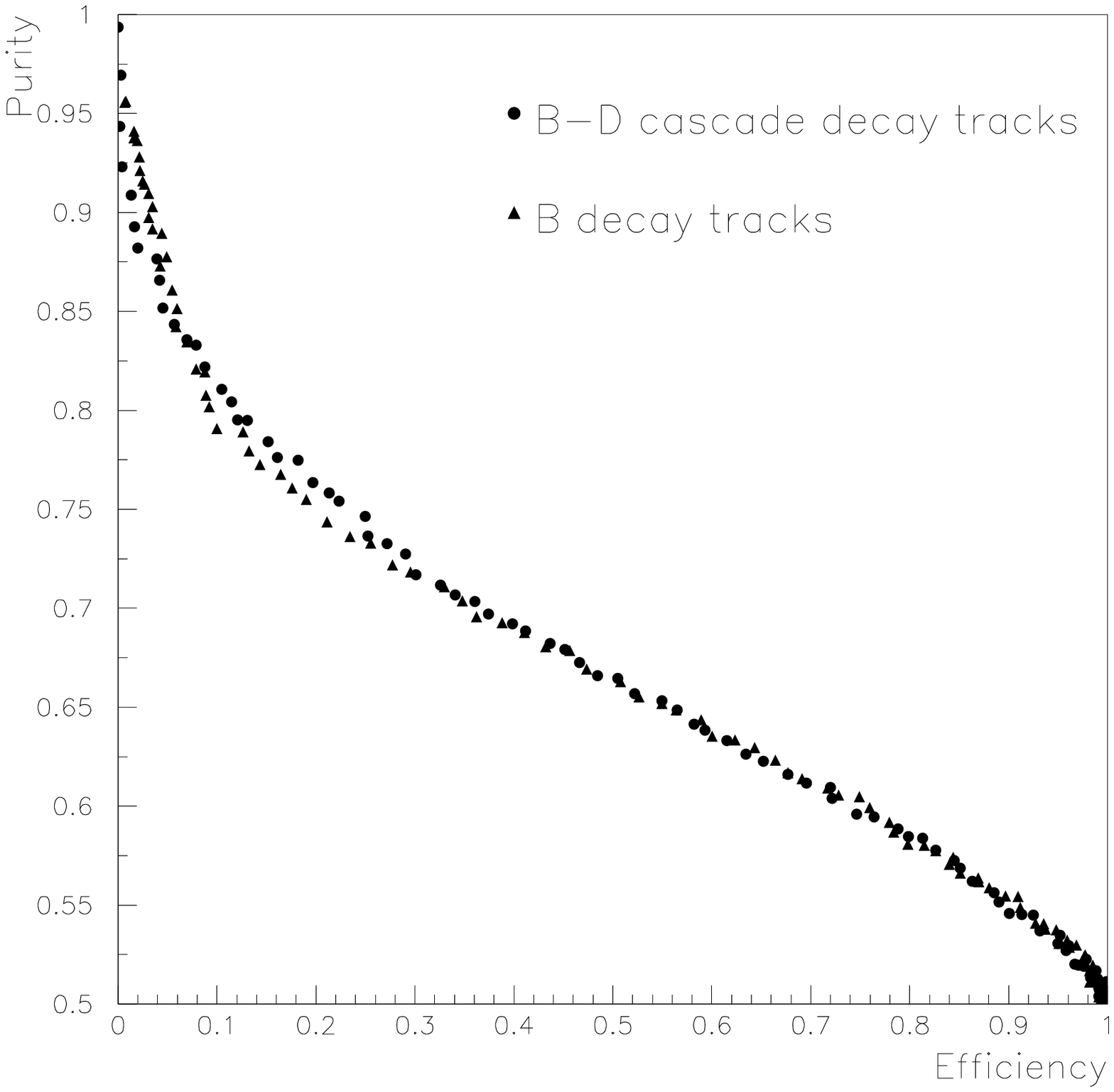,width=16.cm}
\caption[]
{\label{fig:bdperf} The purity against efficiency performance of the BDnet.
The plot is based on equal numbers of `signal' tracks, i.e. B-D cascade 
tracks, and `background' tracks, i.e. all other tracks with TrackNet$>0.5$.
Tracks are selected from events that are contained in the barrel,
with 2-jets and are about 80\% pure in $\ztobb$~events.}   
\end{center}  
\end{figure}

The following discriminating variables are used:
\begin{itemize}  
\item The angle between the track vector and an estimate of the B flight direction
(taken to be  \verb+BSHEM(IBH_BTT,IH)+). 
\item The probability that the track originates from the fitted primary vertex
      (AABTAG algorithm).
\item The probability that the track originates from the fitted secondary vertex
      (AABTAG algorithm).
\item The momentum and angle of the track vector in the B rest frame.
\item The TrackNet output.
\item The kaon network output, described in Section~\ref{sec:partid}.
\item The lepton identification, see Section~\ref{sec:partid}.
\end{itemize}
The remaining variables carry no implicit discriminating power but
are included as gauges of the quality of the other variables:
\begin{itemize} 
\item The track quality word from Section~\ref{sec:qual}.
\item The hemisphere quality word from Section~\ref{sec:qual}.
\item The hemisphere decay length significance, $L/ \sigma_L$.
\item The hemisphere secondary vertex mass.
\item The hemisphere rapidity gap between the track of highest rapidity
below a TrackNet cut at 0.5 and that of smallest rapidity above
the cut at 0.5.
\end{itemize}
The network architecture consisted of 14 input nodes with 15 nodes in a 
hidden layer and with a single output node trained on equal numbers of tracks
directly from B-decay as `signal' or from the following D-decay as `background'.
Figure~\ref{fig:bdout} shows the output of the network for the two 
classes of track used in training and Figure~\ref{fig:bdperf} plots the 
performance of the tag.

\section{Optimal Fragmentation, Decay Flavour and Decay Type Tagging}
\label{sec:flavtag}
In an attempt to use the information available optimally, the BSAURUS
approach to tagging the flavour, i.e. charge of the b-quark, works
by first constructing a probability {\bf per track} and then combining
these to give a probability at the hemisphere level. 

Separate neural networks are trained to tag the 
underlying quark charge for the cases when there was 
a $\Bplus,\Bzero,\Bs$~or B-baryon produced in the hemisphere.
In addition two sets of such networks are produced,
one trained only on tracks originating from the {\bf fragmentation}
process and the other trained only on tracks originating from
weak B-hadron {\bf decay}.

\subsection{The Basis of Optimal Flavour Tagging}
\label{sec:trkflav}

The basis of optimal flavour tagging
is to construct, at the {\bf track level}, the 
conditional probability for the track to have the same charge as
the b-quark in the B-hadron both at the moment of 
fragmentation (i.e. production) and at the moment of decay. In addition, these
probabilities are constructed separately for each of the B-hadron types;  
$\Bplus,\Bzero,\Bs$~and B-baryon.

A neural network is used with a target output value of  +1(-1) if the track 
charge is correlated(anticorrelated) with  the b-quark charge. 
The discriminating input variables used are the following:
\begin{itemize}
\item {\bf Particle identification variables} (see Section~\ref{sec:partid}):
\begin{itemize} 
\item kaon network output, 
\item proton network output,
\item electron network output,
\item muon classification code.
\end{itemize}
\item {\bf B-D vertex separation variables:}
\begin{itemize} 
\item BDnet output,
\item $\frac{BD-BD(min)}{\Delta BD}$, where $BD$~is the BDnet value,
$BD(min)$~is the minimum  BDnet value for all tracks in the hemisphere
above a TrackNet value of 0.5, and $\Delta BD$~is the difference between 
$BD(max)$~and $BD(min)$,
\item the momentum of the track in the B-candidate centre of mass frame.
\end{itemize}
\item {\bf Track-level quality variables:}
\begin{itemize}
\item the helicity angle of the track vector in the B-candidate centre of mass frame,
\item the track quality as defined in Section~\ref{sec:qual}, 
\item the TrackNet output,
\item the track energy.
\end{itemize}
\item {\bf Hemisphere-level quality variables:}
\begin{itemize}
\item the hemisphere rapidity gap between the track of highest rapidity
below a TrackNet cut at 0.5 and that of smallest rapidity above
the cut at 0.5,
\item the hemisphere quality as defined in Section~\ref{sec:qual}, 
\item the number of tracks in the hemisphere passing the standard cuts 
(Section~\ref{sec:trsel}) and with a TrackNet value greater than 0.5,  
\item $\Delta BD$~as defined above,
\item the secondary vertex mass,
\item the secondary vertex fit $\chi^2$~probability, 
\item the ratio of the  B-energy, defined by the \verb+BSHEM(IBH_BTT,IH)+,
to the LEP beam energy, 
\item the error on the vertex charge measurement.
\end{itemize} 
\end{itemize}
In total, the track decay flavour network used all input variables
described above
in a network containing two hidden layers with 20 nodes in the first
and 3 in the second. The track production flavour network 
used the same input variables without the lepton identification
and B-D vertex separation variables, with 
19 nodes in the first hidden layer and 3 in the second.

\begin{figure}[p]
\begin{center}
\epsfig{file=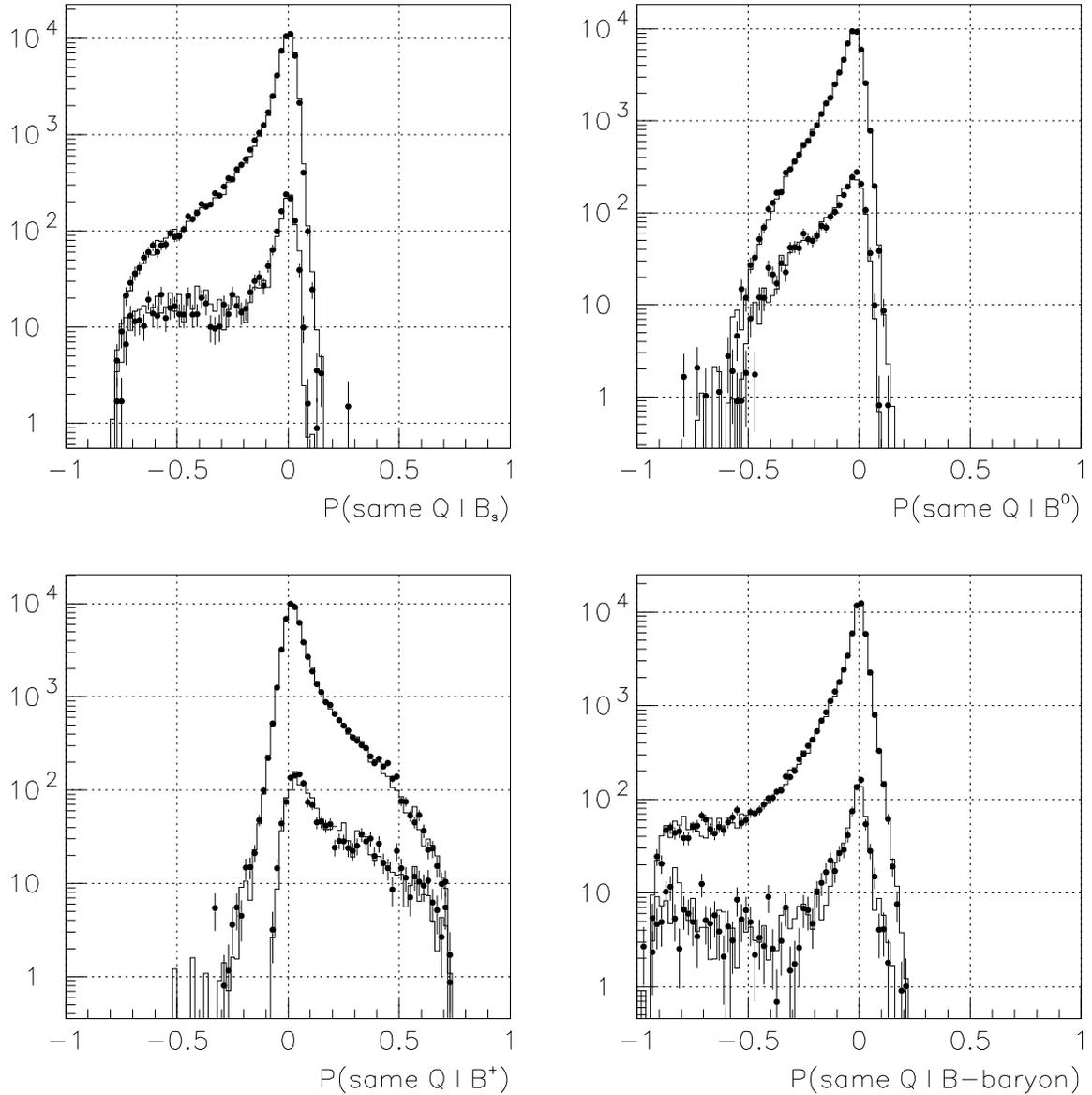,width=18.cm}
\caption[]
{\label{fig:trkptag} The track-level conditional fragmentation flavour
probability for all B-hadron types, comparing data(points) and 
simulation(histogram). The plots with larger normalisation
correspond to the normal mixture of B-hadron types whilst the
distibutions with smaller normalisation are for samples enhanced
in that B-type. The enhancement cuts were chosen to give roughly a
$10\%$ selection efficiency of signal, i.e. BHBN greater than:
0.8 for $\Bplus$, 0.6 for $\Bzero$, 0.2 for $\Bs$ and 0.33 for 
B-baryons. All tracks have TrackNet less than 0.5 and the events 
used were 2-jet events only, contained in the barrel.
Weights were applied as in Appendix A}  
\end{center}  
\end{figure}
\begin{figure}[p]
\begin{center}
\epsfig{file=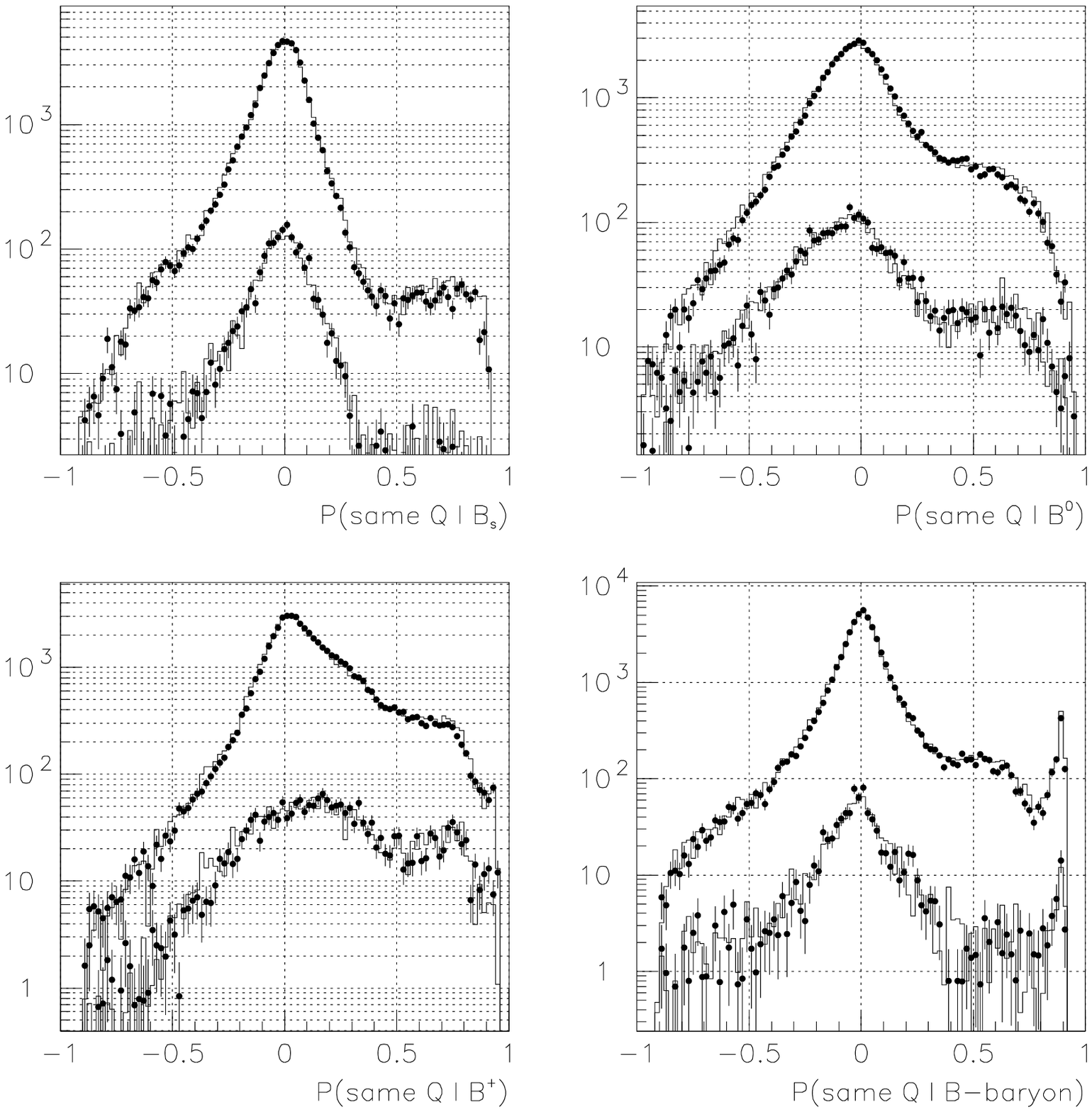,width=18.cm}
\caption[]
{\label{fig:trkdtag} The track-level conditional decay flavour
probability for all B-hadron types, comparing data(points) and 
simulation(histogram). The plots with larger normalisation
correspond to the normal mixture of B-hadron types whilst the
distibutions with smaller normalisation are for samples enhanced
in that B-type. The enhancement cuts were chosen to give roughly a
$10\%$ selection efficiency of signal, i.e. BHBN greater than:
0.8 for $\Bplus$, 0.6 for $\Bzero$, 0.2 for $\Bs$ and 0.33 for 
B-baryons. All tracks have TrackNet greater than 0.5 and the events 
used were 2-jet events only, contained in the barrel. 
Weights were applied as in Appendix A} 
\end{center}  
\end{figure}
\begin{figure}[p]
\begin{center}
\epsfig{file=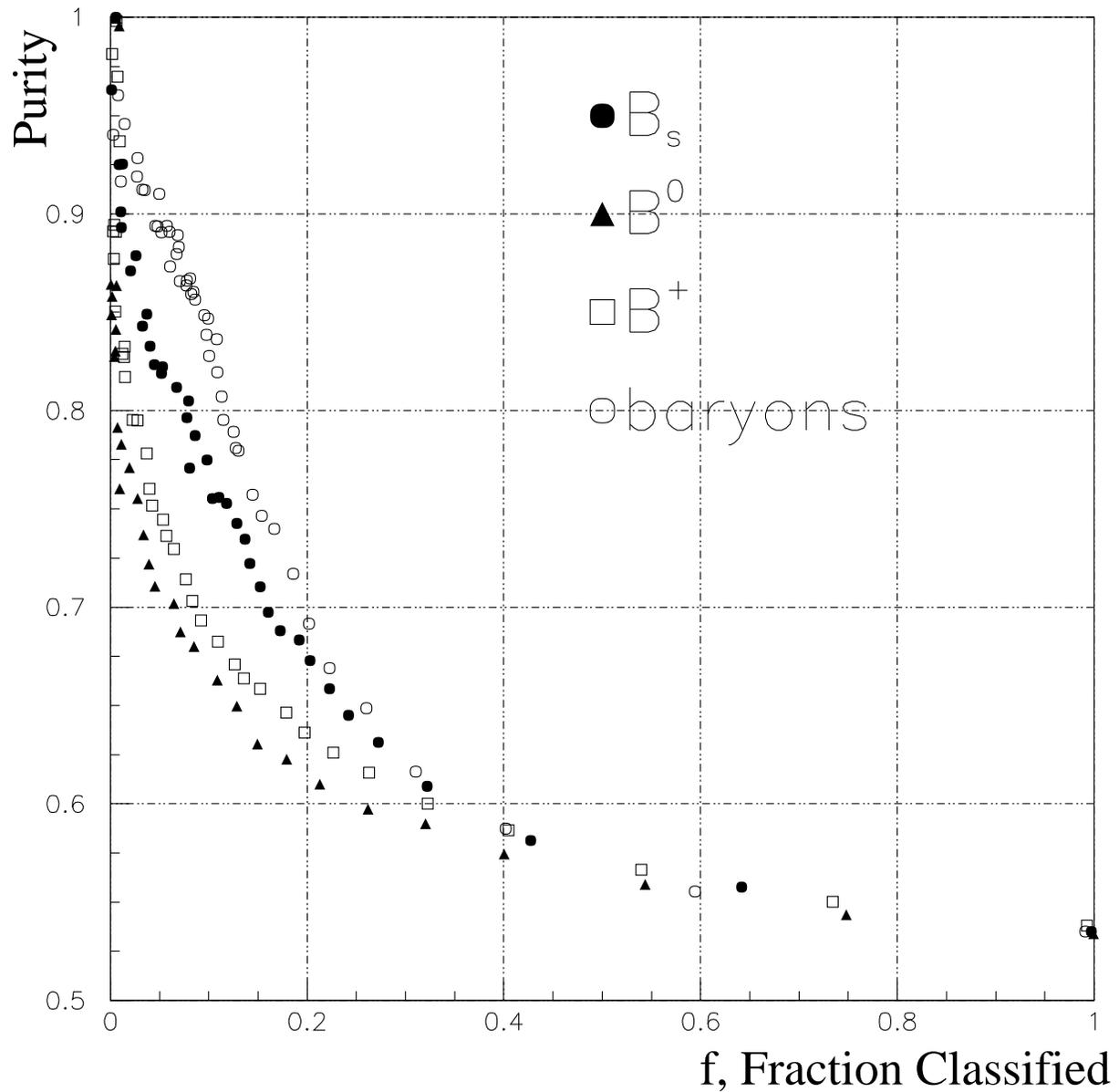,width=16.cm}
\caption[]
{\label{fig:trk_pfltag} The track-level production flavour tag
performance. All tracks have TrackNet less than 0.5 and the events 
used were 2-jet events only, contained in the barrel.}  
\end{center}  
\end{figure}
\begin{figure}[p]
\begin{center}
\epsfig{file=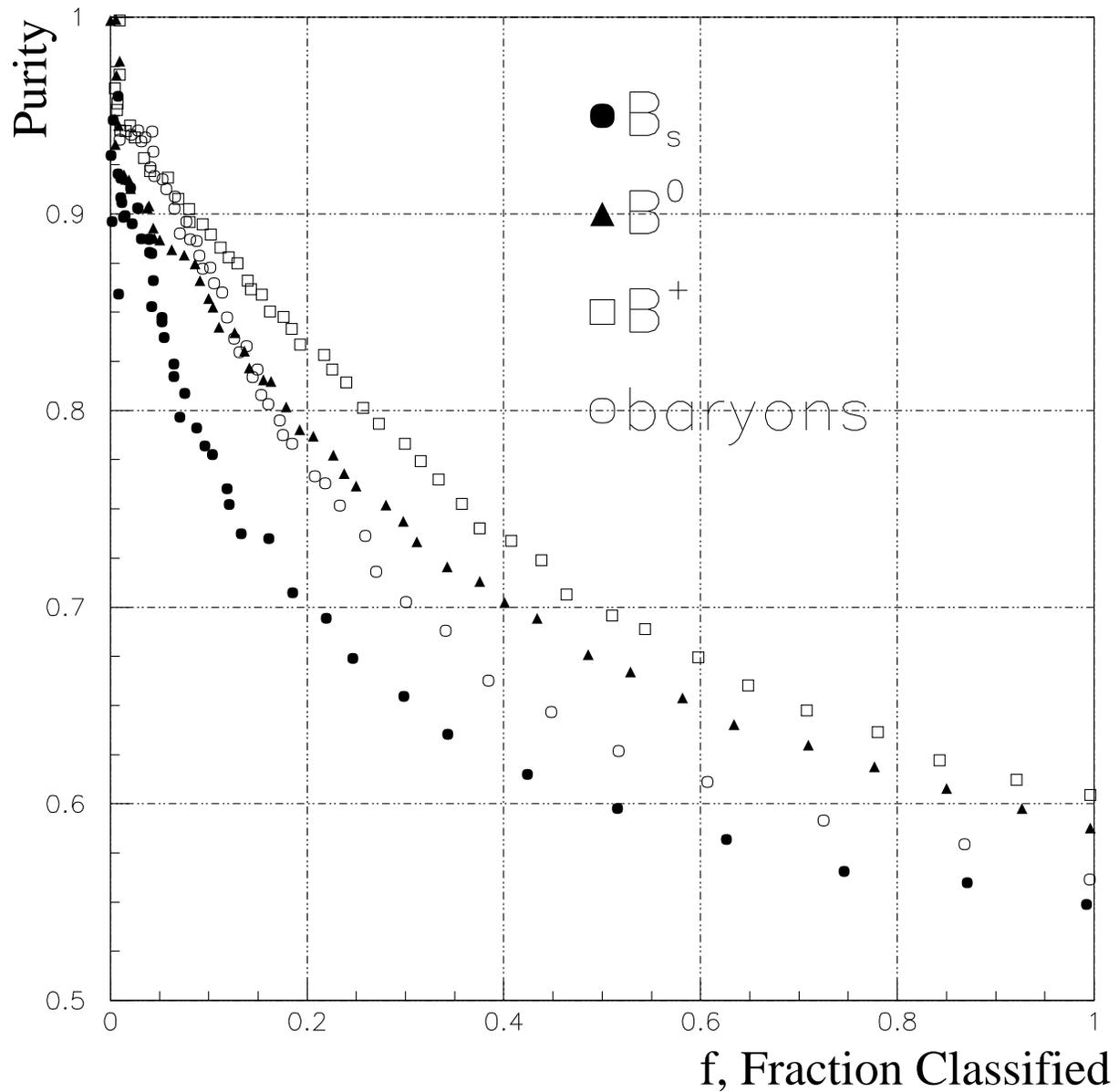,width=16.cm}
\caption[]
{\label{fig:trk_dfltag} The track-level decay flavour tag
performance. All tracks have TrackNet greater than 0.5 and the events 
used were 2-jet events only, contained in the barrel.}  
\end{center}  
\end{figure}

The resulting track charge correlation  conditional probabilities
are plotted in Figure~\ref{fig:trkptag} for the case of fragmentation
flavour and in Figure~\ref{fig:trkdtag} for the case of decay flavour.
Figures~\ref{fig:trk_pfltag} and~\ref{fig:trk_dfltag} show the
corresponding tag performance attained by the
fragmentation and decay flavour  probabilities  respectively. 
These outputs can be found in BSAURUS variables;
\verb+BSPAR(IBP_FFLBS,IT)+,\verb+BSPAR(IBP_FFLB0,IT)+,
\verb+BSPAR(IBP_FFLBP,IT)+ and \verb+BSPAR(IBP_FFLLB,IT)+
for the fragmentation flavour conditional probabilities and in 
\verb+BSPAR(IBP_DFLBS,IT)+,\verb+BSPAR(IBP_DFLB0,IT)+,
\verb+BSPAR(IBP_DFLBP,IT)+ and \verb+BSPAR(IBP_DFLLB,IT)+
for the decay flavour conditional probabilities.
\begin{figure}[p]
\begin{center}
\epsfig{file=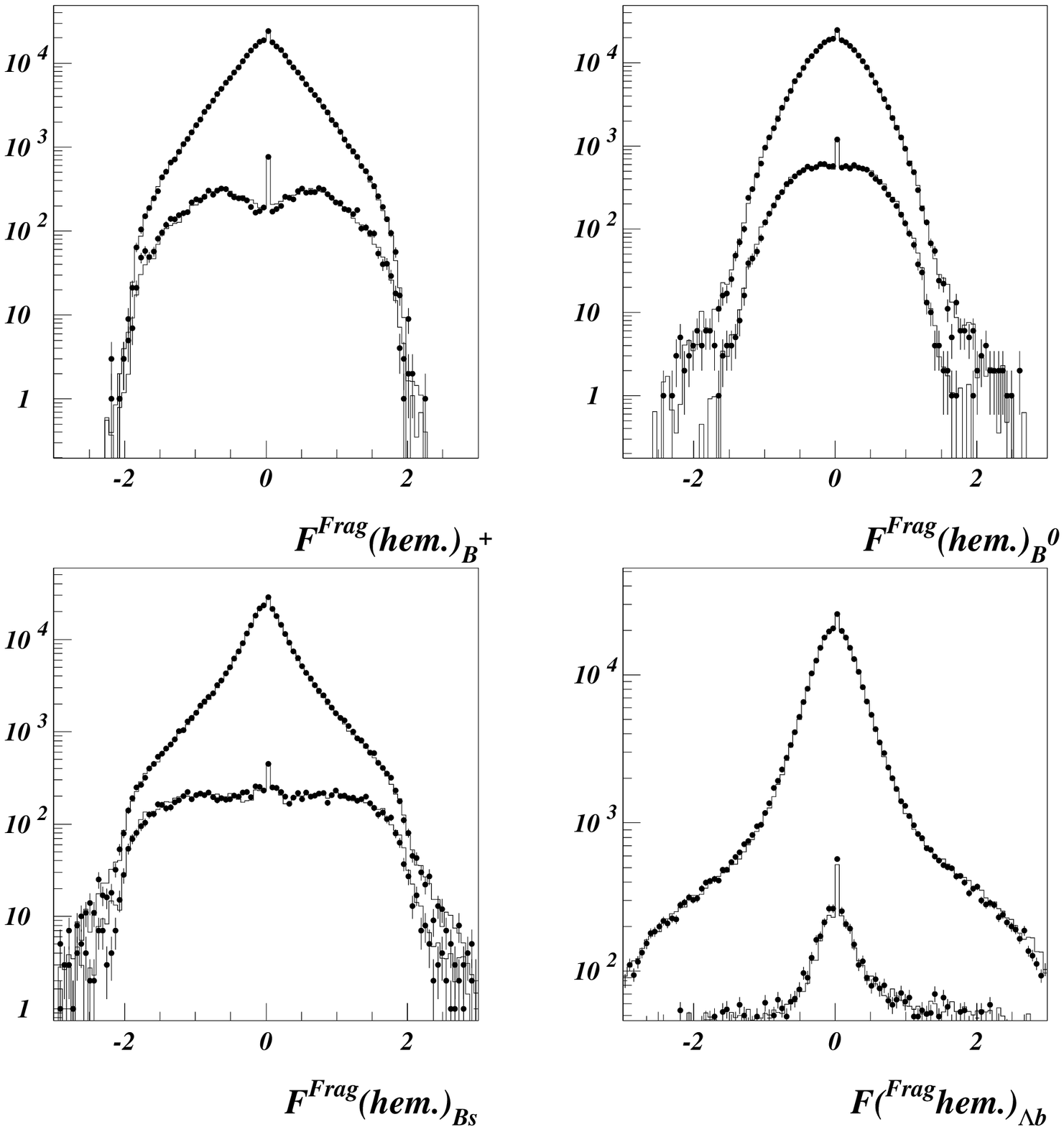,width=16.cm}
\caption[]
{\label{fig:hflprod} The hemisphere-level fragmentation flavour
tag for all B-hadron types, comparing data(points) and 
simulation(histogram). The plots with larger normalisation
correspond to the normal mixture of B-hadron types whilst the
distibutions with smaller normalisation are for samples enhanced
in that B-type. The enhancement cuts were chosen to give roughly a
$10\%$ selection efficiency of signal, i.e. BHBN greater than:
0.8 for $\Bplus$, 0.6 for $\Bzero$, 0.2 for $\Bs$ and 0.33 for 
B-baryons. 
Selection cuts and weights have been applied as detailed in Appendix A.}
\end{center}  
\end{figure}
\begin{figure}[p]
\begin{center}
\epsfig{file=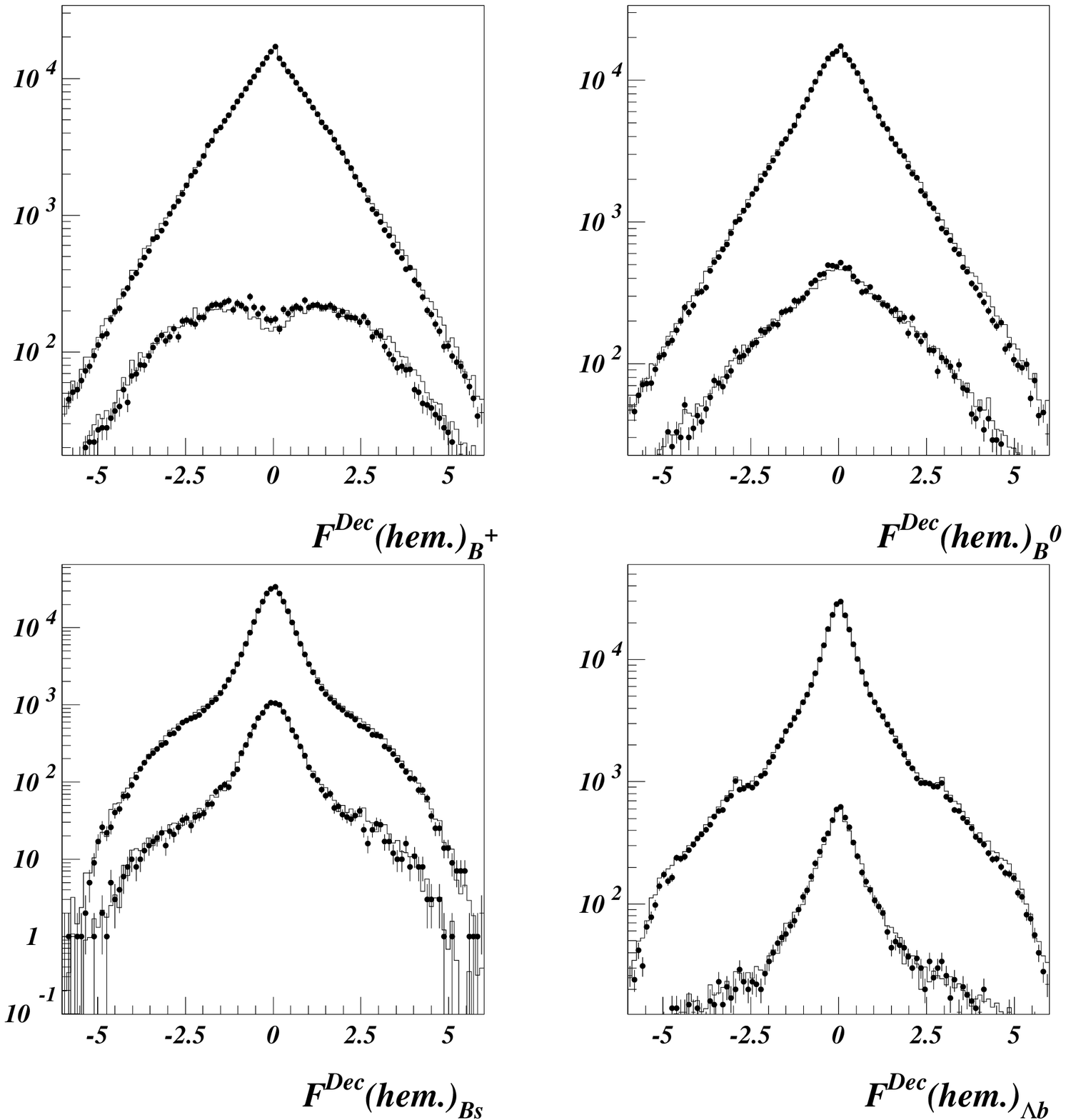,width=16.cm}
\caption[]
{\label{fig:hfldec} The hemisphere-level decay flavour
tag for all B-hadron types, comparing data(points) and 
simulation(histogram). The plots with larger normalisation
correspond to the normal mixture of B-hadron types whilst the
distibutions with smaller normalisation are for samples enhanced
in that B-type. The enhancement cuts were chosen to give roughly a
$10\%$ selection efficiency of signal, i.e. BHBN greater than:
0.8 for $\Bplus$, 0.6 for $\Bzero$, 0.2 for $\Bs$ and 0.33 for 
B-baryons. 
Selection cuts and weights have been applied as detailed in Appendix A.}
\end{center}  
\end{figure}

Finally, to obtain a flavour tag at the hemisphere level, 
the conditional track probabilities described above, 
$P(same\,Q|i)^j$ where 
$i=\Bplus,\Bzero,\Bs$~or B-baryon and $j=fragmentation$~or $decay$, are
combined as the likelihood ratio,
\begin{eqnarray}
F(hem.)_i^j=\sum_{tracks} \ln \left(\frac{1+P(same\,Q|i)^j}{1-P(same\,Q|i)^j} \right).Q(track)
\nonumber
\end{eqnarray}
where $Q(track)$~is the track charge. Which tracks to sum over
depends on the hypothesis considered, i.e. for the fragmentation flavour it is over 
all tracks with TrackNet$<0.5$~whilst for the decay flavour, tracks must
satisfy TrackNet$\geq 0.5$.
These hemisphere flavour tags are compared with the data  in 
Figure~\ref{fig:hflprod} for the fragmentation tag, 
and in Figure~\ref{fig:hfldec} for the decay tag.

\begin{figure}[p]
\begin{center}
\epsfig{file=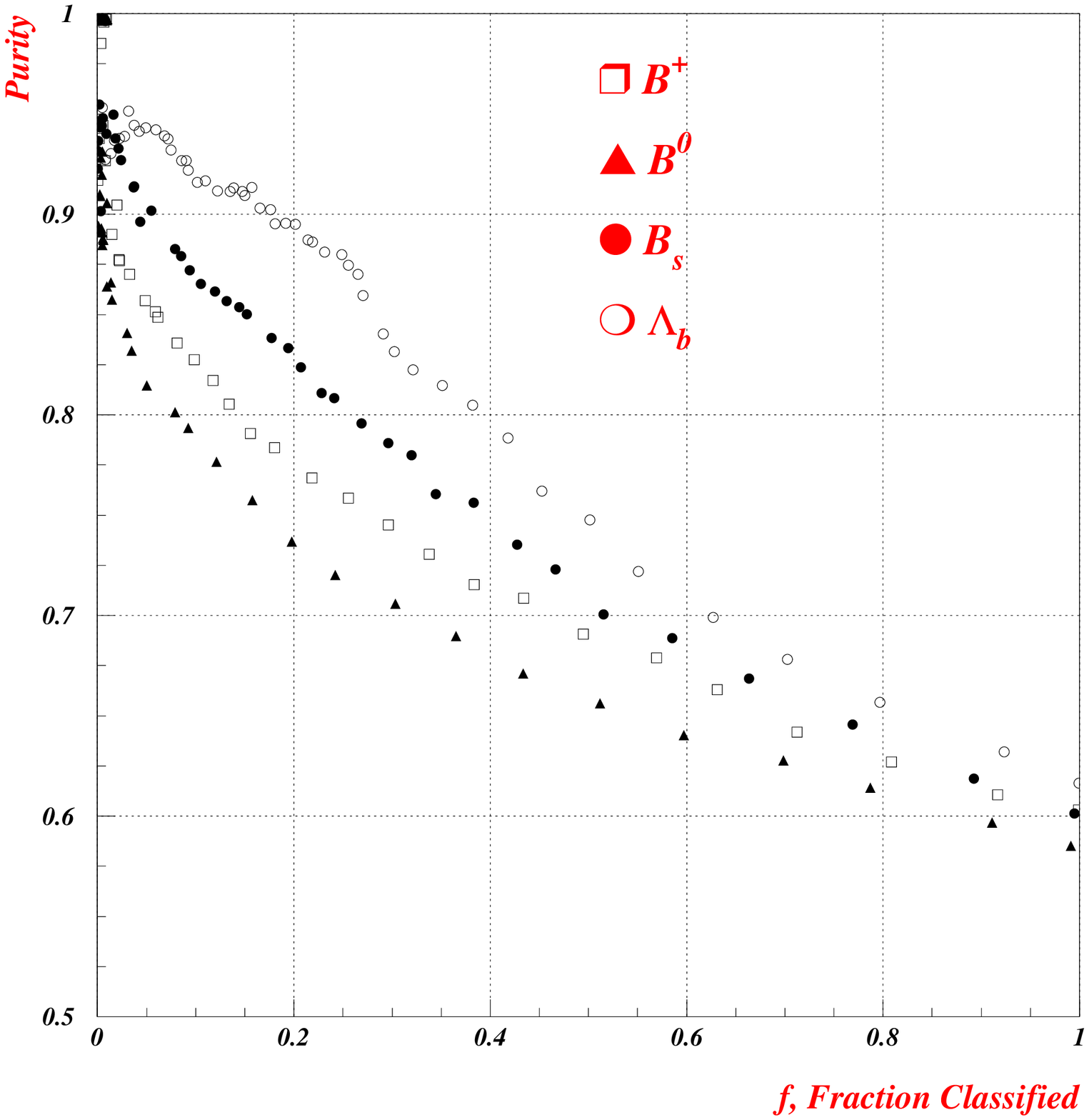,width=16.cm}
\caption[]
{\label{fig:hflprod_perf} The hemisphere-level fragmentation flavour
tag performance for all B-hadron types.
Selection cuts and weights have been applied as detailed in Appendix A.}
\end{center}  
\end{figure}

\begin{figure}[p]
\begin{center}
\epsfig{file=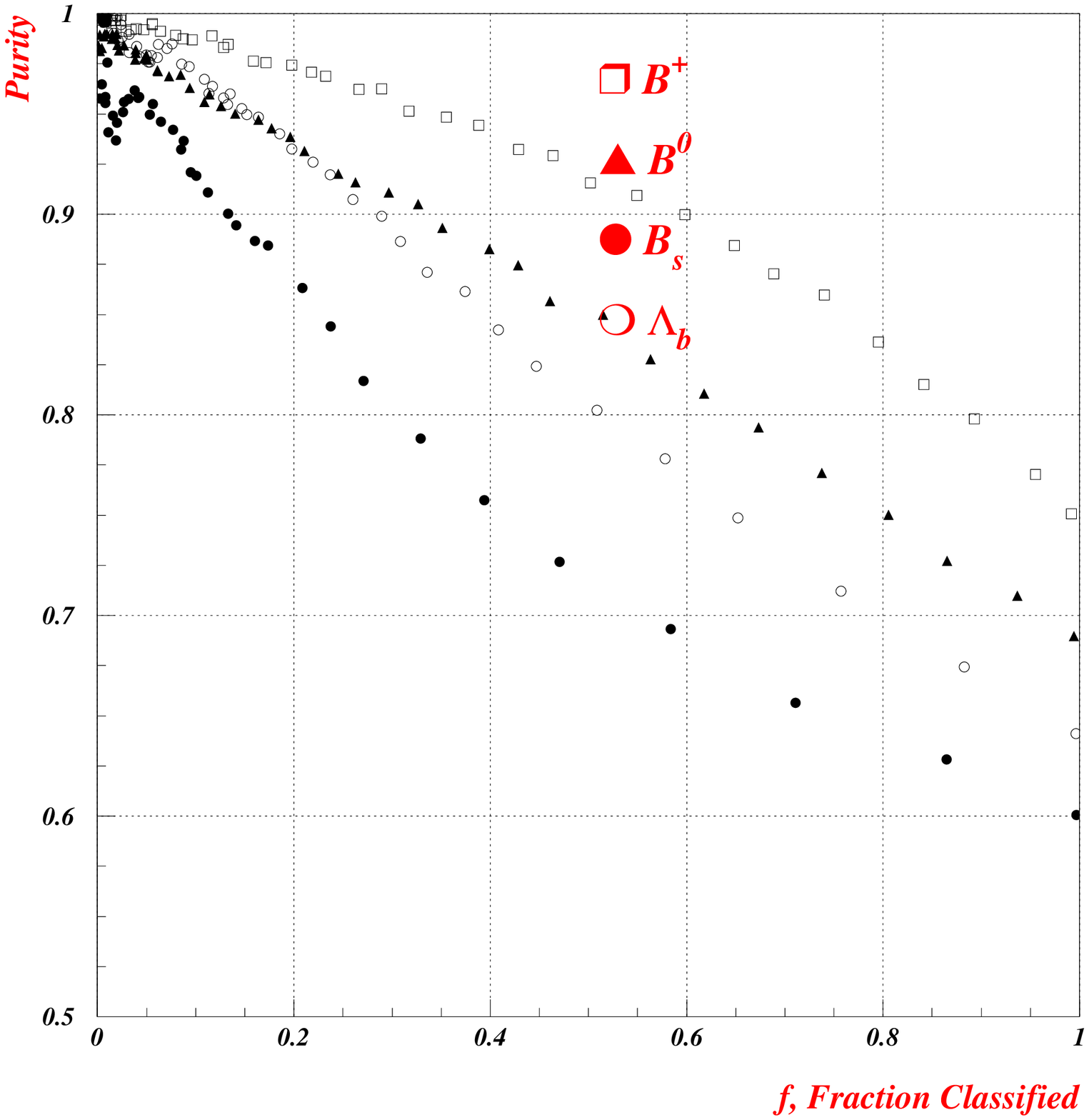,width=16.cm}
\caption[]
{\label{fig:hfldec_perf} The hemisphere-level decay flavour
tag performance for all B-hadron types.
Selection cuts and weights have been applied as detailed in Appendix A.}
\end{center}  
\end{figure}

The purity against efficiency performance of the tags are
shown in Figure~\ref{fig:hflprod_perf} for the fragmentation tag
and in Figure~\ref{fig:hfldec_perf} for the decay tag. 
The tags are located in the following BSAURUS variables;
\verb+BSHEM(IBH_FFLBS,IH)+, \verb+BSHEM(IBH_FFLB0,IH)+,
\verb+BSHEM(IBH_FFLBP,IH)+ and \verb+BSHEM(IBH_FFLLB,IH)+
for the production flavour outputs and in 
\verb+BSHEM(IBH_DFLBS,IH)+, \verb+BSHEM(IBH_DFLB0,IH)+,
\verb+BSHEM(IBH_DFLBP,IH)+ and \verb+BSHEM{IBH_DFLLB,IH)+
for the decay flavour outputs.

\subsection{Applications to Production Flavour Tagging}
\label{sec:prodflav}
\subsubsection{The Same Hemisphere Production flavour Network (SHPN)}
The hemisphere flavour tags described in Section~\ref{sec:trkflav} form 
the major input to a further dedicated network, the 
{\bf S}ame {\bf H}emisphere {\bf P}roduction flavour {\bf N}etwork or {\bf SHPN}.
This network attempts to find the optimal combination of fragmentation
and decay flavour, for each B-species hypothesis, in order to tag the B-hadron
quark flavour at production time.  
The network is constructed
independently of any information from the opposite hemisphere, primarily so
that the output can be used for a $A_{FB}(b\bar{b})$~measurement\cite{ref:afb}
incorporating double-hemisphere flavour tag methods.

The 9 input variables used are (ignoring some details
of variable transformation and re-scaling):
\begin{itemize}
\item $F(hem.)_{B_s}^{Frag.} \cdot P_{same}(B_s)$
\item $\left(F(hem.)_{B^+}^{Dec.}-F(hem.)_{B^+}^{Frag.}\right) \cdot P_{same}(B^+)$
\item $\left(F(hem.)_{bary.}^{Dec.}-F(hem.)_{bary.}^{Frag.}\right) \cdot P_{same}(bary.)$
\item $\left(F(hem.)_{B^0}^{Dec.}\cdot\left( 1-2\sin^2(0.237 \cdot \tau) \right)-
F(hem.)_{B^0}^{Frag.}\right)\cdot P_{same}(B^0)$,
where $\tau$~is the reconstructed B-lifetime. Note this construction
takes account of the $\Bzero$~oscillation frequency found in the simulation. 
This is not possible
for the case of $\Bs$~where the oscillations are so fast that we have essentially a
50-50 mix of $\Bs$~and $\Bsb$. 
\item The jet charge defined as,
\begin{equation} 
      Q_J=\frac{\sum |p_L|_i^{\kappa}Q_i}{\sum |p_L|_i^{\kappa}},
\end{equation}
      where the sum is over all tracks and $p_L$ is the longitudinal momentum
      component to the thrust axis. The optimal choice of the free parameter
      $\kappa$ depends on the type of B-hadron under consideration. For
      this application we choose a range of values
      $\kappa=0.3, 0.6, \infty $ where the last value corresponds to 
      taking the charge of the stiffest track in the hemisphere.     
\item Vertex charge  defined as in Section~\ref{sec:bspec}, equation~\ref{eqn-qv}.
\item Vertex charge significance.
\end{itemize}
Note that the $P_{same}(B^+,B^0,B_s,bary.)$~factors represent the outputs of
the B-species tagging network, SHBN, described in Section~\ref{sec:bspec}.

The treatment of quality variables is slightly different for this
network compared to other BSAURUS networks. In this case, in order
to ensure that the output is inherently symmetric with 
respect to opposite charges, the quality variables are 
used to weight the turn-on gradient (or `temperature')
of the sigmoid function used as the network node transfer function.
This restriction is especially important for analyses such as 
measurements of $A_{FB}(b\bar{b})$~which to first order implicitly
assume that the charge tag used is symmetric with respect to quark and
anti-quark. 
The quality variables used were: 
\begin{itemize}
\item The hemisphere quality word \verb+BSHEM(IBH_QUAL2,IH)+ described in 
Section~\ref{sec:qual}.
\item The hemisphere rapidity gap between the track of highest rapidity
below a TrackNet cut at 0.5 and that of smallest rapidity above
the cut at 0.5.
\item The error on the vertex charge measurement.
\item The ratio of the  B-energy, defined by the \verb+BSHEM(IBH_BTT,IH)+,
to the LEP beam energy. 
\end{itemize}

\begin{figure}[p]
\begin{center}
\mbox{\epsfig{file=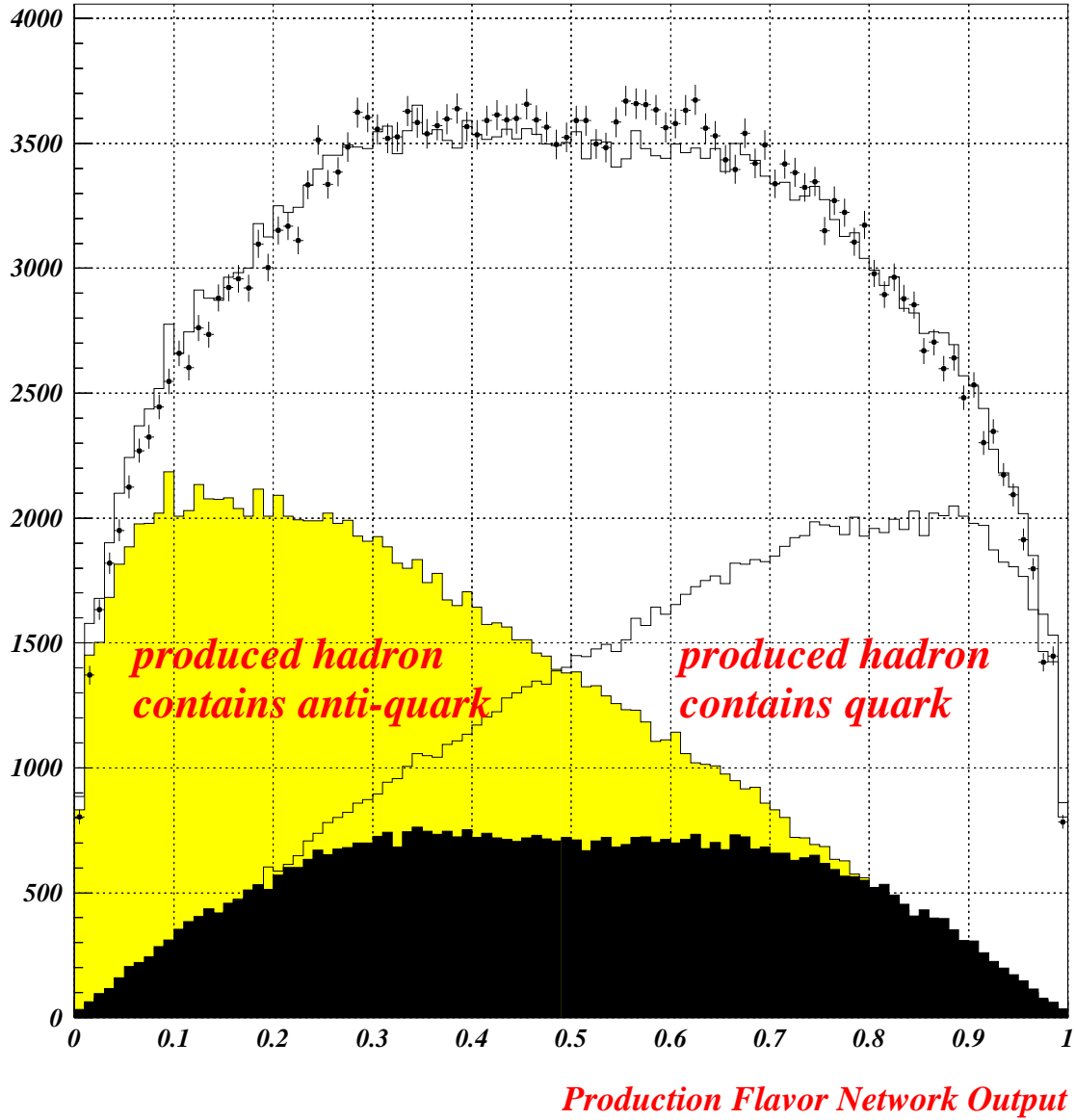,width=16.cm}}   
\caption[]
{\label{fig:prodflav} Output of the production flavour net in 
simulation compared to data. Overlaid are plots from the simulation showing the 
separation attained for the case where a produced B-hadron contains a b or 
$\bar{{\mathrm b}}$~quark.
The black region shows the distibution for the light and charm-quark background.
Selection cuts and weights have been applied as detailed in Appendix A.}
\end{center}
\end{figure}

\begin{figure}[p]
\begin{center}
\epsfig{file=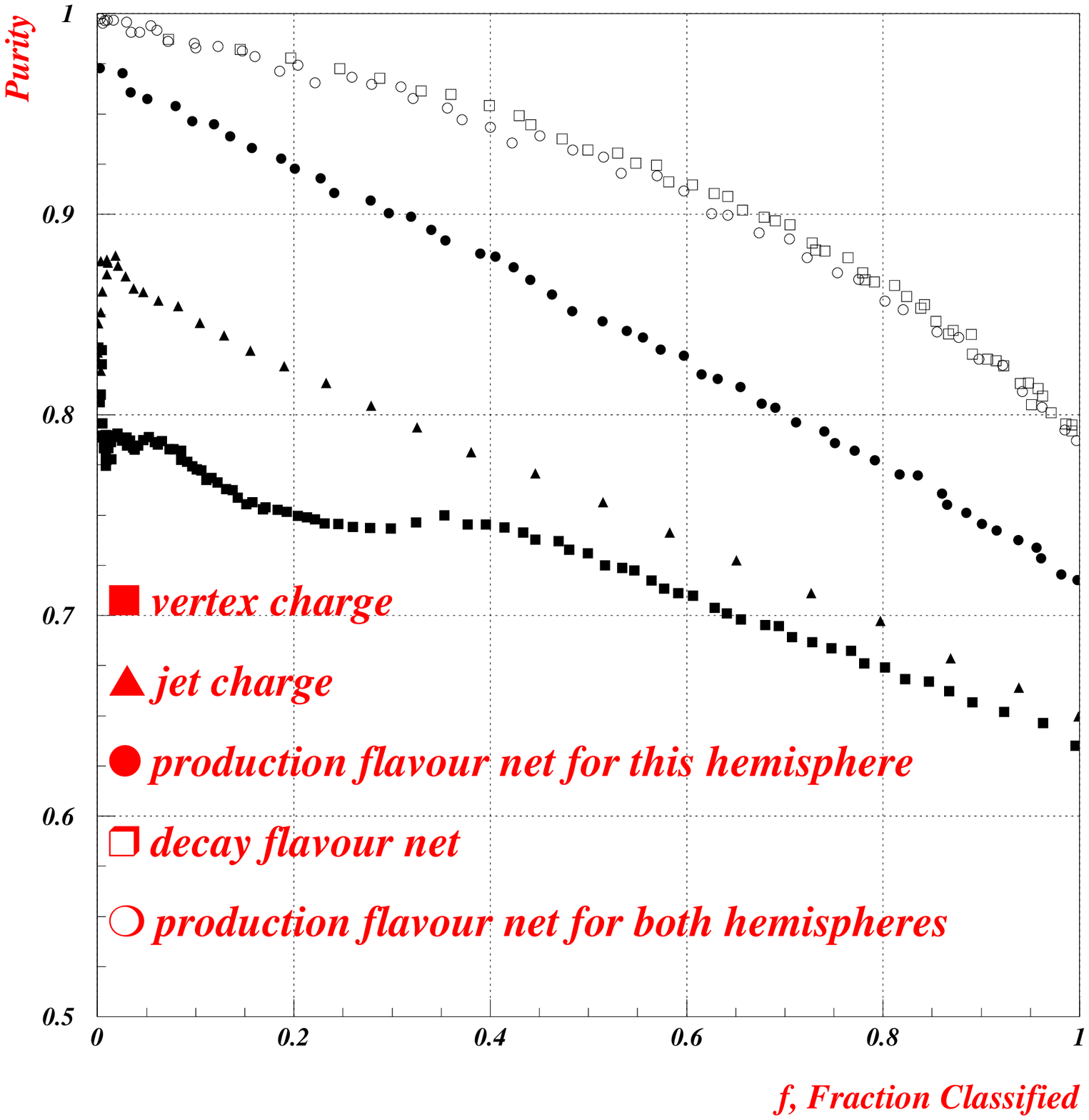,width=16.0cm}
\caption[]
{\label{fig:dec_prod_perf} Perfomance of the production and decay flavour
networks.
The comparison is made to using only the vertex charge or  
jet charge variable (with $\kappa=0.6$) as a production flavour tag.
Selection cuts and weights have been applied as detailed in Appendix A.}
\end{center}
\end{figure}

The SHPN output is stored in BSAURUS variable  
\verb+BSHEM(ISH_FLAV45,IH)+ \\ and is compared to the data in 
Figure~\ref{fig:prodflav}.
The performance of the tag can be seen
in Figure~\ref{fig:dec_prod_perf}(full circles).

\subsubsection{The Both Hemispheres Production flavour Network (BHPN)}
It should be noted that an improved performance for production flavour
tagging can be obtained by including the information from the opposite
side hemisphere. 
The
{\bf B}oth {\bf H}emispheres {\bf P}roduction flavour {\bf N}etwork or {\bf BHPN}
is defined as the ratio,
\begin{eqnarray}
\frac{{\mathrm SHPN}(same) \cdot (1- {\mathrm SHPN}(opp))}
{{\mathrm SHPN}(same) \cdot 
(1- {\mathrm SHPN}(opp))+(1- {\mathrm SHPN}(same)) \cdot {\mathrm SHPN}(opp) }
\end{eqnarray}
where the labels $same$~and $opp$~refer to the production tag
in this hemisphere and in the opposite hemisphere respectively. 
Note that if SHPN tags e.g. the charge of the quark, then 
$(1-{\mathrm SHPN})$ tags the charge of the anti-quark. 
The resulting performance of the BHPN is also shown in 
Figure~\ref{fig:dec_prod_perf}(open circles) and the output
can be found in BSAURUS output \verb+BSHEM(IBH_FLBOTH,IH)+.


\subsection{Applications to Decay Type and Decay Flavour Tagging}
\subsubsection{Both Hemispheres B-species enrichment Network (BHBN)}
\label{sec:BHBN}
The performance of the SHBN (Section~\ref{sec:bspec})
can be further improved by use of the hemisphere flavour tags developed
in Section~\ref{sec:flavtag} and by the introduction  
of charge correlation information from the opposite
hemisphere to form the 
{\bf B}oth {\bf H}emispheres {\bf B}-species enrichment {\bf N}etwork or {\bf BHBN}.

In the same way as for the SHBN, the BHBN is a network with four output target nodes,
one for each B-type, 16 input nodes and two hidden layers consisting of
20 and 10 nodes respectively.
The input variables used, ignoring some details of variable scaling
and transformation, are as follows:
We define the variable, $FL_{OPP}$, which is proportional to the 
production flavour tag SHPN(see Section~\ref{sec:prodflav})
in the opposite hemisphere, but scaled by a 
function in $1-\cos(\theta)$~in an attempt to explicitly account for 
the b-quark forward-backward asymmetry,
\begin{eqnarray*}
FL_{OPP}=\ln \left( SHPN(opp)/(1-SHPN(opp)) \right) - 
\\ \ln \left((0.5-0.12 \cdot \cos(\theta))/(0.5+0.12 \cdot \cos(\theta)) \right)
\end{eqnarray*}
The discriminating variables used are:
\begin{itemize}
\item SHBN probability for $\Bplus$, $P_{same}(B^+)$.
\item SHBN probability for $\Bzero$, $P_{same}(B^0)$.
\item SHBN probability for $\Bs$, $P_{same}(B_s)$.
\item SHBN probability for B-baryon, $P_{same}(bary.)$.
\item $F(hem.)_{B^+}^{Dec.} \cdot (FL_{OPP}-F(hem.)_{B^+}^{Frag.})$
\item $FL_{OPP} \cdot F(hem.)_{B_s}^{Prod}$
\item $F(hem.)_{B_s}^{Dec.} \cdot (FL_{OPP}-F(hem.)_{B_s}^{Frag.})$
\item $F(hem.)_{B^0}^{Dec.} \cdot (FL_{OPP}-F(hem.)_{B^0}^{Frag.})$
\item $F(hem.)_{bary.}^{Dec.} \cdot (FL_{OPP}-F(hem.)_{bary.}^{Frag.})$
\item $F(hem.)_{B^+}^{Dec.} \cdot (FL_{OPP}-F(hem.)_{B^+}^{Frag.})\cdot P_{same}(B^+)$
\item $F(hem.)_{B_s}^{Dec.} \cdot (FL_{OPP}-F(hem.)_{B_s}^{Frag.})\cdot P_{same}(B_s)$
\item $F(hem.)_{B^0}^{Dec.} \cdot (FL_{OPP}-F(hem.)_{B^0}^{Frag.})\cdot P_{same}(B^0)$
\item $F(hem.)_{bary.}^{Dec.} \cdot (FL_{OPP}-F(hem.)_{bary.}^{Frag.})\cdot P_{same}(bary.)$
\end{itemize}
Where the $F(hem.)$~terms are the hemisphere flavour likelihood ratios described in 
Section~\ref{sec:trkflav}. 
The quality variables used are: 
\begin{itemize}
\item The hemisphere quality word \verb+BSHEM(IBH_QUAL2,IH)+ described in 
Section~\ref{sec:qual}.
\item The hemisphere rapidity gap between the track of highest rapidity
below a TrackNet cut at 0.5 and that of smallest rapidity above
the cut at 0.5.
\item The ratio of the B-energy, defined by the \verb+BSHEM(IBH_BTT,IH)+, 
to the LEP beam energy. 
\end{itemize}

\begin{figure}[p]
\begin{center}
\epsfig{file=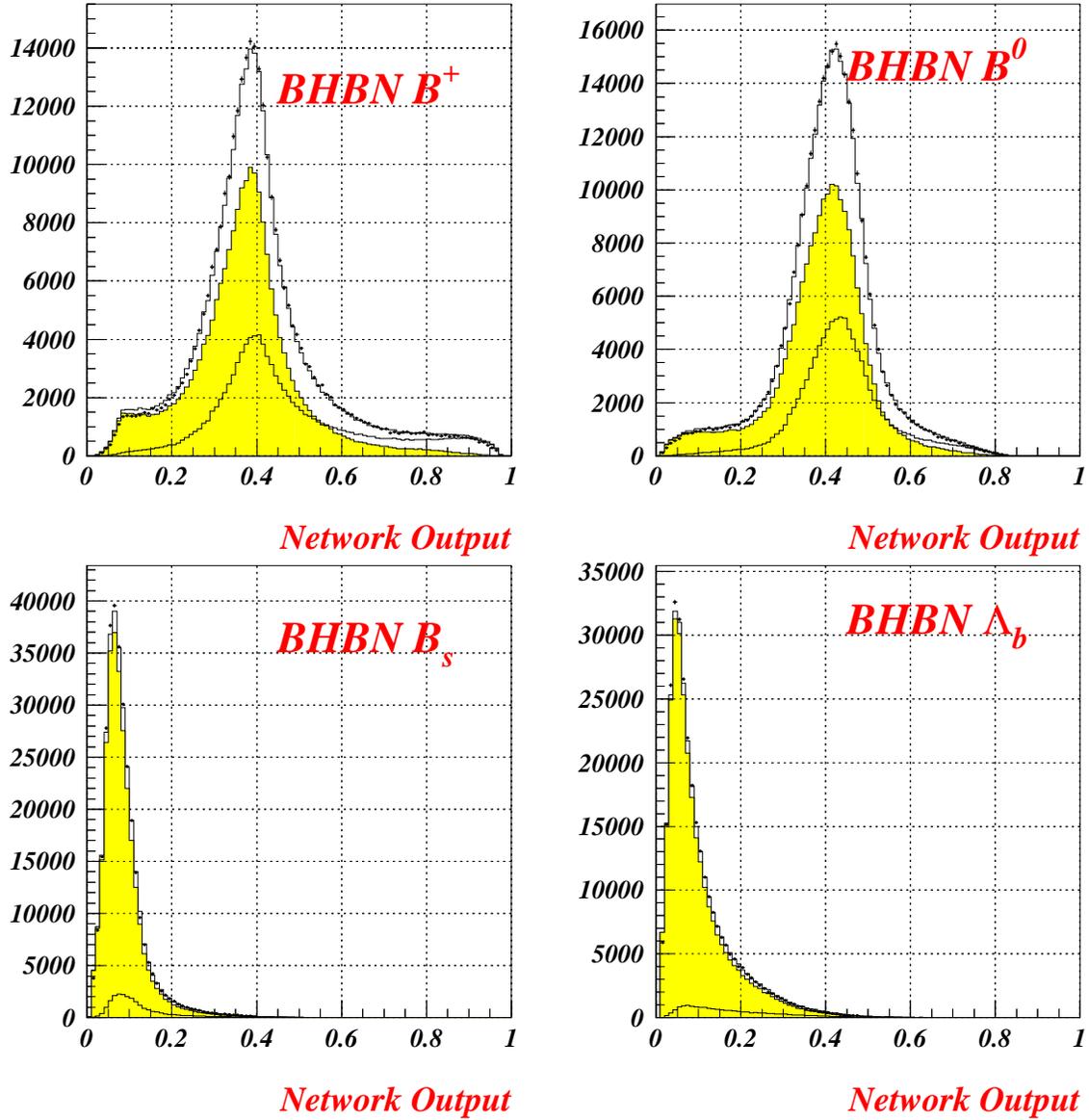,width=16.cm}
\caption[]
{\label{fig:BHBN_out} 
Output of the BHBN network for the $\Bplus,\Bzero,\Bs$~and B-baryon
hypotheses compared to the data (closed points). The two overlaid
curves show the distribution for the hypothesis being considered 
i.e. the `signal' (open histogram), compared
to the distribution for everything else 
i.e. the `background' (shaded histogram).
Selection cuts and weights have been applied as detailed in Appendix A.}
\end{center} 
\end{figure}

\begin{figure}[p]
\begin{center}
\epsfig{file=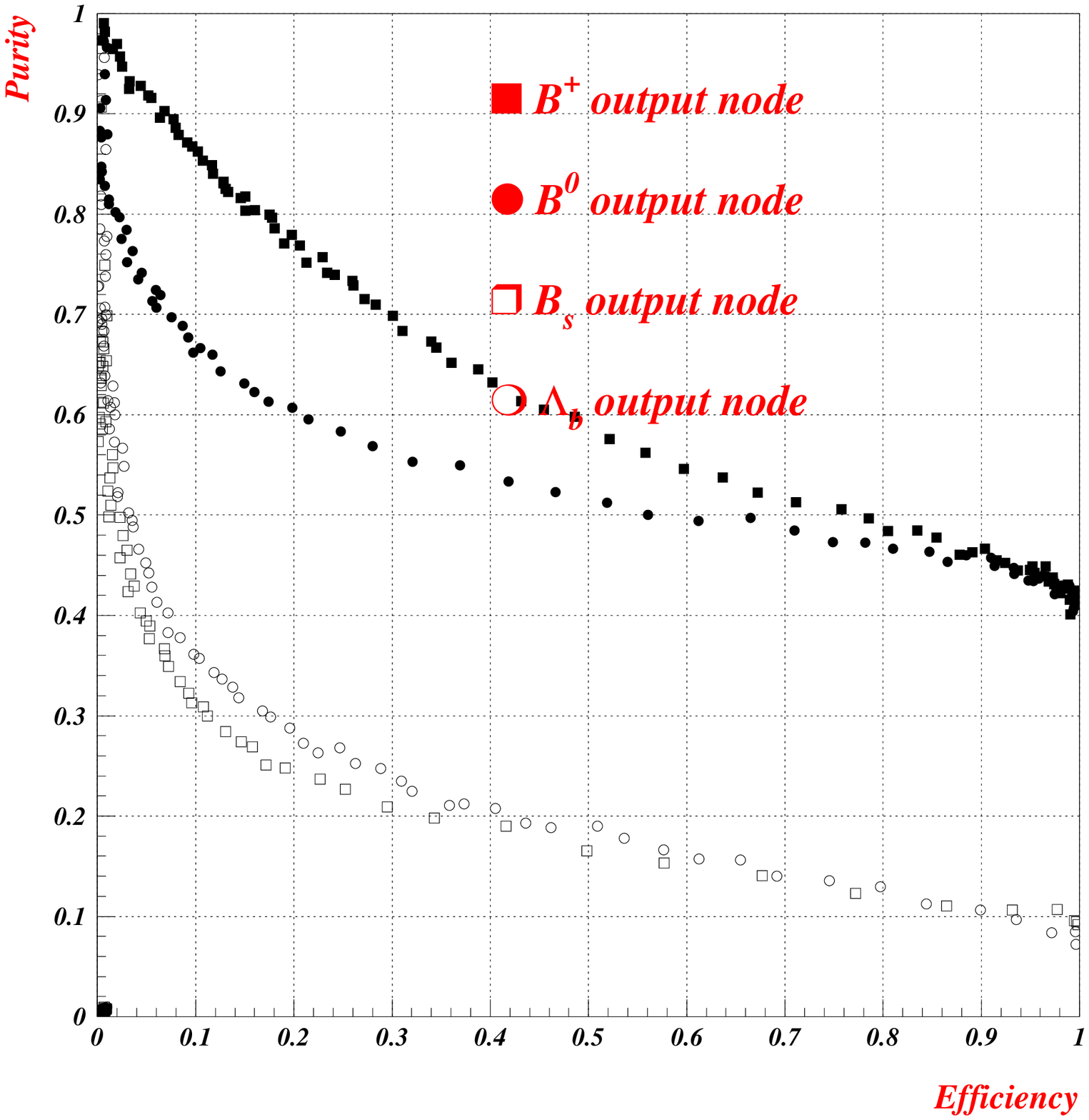,width=16.cm}
\caption[]
{\label{fig:BHBN_perf} Performance of the BHBN for enriching samples in the
various B-hadron types. The plot is based on $\bb$~Monte Carlo
making successive cuts in 
the network output
and where `signal' and `background' are defined as in Figure~\ref{fig:BHBN_out}.
Selection cuts and weights have been applied as detailed in Appendix A.}
\end{center} 
\end{figure}

The outputs of the BHBN are plotted in Figure~\ref{fig:BHBN_out} and can be found 
in BSAURUS variables, 
\verb+BSHEM(IBH_PRBAIBP,IH)+, \verb+BSHEM(IBH_PRBAIB0,IH)+,
\verb+BSHEM(IBH_PRBAIBS,IH)+ and \verb+BSHEM(IBH_PRBAILB,IH)+.
The resulting performance of the network is shown in figure~\ref{fig:BHBN_perf}.

\subsubsection{The Both Hemispheres Decay Flavour Network(BHDN)}
\label{sec:decflav}
The
{\bf B}oth {\bf H}emispheres {\bf D}ecay flavour {\bf N}etwork or BHDN, 
utilises similar constructions for the
discriminating variables as in the SHPN, the main
difference being that opposite side hemisphere information is explicitly
used in order to boost the performance.

We define the variable, $FL_{OPP}$, which is proportional to the 
production flavour tag in the opposite hemisphere, $FL$, but scaled by a 
function in $1-\cos(\theta)$~in an attempt to explicitly account for 
the b-quark forward-backward asymmetry,
\begin{eqnarray}
FL_{OPP}=\ln \left( SHPN(opp)/(1-SHPN(opp)) \right) - \ln \left(1-0.4 \cdot \cos(\theta) \right)
\end{eqnarray}
The discriminating variables used are:
\begin{itemize}
\item $F(hem.)_{B_s}^{Dec.}\cdot P_{both}(B_s)$
\item $\left(F(hem.)_{B^+}^{Dec.}+FL_{OPP}-F(hem.)_{B^+}^{Frag.}\right)\cdot P_{both}(B^+)$
\item $\left(F(hem.)_{bary}^{Dec.}+FL_{OPP}-F(hem.)_{bary}^{Frag.}\right)\cdot P_{both}(bary)$
\item $\left( F(hem.)_{B^0}^{Dec.}+(FL_{OPP}-F(hem.)_{B^0}^{Frag.})
\cdot \left( 1-2\sin^2(0.237*\tau)\right) \right)\cdot P_{both}(B^0)$,
where $\tau$~is the reconstructed B-lifetime.
\end{itemize}
Note that the $P_{both}(B_s,B^+,bary,B^0)$~factors represent the outputs of
the B-species tagging network, BHBN which
includes information from the opposite side hemisphere,
and is described in Section~\ref{sec:bspec}.
The quality variables used were: 
\begin{itemize}
\item Hemisphere quality (word \verb+BSHEM(IBH_QUAL2,IH)+ described in 
Section~\ref{sec:qual}).
\item The hemisphere rapidity gap between the track of highest rapidity
below a TrackNet cut at 0.5 and that of smallest rapidity above
the cut at 0.5.
\item $FL_{OPP}$.
\item  $1-2\sin^2(0.237*\tau)$.
\item \verb+BSHEM(IBH_FLAV45,IH)+,  see Section~\ref{sec:prodflav}.
\end{itemize}

The output of the decay flavour net is stored in BSAURUS word \\
\verb+BSHEM(ISH_DFLAV45,IH)+ 
and the performance of the tag is also shown in 
Figure~\ref{fig:dec_prod_perf}(open squares).

BSAURUS routine, \verb+DECFLA+, gives the user the option to 
recalculate the decay flavour network output based only on
a list of tracks that he/she supplies. This can be useful in analyses
that need to avoid potential biases in the flavour tag.

           
\section*{Summary and Outlook}
The BSAURUS package reconstructs properties of B-hadron decays inside
the jets of $\ztobb$~events in an intrinsically inclusive way 
in order to ensure high efficiency. Exploiting wherever possible, 
physics knowledge of b-quark production, the hadronisation process
and of the subsequent B decay, this information is combined 
to tag quantities of importance for B-physics analyses. In order to
combine information in an optimal way, i.e. to ensure high purity,
neural network techniques are used extensively. 

This note has described how this approach is
applied to DELPHI data in order to form, most importantly, variables
that tag the presence of a $\Bplus,\Bzero,\Bs$~or B-baryon and 
variables that tag the charge or flavour of the b-quark in any
of the B-hadron types at production and decay time separately.
These tags are fundamental inputs to B-physics analyses such as
B-spectroscopy\cite{ref:bspec}, B-species lifetime measurements\cite{ref:blife}, 
${\mathrm B}^0_d-{\mathrm \bar{B}}^0_d$~and ${\mathrm B}^0_s-{\mathrm \bar{B}}^0_s$
oscillations\cite{ref:bsosc}, 
the forward-backward asymmetry in $\bb$~events\cite{ref:afb},
the measurement of B-hadron production fractions\cite{ref:fb} and 
measurements of B-hadron branching ratios\cite{ref:btoddx}. 
Analyses in all these areas
have either already been completed or are currently underway in DELPHI using
the BSAURUS framework. In addition, the list is by no means exhaustive
and new analyses are currently being planned e.g. measuring the 
fragmentation function of B-species and studies of radially excited
B-meson states.

It should be noted that the package is under constant and on-going 
development. New features and improvements are still being added
and this is likely to continue in the future. All news and new
features relating to BSAURUS are documented at the web-site
location:\\
\verb+http://pubxx.home.cern.ch/pubxx/tasks/bcteam/www/inclusive/bsaurus.html.+

\section*{Acknowlegements}
The authors would like to acknowlege the contributions of 
Oliver Podobrin and Christof Kreuter, who were instrumental in developing the
origional version of BSAURUS. 
In addition, Christian Weiser has  made some valuable contributions.
We would also like to thank Wilbur Venus for his detailed reading
of the manuscript.
\section*{Appendix A: Cuts and Weights Applied to BSAURUS performance Plots}
Where indicated, the following cuts have been applied 
to the event sample entering the distribution shown:
\begin{itemize}
\item BSAURUS word, \verb+ISAURUS(IH)+, for hemisphere
number \verb+IH+ must have value 4 or higher
indicating that the secondary vertex fit in this
hemisphere converged normally. 
\item Combined event b-tag$>-0.5$.
\item 2-jet events only.
\item $\left| \cos(\theta_{THRUST}) \right|<0.75$.
\end{itemize}
In addition, to account for known differences between
data and Monte Carlo arising from both detector effects
and physics modelling, the simulation has been 
weighted (via BSAURUS routine \verb+SIMWAIT+, 
stored in output \verb+BSHEM(IBH_WEIGHT,IH)+) to 
account for:
\begin{itemize}
\item The differences between data and Monte Carlo in the
hemisphere quality variable \verb+BSHEM(IBH_QUAL,IH)+
as a function of the hemisphere track multiplicity,
\verb+BSHEM(IBH_NPGOOD,IH)+.
\item Latest world average B-species production fractions.
\item Latest world average B-species lifetime measurements.
\item MARKIII D-topological branching fractions.
\item Any remaining difference between data and Monte Carlo in the
B-energy estimate i.e. variable \verb-BSHEM(IBH_BTT+3,IH)-.
\end{itemize}
Note that the current measurements
of B-physics quantities used to form weights, are
chosen to follow LEP Working Group recommendations. 

\begin{figure}[p]
\begin{center}
{\epsfig{file=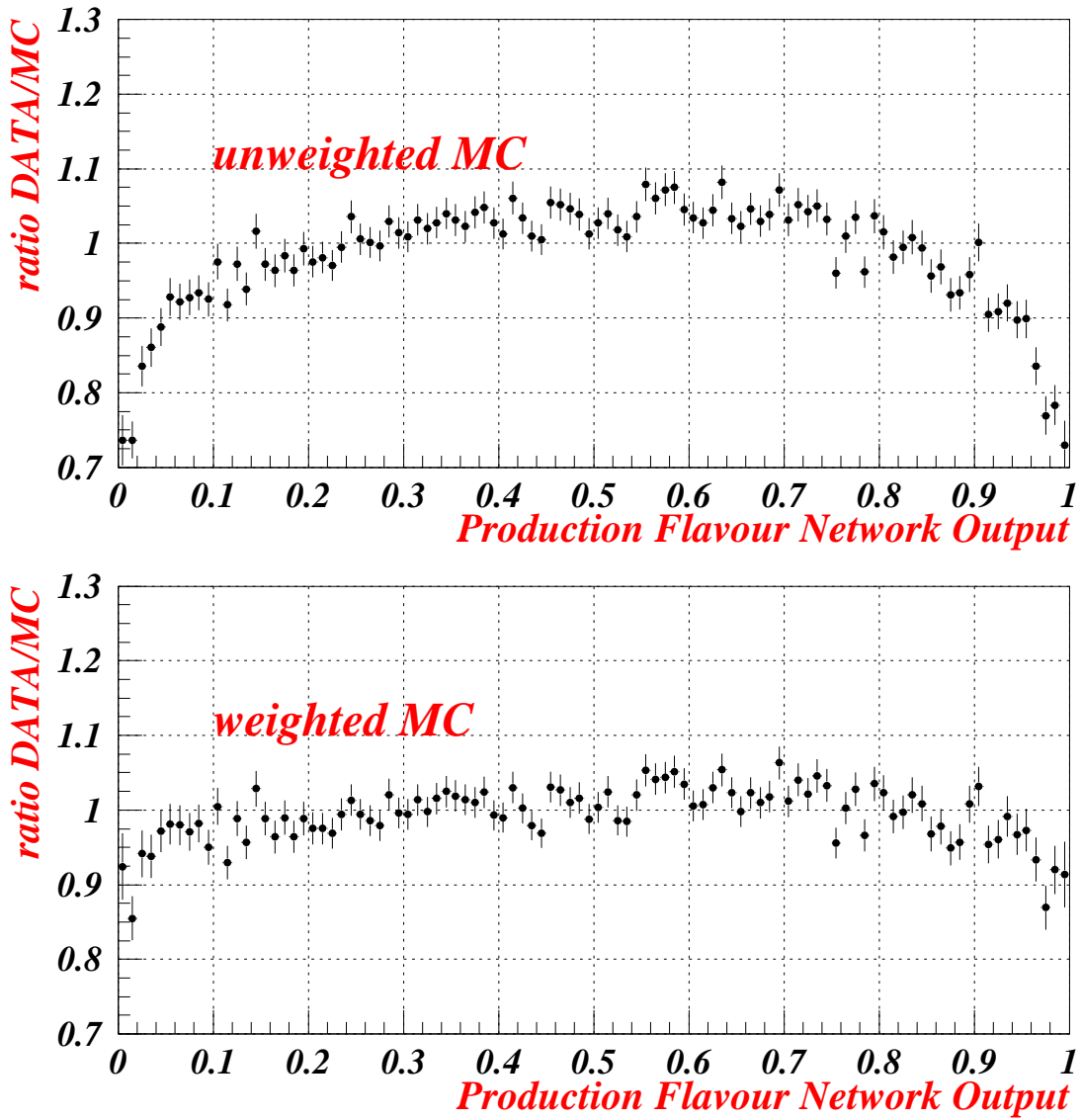,width=16.cm}}   
\caption[]
{\label{fig:weight} The distributions show the ratio of data/simulation 
for the output of the 
production flavour network before(upper plot) and after(lower plot) applying
the weighting procedure of Appendix A.}
\end{center}
\end{figure}

Figure~\ref{fig:weight} shows the effect, for example,
on the production
flavour network output of applying the weight returned by
routine \verb+SIMWAIT+. The much improved agreement between
data and simulation after the weight is applied illustrates
the need for such a correction if the network outputs are
used in analyses to estimate e.g. absolute efficiencies.

\section*{Appendix B: How to Access BSAURUS Outputs}
A call to routine \verb+ BSAURUS+ should be made once per
event from the user routine. A list of the BSAURUS outputs
at track, hemisphere and event level is given in Appendix C.

BSAURUS uses a modified version of the VECSUB package, called
VECTOP, to fill the standard track information array 
\verb+P(NDIM,*)+. 
Of the {\bf utmost importance} to users running the standard VECSUB
package e.g. SKELANA users, is the track output
word, \verb+BSPAR(IBP_IYB,IT)+. 
This gives the matching VECSUB track index for BSAURUS 
(ie VECTOP) track IT. Users failing to do this track matching
will certainly obtain the wrong results from BSAURUS.

The BSAURUS-cards file is released periodically to the
standard DELPHI areas for library creation. A development
version is to be found at:\\
\verb+/afs/cern.ch/user/b/barker/public/delphi/bsaurus.car+.\\
Naturally, the results of using this version cannot
be guaranteed.

\section*{Appendix C: BSAURUS output array definitions}
Quantities calculated in BSAURUS along with other variables, e.g.
truth Monte Carlo information, that are useful for analyses are available
via {\tt COMMON /CSAURUS/} at the event, hemisphere and track level.

1) The event output array:
\begin{verbatim}
*     Array Element     Number of           Description
*                        values     
*    -----------------------------------------------------------------
*     BSEVT(IBE_BTAG)      1     event B-tag
*     BSEVT(IBE_BCONF)     1     event B-confidence (Bill Murray)
*     BSEVT(IBE_BCTAG)     1     event B-tag with charm suppression
*     BSEVT(IBE_BCOMB)     1     event aabtag combined-tag 
*     BSEVT(IBE_PV)        3     primary vertex
*     BSEVT(IBE_PC)        6     primary covariance matrix
*     BSEVT(IBE_NJET)      1     number of jets 
*     BSEVT(IBE_COST)      1     COS(theta) of event thrust axis
*     BSEVT(IBE_RICH)      1     Barrel RICH status: 3=gas+liq.,2=liq. only
*                                ,1=gas only, 0=no liq. or gas 
*     BSEVT(IBEMC_FLAV)    1     MC primary quark flavour (e.g. 12=b)
*     BSEVT(IBEMC_PV)      3     MC primary vertex
\end{verbatim}

2) The hemisphere output array where IHEM is the hemisphere number (1 or 2):
\begin{verbatim}
*     Array Element     Number of           Description
*                        values     
*     -----------------------------------------------------------------
*     BSHEM(IBH_HNUM,IHEM)     1   hemisphere number
*     BSHEM(IBH_BTAG,IHEM)     1   hemisphere B-tag
*     BSHEM(IBH_BCONF,IHEM)    1   hemisphere B-confidence (Bill Murray)
*     BSHEM(IBH_BCTAG,IHEM)    1   hemisphere B-tag with charm suppression
*     BSHEM(IBH_BCOMB,IHEM)    1   hemisphere aabtag combined tag
*     BSHEM(IBH_BJET,IHEM)     4   hemisphere B-candidate jet 4-vector
*     BSHEM(IBH_NJET,IHEM)     1   number jets in hemisphere
 
*     BSHEM(IBH_BF,IHEM)       4   B 4-vector from fit
*     BSHEM(IBH_BT,IHEM)       4   B 4-vector from fit + energy correction
*     BSHEM(IBH_BY,IHEM)       4   B RAW 4-vector: from rapidity algorithm
*                                for 3-jet, net weighted for 2-jet
*     BSHEM(IBH_BTT,IHEM)      4   'optimal' B 4-vector
*     BSHEM(IBH_SV,IHEM)       3   secondary vertex coordinates
*     BSHEM(IBH_SC,IHEM)       6   secondary vertex covariance matrix
*     BSHEM(IBH_SIGPHI,IHEM)   1   Error on direction phi from SV fit
*     BSHEM(IBH_SIGTHET,IHEM)  1   Error on direction theta from SV fit 
*     BSHEM(IBH_DEC,IHEM)      1   B decay length
*     BSHEM(IBH_DECS,IHEM)     1   B decay length error
*     BSHEM(IBH_PROB,IHEM)     1   secondary vertex fit probability
*     BSHEM(IBH_RET,IHEM)      1   ISAURUS - fit return code
*     BSHEM(IBH_YBFIT,IHEM)    1   YBFIT2  - secondary vertex fit return code
*     BSHEM(IBH_NACC,IHEM)     1   no. of particles at sec. vertex
*     BSHEM(IBH_QJET,IHEM)     1   jet charge, kappa=0.6
*     BSHEM(IBH_QJET2,IHEM)    1   jet charge, kappa=0.3
*     BSHEM(IBH_QJET3,IHEM)    1   jet charge, stiffest track in P_L
*     BSHEM(IBH_QVER,IHEM)     1   Vertex charge NN
*     BSHEM(IBH_QVERS,IHEM)    1   Vertex charge NN error
*     BSHEM(IBH_RAWQ,IHEM)     1   Vertex charge (sum trk. charge) 
*     BSHEM(IBH_FLAV,IHEM)     1   Original b flavour NN
*     BSHEM(IBH_FLAV45,IHEM)   1   improved b flavour NN (<0.5->b, >0.5->bbar)
*     BSHEM(IBH_FLBOTH,IHEM)   1   likelihood combination with opp. side info.  
*     BSHEM(IBH_MY,IHEM)       1   rapidity mass
*     BSHEM(IBH_SHEM,IHEM)     1   scaled measured hemisphere energy 
*     BSHEM(IBH_SHEM2,IHEM)    1   scaled measured hemisphere energy (2-BODY) 
*     BSHEM(IBH_QUAL,IHEM)     1   hemisphere quality flag:
*                                  N(rejected trk cuts)+100*N(interactions)+
*                                  1,000N(ambiguous hits)+100,000*N(ISRT=0)
*     BSHEM(IBH_QUAL2,IHEM)    1   Continous hemisphere quality 0-(perfect)10(bad)  
*     BSHEM(IBH_MVER,IHEM)     1   Mass on secondary vertex
*     BSHEM(IBH_BPLUS,IHEM)    1   B+ tagging
*     BSHEM(IBH_BZERO,IHEM)    1   B0 tagging
*     BSHEM(IBH_PRPF,IHEM)     1   prim. vertex fragmentation Borisov prob 
*     BSHEM(IBH_PRPB,IHEM)     1   prim. vertex B Borisov prob 
*     BSHEM(IBH_PRSF,IHEM)     1   sec.  vertex fragmentation Borisov prob 
*     BSHEM(IBH_PRSB,IHEM)     1   sec.  vertex B Borisov prob 
*     BSHEM(IBH_DIPOL,IHEM)    1   sec.  vertex dipole moment
*     BSHEM(IBH_HKST,IHEM)     1   Q*PMIN of tracks in B CMS
*
*     BSHEM(IBH_NPALL,IHEM)    1   # of particles 
*     BSHEM(IBH_NPGOOD,IHEM)   1   # of particles past quality cuts 
*     BSHEM(IBH_NPCUT,IHEM)    1   # particles past IBP_NET > PNET_CUT
*     BSHEM(IBH_NPGC,IHEM)     1   # past IBP_NET>PNET_CUT + qual. requirement
*     BSHEM(IBH_NKS,IHEM)      1   # recon. K_s with y>1.6
*     BSHEM(IBH_NWKSD,IHEM)    1   sum of y-prob. funct for decay K_s 
*     BSHEM(IBH_NWKSF,IHEM)    1   sum of y-prob. funct for frag. K_s
*     BSHEM(IBH_NLAM0,IHEM)    1   # recon. Lambda(uds) with y>1.6
*     BSHEM(IBH_NWLAMBD,IHEM)  1   sum of y-prob. funct for decay lambdas 
*     BSHEM(IBH_NWLAMBF,IHEM)  1   sum of y-prob. functions for frag.lambdas
*     BSHEM(IBH_NNEUT,IHEM)    1   # recon. neutrons with y>1.6
*     BSHEM(IBH_NPION,IHEM)    1   # charged pions
*     BSHEM(IBH_PICHRG,IHEM)   1   # pions charged weighted about SV
*     BSHEM(IBH_NKDECAY,IHEM)  1   Sum of  kaon weights for trknet> 0.5
*     BSHEM(IBH_NKFRAG,IHEM)   1   Sum of  kaon weights for < 0.5
*     BSHEM(IBH_NWKD,IHEM)     1   sum of y-prob. funct for decay K+/- 
*     BSHEM(IBH_NWKF,IHEM)     1   sum of y-prob. functions for frag. K+/-
*     BSHEM(IBH_NPDECAY,IHEM)  1   Sum of proton weights for trknet > 0.5
*     BSHEM(IBH_NPFRAG,IHEM)   1   Sum of proton weights for trknet < 0.5
*     BSHEM(IBH_NWPROTD,IHEM)  1   sum of y-prob. functions for decay protons
*     BSHEM(IBH_NWPROTF,IHEM)  1   sum of y-prob. functions for frag. protons
*     BSHEM(IBH_QKAMAX,IHEM)   1   MAX. q*trknet in hemisphere for tight K.    
*     BSHEM(IBH_PKAON,IHEM)    1   charge correl. coeff. best kaon cand. in hem
*     BSHEM(IBH_QKAON,IHEM)    1   charge of best kaon cand. in hemisphere  
*     BSHEM(IBH_PLEP,IHEM)     1   charge correlation coeff. of best lep 
*     BSHEM(IBH_QLEP,IHEM)     1   charge of best lepton cand. in hemisphere
*
*     BSHEM(IBH_DEC3D,IHEM)    1   3 D decay length 
*     BSHEM(IBH_DEC3DS,IHEM)   1   error on 3 D decay length
*     BSHEM(IBH_BLIFE,IHEM)    1   B lifetime 
*     BSHEM(IBH_BLIFES,IHEM)   1   error on B lifetime 
*     BSHEM(IBH_CHRDIP,IHEM)   1   B-D charge dipole moment from GBDIPOLE
*     BSHEM(IBH_SVVAR1,IHEM)   1   Mean deviation of trks about SV point 
*     BSHEM(IBH_SVVAR2,IHEM)   1   Mean square deviation
*     BSHEM(IBH_DIPPY,IHEM)    1   A dipole moment from the SVVAR-calc
*
*     BSHEM(IBH_QRANK1,IHEM)   1   Charge of leading frag. track
*     BSHEM(IBH_QRANK2,IHEM)   1   Charge of 2ND rank frag track
*     BSHEM(IBH_QRANK3,IHEM)   1   Charge of 3RD rank frag track
*     BSHEM(IBH_YRANK1,IHEM)   1   Rapidity of leading frag track
*     BSHEM(IBH_YRANK2,IHEM)   1   Rapidity of 2ND rank frag track
*     BSHEM(IBH_YRANK3,IHEM)   1   Rapidity of 3RD rank frag track
*     BSHEM(IBH_KRANK1,IHEM)   1   Kaon net for leading rank frag track
*     BSHEM(IBH_KRANK2,IHEM)   1   Kaon net for 2ND rank frag track
*     BSHEM(IBH_KRANK3,IHEM)   1   Kaon net for 3RD rank frag track
*     BSHEM(IBH_PRANK1,IHEM)   1   Proton net for leading rank frag track
*     BSHEM(IBH_PRANK2,IHEM)   1   Proton net for 2ND rank frag track
*     BSHEM(IBH_PRANK3,IHEM)   1   Proton net for 3RD rank frag track
*     BSHEM(IBH_DELTA,IHEM)    1   delta - y-gap above/below trknet=0.5
*     BSHEM(IBH_QFSUM,IHEM)    1   Sum of frag. track charge    
*     BSHEM(IBH_DFLBS,IHEM)   1   decay flavour net for B_s 
*     BSHEM(IBH_DFLB0,IHEM)   1   decay flavour net for B0 
*     BSHEM(IBH_DFLBP,IHEM)   1   decay flavour net for B+ 
*     BSHEM(IBH_DFLLB,IHEM)   1   decay flavour net for LamB 
*     BSHEM(IBH_FFLBS,IHEM)   1   fragmentation flavour net for B_s 
*     BSHEM(IBH_FFLB0,IHEM)   1   fragmentation flavour net for B0 
*     BSHEM(IBH_FFLBP,IHEM)   1   fragmentation flavour net for B+ 
*     BSHEM(IBH_FFLLB,IHEM)   1   fragmentation flavour net for LamB 
*     BSHEM(IBH_PRBBS,IHEM)   1   probability for B_s  from all-flavour net
*     BSHEM(IBH_PRBB0,IHEM)   1   probability for B0   from all-flavour net
*     BSHEM(IBH_PRBBP,IHEM)   1   probability for B+   from all-flavour net
*     BSHEM(IBH_PRBLB,IHEM)   1   probability for LamB from all-flavour net
*     BSHEM(IBH_PRBAIBS,IHEM) 1   B_s  prob. from all-flavour net inc. opp. hem. info.
*     BSHEM(IBH_PRBAIB0,IHEM) 1   B0   prob. from all-flavour net inc. opp. hem. info.
*     BSHEM(IBH_PRBAIBP,IHEM) 1   B+   prob. from all-flavour net inc. opp. hem. info.
*     BSHEM(IBH_PRBAILB,IHEM) 1   LamB prob. from all-flavour net inc. opp. hem. info.
*     BSHEM(IBH_DFLAV45,IHEM) 1   Optimal decay flavour net
*     BSHEM(IBH_LIKBS,IHEM)   1   Likelihood ratio prod. flav. tag for B_s
*     BSHEM(IBH_LIKB0,IHEM)   1   Likelihood ratio prod. flav. tag for B_d
*
*     BSHEM(IBH_WEIGHT,IHEM)  1   SIMWAIT weight for physics/quality differences
*                                 between simulation and data
*
*     BSHEM(IBHMC_BDEC2D,IHEM) 1   MC generated 2-d B decay length
*     BSHEM(IBHMC_BDEC,IHEM)   1   MC generated 3-d B decay length
*     BSHEM(IBHMC_DDEC,IHEM)   1   MC generated 3-d Dbar decay length
*     BSHEM(IBHMC_DPDEC,IHEM)  1   MC generated 3-d D decay length
*     BSHEM(IBHMC_BV,IHEM)     3   MC generated B vertex
*     BSHEM(IBHMC_DV,IHEM)     3   MC generated Dbar vertex
*     BSHEM(IBHMC_DPV,IHEM)    3   MC generated D vertex
*     BSHEM(IBHMC_DST,IHEM)    1   MC D* in hemisphere
*     BSHEM(IBHMC_BP,IHEM)     4   MC B 4-VECTOR
*     BSHEM(IBHMC_BKF,IHEM)    1   MC kf code for weakly decaying B
*     BSHEM(IBHMC_ND,IHEM)     1   MC: # of weakly decaying D
*     BSHEM(IBHMC_OSCB,IHEM)   1   MC flag for oscillating B*
*     BSHEM(IBHMC_BLIFE,IHEM)  1   MC B lifetime
*     BSHEM(IBHMC_BMC,IHEM)1   1   MC B charged decay multiplicity
*     BSHEM(IBHMC_BMTOT,IHEM)  1   MC B charged+neutral decay multiplicity 
*     BSHEM(IBHMC_QTYP,IHEM)   1   MC produced quark flavour
*     BSHEM(IBHMC_PBKF,IHEM)   1   MC kf code for primary B
*     BSHEM(IBHMC_DRSKF,IHEM)  1   MC kf code for right sign D
*     BSHEM(IBHMC_QCDRS,IHEM)  1   MC charged decay mult. of right sign D
*     BSHEM(IBHMC_QNDRS,IHEM)  1   MC charged+neutral decay mult. of RSD
*     BSHEM(IBHMC_DRSP,IH)     1   MC P of the right sign D in the B CMS
*     BSHEM(IBHMC_SLD,IH)      1   MC 1=weak B has decayed to e or mu, 0=not 

\end{verbatim}

3) The track output array where IPART is the track number:
\begin{verbatim}
*     Array Element     Number of           Description
*                        values     
*     -----------------------------------------------------------------
*     BSPAR(IBP_HEM,IPART)   1   hemisphere (or jet) number
*     BSPAR(IBP_IYB,IPART)   1   index of the matching VECSUB track
*     BSPAR(IBP_QUAL,IPART)  1   track quality: 20(if isrt=0) + 10(if from
*                                interaction) + 1(if has ambiguous hits)  
*     BSPAR(IBP_Y,IPART)     1   rapidity with respect to jet 
*     BSPAR(IBP_PRP,IPART)   1   probability to fit primary vertex
*     BSPAR(IBP_PRS,IPART)   1   probability to fit secondary vertex
*     BSPAR(IBP_BRP,IPART)   1   Borisov probability in rphi
*     BSPAR(IBP_BZ,IPART)    1   Borisov probability in z
*     BSPAR(IBP_B3D,IPART)   1   Borisov probability in 3D
*     BSPAR(IBP_BSRP,IPART)  1   Borisov probability in rphi w/r secondary vtx
*     BSPAR(IBP_BSZ,IPART)   1   Borisov probability in z  w/r secodary vtx
*     BSPAR(IBP_BS3D,IPART)  1   Borisov probability in 3D w/r secodary vtx
*     BSPAR(IBP_NET,IPART)   1   net output (-->0 for good fragmentation,
*                                            -->1 for good B decay product)
*     BSPAR(IBP_DNET,IPART)  1   cascade D track net output
*     BSPAR(IBP_NRP,IPART)   1   number of rphi VD hits 
*     BSPAR(IBP_NZ,IPART)    1   number of z VD hits 
*     BSPAR(IBP_KST,IPART)   1   momentum of track in B CMS 
*     BSPAR(IBP_THETST,IPART)1   helicity angle of the track
*     BSPAR(IBP_BSAU,IPART)  1   pointer to BSAURUS internal particle array
*     BSPAR(IBP_SV,IPART)    1   1--> trk in sec. vertex 0--> trk not in SV
*     BSPAR(IBP_TRKE,IPART)  1   Track energy
*     BSPAR(IBP_TRKM,IPART)  1   Track mass
*     BSPAR(IBP_TRKP,IPART)  1   Track momentum
*     BSPAR(IBP_TRKQ,IPART)  1   Track charge
*     BSPAR(IBP_TRKL,IPART)  1   Track length
*     BSPAR(IBP_ERRE,IPART)  1   Error on track energy
*     BSPAR(IBP_DELPP,IPART) 1   Delta(P)/P
*     BSPAR(IBP_DFLBS,IPART) 1   charge/decay flavour correlation net for B_s 
*     BSPAR(IBP_DFLB0,IPART) 1   charge/decay flavour correlation net for B0 
*     BSPAR(IBP_DFLBP,IPART) 1   charge/decay flavour correlation net for B+ 
*     BSPAR(IBP_DFLLB,IPART) 1   charge/decay flavour correlation net for LamB 
*     BSPAR(IBP_FFLBS,IPART) 1   charge/fragm. flavour correlation net for B_s 
*     BSPAR(IBP_FFLB0,IPART) 1   charge/fragm. flavour correlation net for B0 
*     BSPAR(IBP_FFLBP,IPART) 1   charge/fragm. flavour correlation net for B+ 
*     BSPAR(IBP_FFLLB,IPART) 1   charge/fragm. flavour correlation net for LamB
*     BSPAR(IBP_MACK,IPART)  1   MACRIB kaon net output
*     BSPAR(IBP_MACP,IPART)  1   MACRIB proton net output
*
*     BSPAR(IBPMC_KF,IPART)  1   MC Lund KF code
*     BSPAR(IBPMC_KFP,IPART) 1   MC Lund KF parent code
*     BSPAR(IBPMC_TYP,IPART) 1   MC code
*                                0 : unidentified
*                                1 : ordinary fragmentation
*                                2 : fragmentation partner of B-hadron
*                                3 : decay of primary (not weak) C
*                                4 : decay of D*
*                                5 : decay of weak B
*                                6 : decay of excited B
*                                7 : decay of weak C
*                                1x : same as for x but decay of 
*                                     long living particles (e.g. K_l)
*                                2x : same as for x but from 
*                                     hadronic interaction or photo conv.
*                                -/+ : b/bbar quark
*                                      6       : content of primary B
*                                      3,4,5,7 : content of weak B
*
*     BSPAR(IBPMC_KST,IPART)     1 MC momentum of track in MC B CMS
*     BSPAR(IBPMC_THETST,IPART)  1 MC helicity angle of the track
*     BSPAR(IBPMC_TRKP,IPART)    1 truth track momentum
*     BSPAR(IBPMC_TRKE,IPART)    1 truth track energy

\end{verbatim}

4) Other:
\begin{verbatim}
*     Array Element     Number of           Description
*                        values     
*     -----------------------------------------------------------------
*     NTOT= total number of particles
*     NCH= number of charged tracks
*
*     Hemisphere status word ISAURUS
*     ==============================
*     ISAURUS(IHEM) = 0  ! event not processed
*     ISAURUS(IHEM) = 1  ! event accepted by user routine FORUSE
*     ISAURUS(IHEM) = 2  ! rapidity algorithm successful
*     ISAURUS(IHEM) = 3  ! rapidity energy > EYMIN (10 GeV)
*     ISAURUS(IHEM) = 4  ! sec. vertex fit successful 
*     ISAURUS(IHEM) = 5  ! sec. vertex fit with convergence

\end{verbatim}



\clearpage

\begin{thebibliography}{99}
\bibitem{ref:ELEPHANT} 
M. Feindt, C. Kreuter, O. Podobrin, `ELEPHANT reference manual', DELPHI internal note, 96-82 PROG 217 (1996). 
\bibitem{ref:MAMMOTH} 
M. Feindt, W. Oberschulte gen. Beckmann, C. Weiser, 
`How to use the MAMMOTH program', DELPHI internal note, 96-52 PROG 216 (1996).
\bibitem{ref:MACRIB}
Z. Albrecht, M. Feindt, M.Moch, `MACRIB- High Efficiency, High Purity Hadron
Identification for DELPHI', DELPHI internal note, 99-150 RICH 95 (1999).
\bibitem{ref:jetset}
T. Sj\"{o}strand, Computer Physics Communications, {\bf 82} (1994) 74. 
\bibitem{ref:AABTAG}
 G. V. Borisov,
`Lifetime Tag of Events $Z^0 \rightarrow b\bar{b}$ with the DELPHI detector. AABTAG program.',
DELPHI internal note, 94-125 PROG 208 (1994);\\
G. V. Borisov, C. Mariotti, Nucl. Inst. Meth {\bf A372} (1996) 181; \\  
G. V. Borisov, `Combined b-tagging', DELPHI internal note, 97-94 PHYS 716  (1997).
\bibitem{ref:jetnet}
L. L\"{o}nnblad, C. Peterson, T. R\"{o}gnvaldsson, Pattern Recognition
in High Energy Physics with Artificial Neural Networks -  
JETNET 2.0, Computer Physics Communications, {\bf 70 } (1992) 167.

\bibitem{ref:bspec}
DELPHI Collab., `Observation of Orbitally Excited B Mesons', Phys. Lett. {\bf B345} (1995) 598;\\ 
DELPHI Collab., `$B^*$~Production in Z Decays', Zeit. Phys. {\bf C68} (1995) 353; \\ 
M. Feindt, W. Oberschulte gen. Beckmann, O. Podobrin,
`First Observation of the $B^*$~Dalitz Decay $B^*\rightarrow  B e^+e^-$',  DELPHI conference report,
97-99 CONF 81; \\
T. Albrecht, `Neural networks for particle identification and $B_s^{**}$~enrichment at LEP', Karlsruhe
Diploma Thesis, IEKP-KA/99-11.                 

\bibitem{ref:blife}
G. Barker, M. Feindt, C. Haag, `A precise measurement of the $\Bplus$~and $\Bzero$~lifetimes with the DELPHI detector',
 DELPHI conference report, 2000-109 CONF 408. 
\bibitem{ref:bsosc}
T. Allmendinger, G. Barker, M. Feindt, `Fully inclusive search for $\Bs$~oscillations
with BSAURUS', DELPHI conference report, 2000-104 CONF 403.
\bibitem{ref:afb}
K. M\"{u}nich et al., `Determination of the forward-backward asymmetry of b-quarks
using inclusive charge reconstruction and lifetime tagging at LEP I', DELPHI conference report,
2000-102 CONF 401.
\bibitem{ref:fb} M. Feindt, C. Weiser, `A measurement of the branching fractions of
the b-quark into strange, neutral and charged B-mesons', DELPHI conference report,
99-104 CONF 291.
\bibitem{ref:btoddx}
C. Schwanda, `Observation of wrong sign D in B decays and 
measurement of B$\rightarrow$DX', DELPHI conference report, 2000-105 CONF 404.
\end{thebibliography}
\end{document}